\documentclass[11pt]{article}
\usepackage{geometry}
\usepackage{verbatim}
\usepackage{prettyref}
\usepackage{amsmath}
\usepackage{amssymb}
\usepackage{graphicx}
\usepackage{setspace}
\usepackage{esint}
\usepackage{marvosym}
\usepackage{feynmp-auto}
\usepackage{slashed}
\usepackage{bbm}
\usepackage{color}
\definecolor{darkgreen}{rgb}{0,0.5,0}
\definecolor{darkblue}{rgb}{0,0,0.6}
\definecolor{purple}{rgb}{0.4,.2,0.7}
\usepackage[colorlinks=true,citecolor=darkgreen,linkcolor=darkblue,urlcolor=purple]{hyperref}

\newcommand{\vev}[1]{{\left< {#1} \right>}}

\newcommand{\tr}{{\rm tr\,}}

\newcommand{\Tr}{{\rm Tr\,}}

\numberwithin{equation}{section}
\numberwithin{figure}{section}
\numberwithin{equation}{section}
\numberwithin{table}{section}

\usepackage{slashed}

\pdfoutput=1
\usepackage{putex}
\usepackage[vcentermath]{youngtab}
\usepackage{subfig}
\usepackage{lscape}

\usepackage{graphicx}
\usepackage{epstopdf}
\usepackage{enumerate}
\usepackage{cite}
\usepackage{tensor}
\usepackage{slashed}
\usepackage{amsmath}
\usepackage{amssymb}
\usepackage{mathrsfs}

\usepackage{bbm}

\usepackage{hyperref}

\numberwithin{equation}{section}

\newcommand {\be} {\begin {equation}}
\newcommand {\ee} {\end {equation}}

\newcommand {\bes} {\begin {equation*}}
\newcommand {\ees} {\end {equation*}}

\newcommand{\eps}{\epsilon}
\newcommand{\intdd}[1]{\int \frac{d^d #1}{(2\pi)^d}}

\def\CH{{\cal H}}

\newcommand{\beq}{\begin{equation}}
\newcommand{\eeq}{\end{equation}}

\def\be{ \begin{equation} }
\def\ee{ \end{equation} }

\def\Tr{{\textrm{Tr}}}

\def\XXint#1#2#3{{\setbox0=\hbox{$#1{#2#3}{\int}$}
\vcenter{\hbox{$#2#3$}}\kern-.5\wd0}}

\def\Tr{{\textrm{Tr}}}

\def\tr{{\textrm{tr}}}

\def\???th{{\textrm{\,???th}}}
\def \eps {\epsilon}

\def\XXint#1#2#3{{\setbox0=\hbox{$#1{#2#3}{\int}$}
     \vcenter{\hbox{$#2#3$}}\kern-.5\wd0}}

\newcommand{\ignore}[1]{}

\begin{document}

\preprint{PUPT-2496}

\institution{PU}{Department of Physics, Princeton University, Princeton, NJ 08544}
\institution{PCTS}{Princeton Center for Theoretical Science, Princeton University, Princeton, NJ 08544}

\title{On $C_J$ and $C_T$ in the Gross-Neveu and $O(N)$ Models}

\authors{Kenan Diab,\worksat{\PU} Lin Fei,\worksat{\PU} Simone Giombi,\worksat{\PU} Igor R.~Klebanov,\worksat{\PU,\PCTS} Grigory Tarnopolsky\worksat{\PU}  }

\abstract{We apply large $N$ diagrammatic techniques for theories with double-trace interactions to the leading corrections to $C_J$, the coefficient
of a conserved current two-point function, and $C_T$, the coefficient of the stress-energy tensor two-point function. We study in detail two famous conformal field theories in continuous dimensions, the scalar $O(N)$ model and the Gross-Neveu model. For the $O(N)$ model, where the answers for the leading
large $N$ corrections to $C_J$ and $C_T$ were derived long ago using analytic bootstrap, we show that the diagrammatic approach reproduces them correctly. We also carry out a new
perturbative test of these results using the $O(N)$ symmetric cubic scalar theory in $6-\epsilon$ dimensions.
We go on to apply the diagrammatic
method to the Gross-Neveu model, finding explicit formulae for the leading corrections to $C_J$ and $C_T$ as a function of dimension. We check these large $N$ results using regular
perturbation theory for the Gross-Neveu model in $2+\epsilon$ dimensions and the Gross-Neveu-Yukawa model in $4-\epsilon$ dimensions. 
For small values of $N$, we use Pad\' e approximants based on the $4-\epsilon$ and $2+\epsilon$ expansions to
estimate the values of $C_J$ and $C_T$ in $d=3$. For the $O(N)$ model our estimates are close to those found using the conformal bootstrap. For the GN model, our estimates
suggest that, even when $N$ is small, $C_T$ differs by no more than $2\%$ from that in the theory of free fermions.
We find that the inequality $C_T^{\textrm{UV}} > C_T^{\textrm{IR}}$ applies both to the GN
and the scalar $O(N)$ models in $d=3$.
}

\date{}
\maketitle

\tableofcontents

\section{Introduction and Summary}

The essential data characterizing a $d$-dimensional conformal field theory (CFT) includes the scaling dimensions of conformal primary operators and their operator product coefficients 
\cite{Polyakov:1970xd,Polyakov:1974gs}. 
In general, the normalizations of operators may be chosen arbitrarily; therefore, the normalizations of their two-point functions are not physical observables. 
Exceptions to this are provided by the conserved currents: their insertions into correlations functions of other operators are determined by the Ward identitites
which fix the normalizations of the currents. Therefore, the coefficients of the two-point functions of conserved currents are physically meaningful.
The most commonly encountered ones are $C_J$, which refers to the conserved spin-1 currents $J_\mu^a$, $a=1, \ldots {\rm dim} (G)$, associated with a global symmetry of the theory with group $G$, and
$C_T$, which refers to the stress-energy tensor $T_{\mu\nu}$ \cite{Osborn:1993cr}:
\begin{align}
&\langle J_{\mu}^{a}(x_{1}) J_{\nu}^{b}(x_{2})\rangle = C_{J} \frac{I_{\mu\nu}(x_{12})}{(x_{12}^{2})^{d-1}} \delta^{ab}\,,  \\
&\langle T_{\mu\nu}(x_{1}) T_{\lambda \rho}(x_{2})\rangle = C_{T} \frac{I_{\mu\nu,\lambda\rho}(x_{12})}{(x_{12}^{2})^{d}} \,,
\label{JJ-TT}
\end{align}
where 
\begin{align}
&I_{\mu\nu}(x)\equiv \delta_{\mu\nu}-2\frac{x_{\mu}x_{\nu}}{x^{2}}
\,, \notag\\
&I_{\mu\,\nu,\lambda\rho}(x) \equiv \frac{1}{2}(I_{\mu\lambda}(x)I_{\nu\rho}(x)+I_{\mu\rho}(x)I_{\nu\lambda}(x))-\frac{1}{d}\delta_{\mu\nu}\delta_{\lambda\rho}\,.
\end{align}
These quantities have various applications: $C_J$ determines the universal charge or spin conductivity
\cite{Huh:2013vga,Huh:2014eea}; $C_T$ appears in many contexts, including some properties of the R\' enyi  and entanglement entropies 
\cite{Perlmutter:2013gua,Mezei:2014zla}. For example, 
$C_T$ determines the leading response of the entanglement entropy across a sphere to small variations in its shape 
\cite{Mezei:2014zla}; in particular, in $d=3$ it determines its limiting behavior for entangling contours with cusps \cite{Bueno:2015rda}. $C_T$ is also one of the natural measures of the number of degrees of freedom, and in two dimensions it satisfies the famous Zamolodchikov 
theorem \cite{Zamolodchikov:1986gt}. In higher dimensions there are counter-examples to the monotonicity of $C_T$ \cite{Nishioka:2013gza,Cappelli:1990yc,Fei:2014yja}, 
but it is still interesting to study its behavior under RG flow.
  
A number of results about $C_J$ and $C_T$ are available for CFTs in $d>2$ \cite{Cappelli:1990yc,PhysRevB.44.6883,Petkou:1995vu,Huh:2013vga,Huh:2014eea}. 
Of special interest to us is the work by Petkou \cite{Petkou:1995vu}, who used 
large $N$ methods and operator product expansions 
to determine the leading $1/N$ corrections to $C_J$ and $C_T$ for the critical scalar $O(N)$ model with quartic interaction $(\phi^i \phi^i)^2$. 
Defining 
\begin{align}
&C_{J} = C_{J0}\Big(1+\frac{C_{J1}}{N}+\frac{C_{J2}}{N^{2}}+\mathcal{O}(1/N^{3})\Big)\,,\notag\\
&C_{T} = C_{T0}\Big(1+\frac{C_{T1}}{N}+\frac{C_{T2}}{N^{2}}+\mathcal{O}(1/N^{3})\Big)\,,
\end{align}
Petkou found  \cite{Petkou:1995vu}
\begin{align}
& C^{\textrm{O(N)}}_{J1} =-\frac{8(d-1)}{d(d-2)}\eta^{\textrm{O(N)}}_{1}, \label{largeNCJ}\\
& C^{\textrm{O(N)}}_{T1}=-2 \left(\frac{2 \mathcal{C}_{\textrm{O(N)}}(d)}{d +2}+
\frac{d^2+6 d -8}{d  \left(d^2-4\right)}\right) \eta^{\textrm{O(N)}}_{1}\ .  \label{largeNCT}
\end{align}
Here
\begin{equation}
\eta^{\textrm{O(N)}}_{1}=\frac{2  \Gamma (d -2)\sin (\pi  \frac{d}{2} )}{\pi  \Gamma (\frac{d}{2} -2) \Gamma (\frac{d}{2} +1)}
\label{etaoneon}
\end{equation}
is the $1/N$ correction to the dimension of the fundamental scalar field $\phi^i$, and
\begin{equation}
\mathcal{C}_{\textrm{O(N)}}(d) =\psi  (3-\frac{d}{2} )+\psi(d-1)-\psi(1)-\psi  (\frac{d}{2}  )\ ,
\end{equation}
where $\psi(x)=\Gamma'(x)/\Gamma(x)$ is the digamma function. In $d=3$, these results yield
\begin{equation}
\begin{aligned}
&C_J^{\textrm{O(N)}}|_{d=3}=C_{J0}^{\textrm{O(N)}}\left(1-\frac{64}{9\pi^2 N}+\mathcal{O}(1/N^2)\right)\,,\\
&C_T^{\textrm{O(N)}}|_{d=3}=C_{T0}^{\textrm{O(N)}}\left(1-\frac{40}{9\pi^2 N}+\mathcal{O}(1/N^2)\right)\,.
\label{CTCJON-3d}
\end{aligned}
\end{equation}

The critical $O(N)$ model with the quartic interaction $(\phi^i \phi^i)^2$ is weakly coupled in $4-\eps$ dimensions \cite{Wilson:1971dc}, and the
results (\ref{largeNCJ}), (\ref{largeNCT}) agree with the $\eps$ expansions found from conventional perturbation theory \cite{Jack:1983sk, Cappelli:1990yc}.
In recent works \cite{Fei:2014yja, Fei:2014xta, Fei:2015kta} it was shown that, for sufficiently large $N$, the $O(N)$ model has another weakly coupled description in
$6-\epsilon$ dimensions. It involves an additional scalar field $\sigma$ with the action
\begin{equation}
\int d^d x \left (\frac{1}{2}(\partial_\mu\phi^i)^2+ \frac 1 2 (\partial_\mu \sigma)^2 +
 \frac{1}{2}g_1\sigma\phi^i\phi_i + \frac{1}{6}g_2 \sigma^3 \right )\ .
\label{cubicact}
\end{equation} 
In section \ref{section:scalar} we will use this cubic $O(N)$ symmetric theory to develop the $6-\eps$ expansion of $C_J$ and $C_T$, providing additional checks of the large $N$ results  (\ref{largeNCJ}), (\ref{largeNCT}).
In particular, for $d=6$ the large $N$ result (\ref{largeNCT}) 
yields \cite{Fei:2014yja}
\begin{equation}
C_{T1}^{\textrm{O(N)}}|_{d=6} =1\ ,
\end{equation}
which precisely reproduces the contribution of a 6d canonical scalar field.
More generally, in even dimensions $d$, generalizing the arguments leading to (\ref{cubicact}), we expect to find a (non-unitary) free theory 
of $N$ canonical scalars $\phi^i$ and a $\Delta=2$ scalar with local kinetic term $\sim \sigma (\partial^2)^{\frac{d}{2}-2}\sigma$. 
For instance, for $d=8$ this was recently discussed in 
\cite{Gracey:2015xmw}. 
Here
\begin{equation}
C_{T1}^{\textrm{O(N)}}|_{d=8} = -4\, .
\end{equation}
This implies that the ratio of the $C_T$ of a free 4-derivative scalar to that of a canonical scalar is $-4$.  
The value of $C_{T1}^{\textrm{O(N)}}$ for general even $d$ is given in \cite{Stergiou:2015roa} and in eq.~(\ref{CTON-2n}).

In section \ref{section:Fermions} we will derive formulae for $C_J$ and $C_T$ in the $d$-dimensional Gross-Neveu (GN) model \cite{Gross:1974jv}, which has the action
\begin{equation}
S_{\textrm{GN}}=-\int d^d x\left(\bar\psi_i \gamma^{\mu}\partial_{\mu}\psi^i +\frac{g}{2}(\bar\psi_i \psi^i)^2\right)\ .
\label{GNact}\end{equation}
We will take $\psi^i$ with $i=1, 2, \ldots \tilde N$ to be a collection of $\tilde N$ Dirac fermions, and we will 
denote $N=\tilde N {\rm Tr}{\bf 1}$, where ${\rm Tr}{\bf 1}$ is the trace of the identity operator
on the vector space on which the Dirac matrices act. Since 
this factor can be absorbed into the expansion parameter $N$, one may keep it arbitrary in intermediate steps of the calculation, 
and set it to the desired value at the end. For instance, for the case 
of $\tilde N$ 2-component Dirac fermions in $d=3$, one should take ${\rm Tr}{\bf 1}=2$, i.e $N=2\tilde N$. 
In $2\le d\le 4$, it is natural to take $\psi^i$ to be 4-component fermions, i.e. $N=\tilde N {\rm
Tr}{\bf 1}=4\tilde N$. This allows us to 
smoothly connect to the GNY model in $d=4-\epsilon$ described below. The 4-component fermion notation also appears naturally in $d=3$ in 
the condensed matter applications of models involving fermions, see for instance \cite{Pisarski:1984dj, Appelquist:1986fd, Appelquist:1988sr, Scherer:2013pda, Braun:2014wja}. 

The perturbing operator $O(x)=\frac{1}{2}(\bar\psi_i \psi^i)^2$ in (\ref{GNact}) has dimension $\Delta=2(d-1)$ in the
free theory. In $d=2$ the GN model is asymptotically free, while for $d>2$ it is free in the IR and has an interacting UV fixed point (it is unitary for $2<d<4$). 
For this interacting CFT we will find, after lengthy calculations,\footnote{
Besides their intrinsic interest, formulae (\ref{largeNCJ}),  (\ref{largeNCT}), (\ref{GNCJ}),  (\ref{GNCT}) may have applications to the 
higher-spin AdS/CFT dualities which relate the $d$-dimensional $O(N)$ \cite{Klebanov:2002ja} 
or Gross-Neveu models  \cite{Leigh:2003gk,Sezgin:2003pt}
to Vasiliev theories  \cite{Vasiliev:1990en,Vasiliev:1992av} in AdS$_{d+1}$ (for a review, see \cite{Giombi:2012ms}).
}
\begin{align}
&  C^{\textrm{GN}}_{J1} =-\frac{8(d-1)}{d(d-2)}\eta^{\textrm{GN}}_{1}\,, \label{GNCJ}\\
&  C^{\textrm{GN}}_{T1}=
- 4 \eta^{\textrm{GN}}_{1}\left(\frac{\mathcal{C}_{\textrm{GN}}(d)}{d +2}+\frac{(d-2)}{d(d+2)(d-1)}\right)\,,  \label{GNCT} 
\end{align}
where 
\begin{equation}
\eta_{1}^{\textrm{GN}}=\frac{ \Gamma (d -1)(d-2)^2}{4\Gamma (2-\frac{d}{2}) \Gamma (\frac{d}{2} +1) \Gamma (\frac{d}{2})^2}
\end{equation}
is the $1/N$ correction to the dimension of the fundamental fermion field $\psi^i$, and
\begin{equation}
\mathcal{C}_{\textrm{GN}}(d) =\psi(2-\frac{d}{2})+\psi(d-1)-\psi(1)-\psi(\frac{d}{2}) \ .
\end{equation}
In $d=3$, we find
\begin{equation}
\begin{aligned}
&C_J^{\textrm{GN}}|_{d=3}=C_{J0}^{\textrm{GN}}\left(1-\frac{64}{9\pi^2 N}+\mathcal{O}(1/N^2)\right)\,,\\
&C_T^{\textrm{GN}}|_{d=3}=C_{T0}^{\textrm{GN}}\left(1+\frac{8}{9\pi^2 N}+\mathcal{O}(1/N^2)\right)\,.
\label{CTCJGN-3d}
\end{aligned}
\end{equation}

We will derive these results using a large $N$ diagrammatic approach similar to that used in 
\cite{Vasiliev:1975mq,Vasiliev:1981yc,Vasiliev:1981dg,Gracey:1992cp,Huh:2013vga,Huh:2014eea}
(for a review, see \cite{Moshe:2003xn}). We will also use the diagrammatic method to rederive the formulae (\ref{largeNCJ}), (\ref{largeNCT}) for the scalar $O(N)$ model, finding complete agreement with the 
bootstrap method of \cite{Petkou:1995vu}; these calculations are presented in section \ref{cjlargen}. 
The diagrammatic approach has also been used to calculate $C_{J1}$ and $C_{T1}$ in 3-dimensional QED \cite{Huh:2013vga,Huh:2014eea}. 
A paper \cite{Giombi:2016fct}, which is a follow-up to the present one, will use the diagrammatic approach to calculate
the $C_{J1}$ and $C_{T1}$ in $d$-dimensional conformal QED and compare the results with the $\epsilon$ expansions. An important feature of the diagrammatic approach, which we will uncover, is the necessity of a divergent multiplicative ``renormalization" $Z_T$ for the stress-energy tensor
(for the conserved current such a renormalization is not needed). Despite this renormalization, the anomalous dimension of the stress-tensor 
is, of course, exactly zero.

The interacting Gross-Neveu CFT has different perturbative $\epsilon$ expansions near 2 and 4 dimensions. In $2+\eps$ dimensions, where the theory has a weakly coupled UV fixed point, it involves the original GN formulation (\ref{GNact}) with the quartic interaction. There is an alternate, Gross-Neveu-Yukawa (GNY) 
formulation of the theory \cite{Hasenfratz:1991it,ZinnJustin:1991yn} which contains
an additional real scalar field $\sigma$ with a Yukawa coupling to the $\tilde N$ Dirac fermions:
\begin{equation}
S_{\textrm{GNY}} =\int d^d x \left ( -\bar{\psi}_i(\slashed{\partial}+g_1\sigma)\psi^i + \frac{1}{2}(\partial_\mu \sigma)^2
+ \frac{g_2}{24}\sigma^4\, \right ).
\label{GNYact}\end{equation}
 This theory, which may be regarded as the UV completion of the GN model, has a weakly coupled IR fixed point in $d=4-\eps$. Using these tools, we develop the $2+\eps$ and $4-\eps$ expansions of $C_T$ and $C_J$ for the GN. In the large $N$ limit these expansions agree with
 (\ref{GNCJ}) and (\ref{GNCT}), providing their important perturbative checks. In particular, we see that for $d=4$, the large $N$ result (\ref{GNCT}) 
yields
\begin{equation}
C_{T1}^{\textrm{GN}}|_{d=4} =\frac{2}{3}\,,
\end{equation}
which precisely reproduces the contribution of a 4d free scalar field.\footnote{Recall that in dimension $d$ a free scalar 
has $C_T^{\rm sc} = \frac{d}{(d-1)S_d^2}$ and a free fermion $C_T^{\rm fer} = {\rm Tr}{\bf 1} \frac{d}{2 S_d^2}$ \cite{Osborn:1993cr}. In $d=4$, we then have $C_{T}^{\rm sc}/(\tilde N C_{T}^{\rm fer}) = \frac {2} {3N}$.} 
More generally, in even dimensions $d$, generalizing the arguments leading to (\ref{GNYact}), we expect to find a (non-unitary) free theory 
of $\tilde N$ Dirac fermions and a free scalar with $\Delta=1$ and local kinetic term $\sim \sigma (\partial^2)^{\frac{d}{2}-1}\sigma$. For instance, in $d=6$ 
we find 
\begin{equation}
C_{T1}^{\textrm{GN}}|_{d=6} = -2\,,
\end{equation}
which implies that $C_T =-6/S_6^2$ for the 4-derivative scalar field in $d=6$ (in units where $C_T = 6/(5 S_6^2)$ for the ordinary 2-derivative 
scalar). 
The ratio of the $C_T$ of a free $(d-2)$-derivative scalar to that of a canonical scalar  
 is given in all even dimensions in eq.~(\ref{CTGN-2nratio}). Interestingly, it is always an integer. 

Using the $2+\eps$ and $4-\eps$ expansions, in section \ref{padeGN} we carry out two-sided Pad\' e extrapolations and find
estimates for $C_T$ and $C_J$ in $d=3$ for small values of $\tilde N$. The values of $C_T$ we find are typically just $1-2 \%$ above those for the theory of free fermions.  
Our estimates suggest that, as the $d=3$ theory flows from the interacting GN fixed point to the free fermion theory, $C_T$ decreases for all $\tilde N$.
There is a supersymmetric counter-example to the $d=3$ ``$C_T$-theorem'' \cite{Nishioka:2013gza}, but we find that the inequality $C_T^{\textrm{UV}} > C_T^{\textrm{IR}}$ applies both to the GN
and the scalar $O(N)$ models in $d=3$. However, as we discuss in section \ref{GNcjct},
for the $GN$ model with large $\tilde N$  it is violated for $2< d \lesssim 2.3$.

\section{Change of $C_J$ and $C_T$ under Double-Trace Perturbations}
\label{Double-trace}

In this section we work out the general structure of the change in the $C_J$ and $C_T$ coefficients under RG flows in large $N$ theories, which are induced by double-trace operators $O^2$.
Both the critical scalar and the GN model are of this type, and in later sections we will carry out specific calculations for these models. 

Before proceeding, let us introduce some useful notation that we will use in the rest of the paper. To deal efficiently with the tensor structures 
in stress-energy tensor and current correlators, it is convenient to introduce an auxiliary null vector $z^{\mu}$, satisfying
\begin{align}
z^{2}= z^{\mu}z^{\nu}\delta_{\mu\nu}=0\ .
\end{align}
We work in flat $d$-dimensional Euclidean space, so such a null vector is complex, but we will never need to specify an explicit form of $z^{\mu}$. 
It is convenient to define the stress-energy tensor and current projected onto the auxiliary null vector
\begin{equation}
\begin{aligned}
T(x) \equiv z^{\mu}z^{\nu} T_{\mu\nu}\,,\qquad J(x) \equiv z^{\mu} J_{\mu}\,. 
\end{aligned}
\end{equation}
From (\ref{JJ-TT}), we see that the two-point functions of $T$ and $J$ take the simple form
\begin{align}
&\langle T (x)T (0) \rangle= \frac{4C_T}{(x^{2})^{d}}\frac{x_z^4}{x^4} \,, \notag\\
&\langle J^a (x) J^b (0)\rangle = \delta^{ab}\frac{-2C_J}{(x^{2})^{d-1}}\frac{x_z^2}{x^2}\,, \label{eq:cj-def}
\end{align}
where we have introduced the notation  $x_z\equiv  z^\mu x_\mu$.   
Using the Fourier transform 
\begin{align}
&\int \frac{d^{d}p}{(2\pi)^{d}} \frac{e^{i p x}}{(p^{2})^{\alpha}} =\frac{ \Gamma (\frac{d}{2} -\alpha )}{4^{\alpha } \pi ^{\frac{d}{2} }\Gamma (\alpha )} \frac{1}{(x^{2})^{\frac{d}{2}-\alpha}},\,  \\
&\int d^{d}x \frac{e^{-i p x}}{(x^{2})^{\alpha}} = \frac{(4\pi) ^{\frac{d}{2} }  \Gamma (\frac{d}{2} -\alpha )}{4^{\alpha  }\Gamma (\alpha )}\frac{1}{(p^{2})^{\frac{d}{2}-\alpha}} \, , 
\end{align}
we find in momentum space
\begin{align}
&\langle T_{\mu\nu}(p) T_{\lambda \rho}(-p)\rangle =C_{T} \frac{\pi ^{\frac{d}{2} } \Gamma (1-\frac{d}{2} )}{2^{d -2} \Gamma (d +2)}\,   
(p^{2})^{\frac{d}{2}}\tilde{I}_{\mu\nu,\lambda\rho}(p)\,, \notag \\
&\langle J_{\mu}^{a}(p)J_{\nu}^{b}(-p)\rangle =- C_{J} \frac{\pi ^{\frac{d}{2}} \Gamma (2-\frac{d}{2} ) }{2^{d -3} \Gamma (d )}\,(p^2)^{\frac{d}{2} -1} \Pi_{\mu\nu}(p)\delta^{ab}\,,
\end{align}
where $\Pi_{\mu\nu}(p)=\delta_{\mu\nu}-p_{\mu}p_{\nu}/p^{2}$ and
\begin{align}
&\tilde{I}_{\mu\nu,\lambda\rho}(p) \equiv \frac{1}{2} \Pi_{\mu\nu}(p)\Pi_{\lambda\rho}(p)-\frac{d-1}{4}\big(\Pi_{\mu\lambda}(p)\Pi_{\nu\rho}(p)+\Pi_{\mu\rho}(p)\Pi_{\nu\lambda}(p)\big)\,
\ .
\end{align}
Therefore, 
\begin{align}
&\langle T(p)T(-p) \rangle = C_{T} \frac{\pi ^{\frac{d}{2}  } \Gamma (2-\frac{d}{2}  )}{2^{d -2} \Gamma (d +2)}\,\frac{p_{z}^{4}}{(p^{2})^{2-\frac{d}{2} }}\,, \notag\\
&\langle J^{a}(p)J^{b}(-p) \rangle=C_{J} \frac{\pi ^{\frac{d}{2} } \Gamma (2-\frac{d}{2}  ) }{2^{d -3} \Gamma (d )}\,\frac{p_{z}^{2}}{(p^{2})^{2-\frac{d}{2} }}\delta^{ab}\ ,
\end{align}
where $p_{z}\equiv z^{\mu}p_{\mu}$.

Let us consider a general CFT$_0$ in $d$ Euclidean dimensions, and assume that it admits a large $N$ expansion with the usual properties. Given 
a single trace operator $O(x)$ of dimension $\Delta_O$ in the spectrum of the CFT, we can consider the double-trace deformation
\begin{equation}
S_{\lambda} = S_{{\rm CFT}_0}+\lambda \int d^d x O(x)^2\,.
\end{equation}
When $\Delta_O < d/2$, the deformation is relevant and there is a RG flow from CFT$_0$ to a new CFT where $\Delta_O^{\textrm{IR}} = d-\Delta_O +\mathcal{O}(1/N)$ 
\cite{Witten:2001ua,Gubser:2002vv}. When $\Delta_O >d/2$, the deformation is irrelevant, but one may show that there is a large $N$ UV fixed point, where 
$\Delta_O^{\textrm{UV}} =d-\Delta_O +\mathcal{O}(1/N)$, and the RG flow leads to CFT$_0$ in the IR.
A well-known example of the IR fixed point is the scalar $O(N)$ model, i.e.~the theory of $N$ massless scalar fields $\phi^i$ perturbed by the $(\phi^i\phi^i)^2$ operator; we will discuss the calculation of $C_T$ in this theory
in  section \ref{section:scalar}.
A well-known example of the UV fixed point is the Gross-Neveu model (\ref{GNact});
it will be discussed in section \ref{section:Fermions}.
To be definite when writing powers 
of $N$, we will assume below that the unperturbed CFT$_0$ is vector-like, i.e. $C_O \sim  N$ and $\langle TT\rangle_0 \sim N$. 

The $1/N$ expansion in the perturbed CFT may be developed with the aid of a Hubbard-Stratonovich auxiliary field. We may rewrite the perturbed action as
\begin{equation}
S_{\lambda} = S_{{\rm CFT}_0}+\int d^d x \sigma O-\frac{1}{4\lambda} \int d^d x \sigma^2\,.
\label{SHS}
\end{equation} 
The equation of motion of $\sigma$ imposes $\sigma =2\lambda O$ and leads to the original action.
However, by performing the path integral in the CFT$_0$,
one may derive an effective action for $\sigma$. At large $N$, we have
\begin{equation}
\langle e^{-\int d^d x \sigma O} \rangle_0 
\approx  e^{\int d^d x d^d y~\frac{1}{2}\sigma(x)\sigma(y) \langle O(x)O(y)\rangle_0 +\mathcal{O}(\sigma^3)}\,,
\end{equation} 
so the quadratic term in the $\sigma$ effective action is
\begin{eqnarray}
S[\sigma] &=& -\frac{1}{2} \int d^d x d^d y~\sigma(x)\sigma(y) \langle O(x)O(y)\rangle_0-\frac{1}{4\lambda} \int d^d x \sigma^2 \\
&=& -\frac{1}{2}\int \frac{d^dp}{(2\pi)^d} \sigma(p)\sigma(-p)\left(
C_O \frac{(4\pi)^{d/2} \Gamma\left(d/2-\Delta_O\right)}{4^{\Delta_O}\Gamma\left(\Delta_O\right)}(p^2)^{\Delta_O-d/2}+\frac{1}{2\lambda}\right)\,, 
\label{S2sig}
\end{eqnarray}
where we have used
\begin{equation}
\langle O(x) O(y)\rangle_0 = \frac{C_O}{|x-y|^{2\Delta_O}} = 
C_O \frac{(4\pi)^{d/2} \Gamma\left(d/2-\Delta_O\right)}{4^{\Delta_O}\Gamma\left(\Delta_O\right)} 
\int \frac{d^dp}{(2\pi)^d} e^{ip(x-y)}(p^2)^{\Delta_O-d/2}\,.
\end{equation}
When $\Delta_O <d/2$, we see that the second term in (\ref{S2sig}) can be dropped in the IR limit
(and when $\Delta_O>d/2$, it can be dropped in the 
UV limit), and so at the perturbed fixed point we get the two-point function of $\sigma$, at leading order in $1/N$, to be
\begin{equation}
G_{\sigma}(p) = \langle \sigma(p)\sigma(-p)\rangle = -\frac{4^{\Delta_O}\Gamma\left(\Delta_O\right)}{C_O(4\pi)^{d/2} \Gamma\left(d/2-\Delta_O\right)}(p^2)^{d/2-\Delta_O}\equiv 
\tilde C_{\sigma} (p^2)^{d/2-\Delta_O}
\end{equation}
or, in coordinate space,
\begin{equation}
G_{\sigma}(x,y) = \frac{(d/2-\Delta_O)\sin\left((d/2-\Delta_O)\pi\right)\Gamma\left(d-\Delta_O\right)\Gamma\left(\Delta_O\right)}
{\pi^{d+1}C_O|x-y|^{2(d-\Delta_O)}}\equiv \frac{C_{\sigma}}{|x-y|^{2(d-\Delta_O)}}\,.
\label{sig-prop}
\end{equation}
This shows that the scalar operator $\sigma\sim O$ now has dimension $d-\Delta_O +\mathcal{O}(1/N)$. At the perturbed fixed 
point, we may hence omit the last term in (\ref{SHS}) and work with the action
\begin{equation}
S_{\rm crit} = S_{{\rm CFT}_0}+\int d^d x \sigma O\,.
\label{Scrit}
\end{equation}
A $1/N$ diagrammatic expansion can be obtained using this action and the effective $\sigma$ propagator (\ref{sig-prop}) (with the prescription that 
the planar bubble diagrams contributing to $\langle \sigma \sigma \rangle$ should not be included as they are already taken into account by 
the effective propagator). 

The two-point function of the stress-energy tensor may be then computed as
\begin{eqnarray}
&&\langle T(x) T(0)\rangle_{\rm crit}= 
\int D\sigma \langle T(x) T(0)e^{-\int \sigma O}\rangle_{0}  \cr 
&&=\langle T(x)T(0)\rangle_0+\frac{1}{2}\int d^d z_1 d^d z_2 G_{\sigma}(z_1,z_2)\langle T(x) T(0)O(z_1)O(z_2)\rangle_0\\
&&+\frac{1}{2}\int d^d z_1 d^d z_2 d^d z_3 d^d z_4 G_{\sigma}(z_1,z_3)G_{\sigma}(z_2,z_4)\langle T(x) O(z_1)O(z_2) \rangle_0 
\langle T(0) O(z_3)O(z_4) \rangle_0+ \mathcal{O}\left(1/N\right)\,,\nonumber 
\end{eqnarray}
where to obtain the ``Aslamazov-Larkin term" \cite{Aslamasov1968238} in the last line we have used the large $N$ approximation to rewrite the 6-point function as a 
product of 3-point functions. Note that since $C_O \sim N$, both of the contributions above are of order $N^0$. By conformal invariance, we may write 
\begin{equation}
\begin{aligned}
&\frac{1}{2}\int d^d z_1 d^d z_2 G_{\sigma}(z_1,z_2)\langle T(x) T(0)O(z_1)O(z_2)\rangle_0  
= I_{\langle TTOO\rangle} \frac{(x_z)^4}{(x^2)^{d+2}}\,,\\
&\frac{1}{2}\int d^d z_1 \cdots d^d z_4 G_{\sigma}(z_1,z_3)G_{\sigma}(z_2,z_4)\langle T(x) O(z_1)O(z_2) \rangle_0 
\langle T(0) O(z_3)O(z_4) \rangle_0 = I_{\langle TOO\rangle^2}\frac{(x_z)^4}{(x^2)^{d+2}}
\label{I4-I33}
\end{aligned}
\end{equation}
and so
\begin{align}
\langle T(x) T(0)\rangle_{\rm crit} = \left(4 C_{T0}+I_{\langle TTOO\rangle}+I_{\langle TOO\rangle^2}+\mathcal{O}(1/N)\right) \frac{(x_z)^4}{(x^2)^{d+2}}\,.
\label{TTcrit}
\end{align}
Thus, we see that the change in $C_T$ to leading order in $1/N$ receives contributions from both integrated 4-point and 3-point 
functions in the unperturbed CFT. While $\langle TOO\rangle$ has a universal form that only depends on $\Delta_O$ due to the 
conformal Ward identity, the 4-point function $\langle TTOO\rangle$ does not have a universal form. Therefore, unlike the 
sphere free energy \cite{Gubser:2002vv,Diaz:2007an}, 
we do not expect a simple universal formula for the change in $C_T$ that only depends on the dimension of the perturbing operator.

So far we have ignored the issues of regularization, but in fact the result (\ref{TTcrit}) by itself is not well-defined, since 
the contributions $I_{\langle TTOO\rangle}$ and $I_{\langle TOO \rangle^2}$ are divergent and require regularization. The usual dimensional continuation does not work 
in this case, because the vertex in (\ref{Scrit}) is critical for all $d$ within the $1/N$ expansion. One 
may use a simple momentum cutoff, however this makes the integrals hard to compute in general $d$. A regulator that is often employed, 
and which we will use in the paper, is to formally shift the dimension of $\sigma$ by a small parameter $\Delta$ that is taken to zero at 
the end of the calculation \cite{Vasiliev:1975mq, Vasiliev:1981yc, Vasiliev:1981dg, Derkachov:1997ch}. Explicitly, we take the propagator in the regularized theory to be
\begin{equation}
G_{\sigma}(p) = \tilde C_{\sigma} (p^2)^{d/2-\Delta_O-\Delta}\,,\qquad \Delta \rightarrow 0\,.
\end{equation} 
This makes the vertex dimensionful, $S_{\rm vertex} = \mu^{\Delta} \int \sigma O$, where we introduced an arbitrary renormalization scale $\mu$
to compensate dimensions. Then, the integrals (\ref{I4-I33}) in the regularized theory take the form
\begin{equation}
\begin{aligned}
&I_{\langle TTOO\rangle} = \left(x^2\mu^2\right)^{\Delta} \left(\frac{1}{\Delta} I_{\langle TTOO \rangle}^{(1)}+I_{\langle TTOO \rangle}^{(0)}+\mathcal{O}(\Delta) \right)\,,\\
&I_{\langle TOO \rangle^2} = \left(x^2\mu^2\right)^{2\Delta} \left(\frac{1}{\Delta} I_{\langle TOO \rangle^2}^{(1)}+I_{\langle TOO\rangle ^2}^{(0)}+\mathcal{O}(\Delta) \right)\,.
\end{aligned} 
\end{equation}
Importantly, we see that the two contributions carry a different power of the renormalization scale, since they involve two and four vertices 
respectively. Then, we find 
\begin{equation}
\begin{aligned}
 I_{\langle TTOO\rangle }+I_{\langle TOO \rangle^2} = & \frac{1}{\Delta}\left(I_{\langle TTOO\rangle }^{(1)}+I_{\langle TOO\rangle^2}^{(1)}\right) \\
& +\log(\mu^2 x^2)\left(I_{\langle TTOO\rangle }^{(1)}+2 I_{\langle TOO\rangle^2}^{(1)}\right)+I_{\langle TTOO\rangle }^{(0)}+I_{\langle TOO\rangle^2}^{(0)}
+\mathcal{O}(\Delta)\,.
\label{regSum}
\end{aligned} 
\end{equation}
Absence of an anomalous dimension for $T$ requires $I_{\langle TTOO\rangle }^{(1)}+2 I_{\langle TOO\rangle^2}^{(1)}=0$, so that the logarithmic term vanishes. We will 
see in the explicit examples below that this is indeed the case, as expected. However, we see that the $1/\Delta$ pole cannot cancel by itself, 
since it involves a different combination of the coefficients (unless both contributions are finite by themselves, but 
in all examples we studied, this does not appear to be the case). A resolution of this issue is to allow for a divergent ``$Z$-factor" renormalization of 
the stress tensor so that the poles are cancelled
\begin{equation}
T^{{\rm ren}}(x) = Z_T ~T(x)\,,\qquad Z_T = 1+\frac{1}{N}\left(\frac{Z_{T1}}{\Delta}+Z_{T1}' +\mathcal{O}(\Delta)\right)+\mathcal{O}(1/N^{2})\,.
\label{Tren}
\end{equation}
The pole coefficient $Z_{T1}$ is fixed by cancellation of the $1/\Delta$ divergence in (\ref{regSum}). In addition, we will find that a non-trivial 
finite shift $Z_{T1}'$ is required in order for the conformal Ward identity to hold. This peculiar stress tensor ``renormalization" is presumably due 
to the unusual features of the regularized $1/N$ perturbation theory, at least within the regularization scheme we employ. Putting everything together, 
one arrives at the following final answer for the shift in $C_T$ to leading order at large $N$ (recall that $C_{TO}\sim N$):
\begin{eqnarray}
C_T = C_{T0} + \frac{1}{4}\left(I_{\langle TTOO\rangle }^{(0)}+I_{\langle TOO\rangle^2}^{(0)} + \frac{8}{N} C_{T0} Z_{T1}'\right)+\mathcal{O}(1/N)\,.
\end{eqnarray}
As we will see below, the shift proportional to $Z_{T1}'$ is essential for reproducing the result of \cite{Petkou:1994ad}  for the scalar $O(N)$ model, and also for matching the $4-\epsilon$ and $2+\epsilon$ expansions for the GN model. 

One may study in a similar way the current two point function $\langle JJ\rangle$. Assuming for simplicity that the perturbing operator is neutral 
under the symmetry generated by $J$, following analogous steps as above, one ends up with 
\begin{eqnarray}
&&\langle J^a(x) J^b(0)\rangle_{\rm crit}= 
\int D\sigma \langle J^a(x) J^b(0)e^{-\int \sigma O}\rangle_{0}  \cr 
&&= \langle J^a(x)J^b(0)\rangle_0+\frac{\mu^{2\Delta}}{2}\int d^d z_1 d^d z_2 G_{\sigma}(z_1,z_2)\langle J^a(x) J^b(0)O(z_1)O(z_2)\rangle_0+\mathcal{O}(1/N)\,.
\end{eqnarray}
This yields
\begin{equation}
\langle J^a(x)J^b(0)\rangle_{\rm crit} = \delta^{ab}\left(-2 C_{J0}+(x^2\mu^2)^{\Delta}\left(\frac{1}{\Delta} I_{\langle JJOO\rangle}^{(1)}
+I_{\langle JJOO\rangle}^{(0)}+\mathcal{O}(\Delta)\right)\right)\frac{(x_z)^2}{(x^2)^d}\,.
\end{equation}
In this case, since the only contribution is given by the integrated 4-point function, the absence of the anomalous dimension of $J$ requires 
that $I_{\langle JJOO\rangle}^{(1)}=0$. Therefore, no ``$Z$-factor" is needed, at least to this order in the $1/N$ expansion 
(examining the Ward identities for $J$, we will find that a finite shift analogous to the one in (\ref{Tren}) is not needed either).\footnote{One may study a different model where double-trace perturbations include the product $O O^*$ of an operator that is charged under the symmetry associated to $J$ and its conjugate. In this 
case, an Aslamazov-Larkin contribution will be present, and one will need a ``$Z_J$-factor"
analogous to the $Z_T$ discussed above.}
Then, the final result is
\begin{equation}
C_J = C_{J0}-\frac{1}{2}I_{\langle JJOO\rangle}^{(0)}+\mathcal{O}(1/N)\,.
\end{equation}

\section{Scalar $O(N)$ Model}
\label{section:scalar}

\subsection{Scalar with cubic interaction in $6-\epsilon$ dimensions}

In this section, we will consider a theory of $N$ scalar fields $\phi^i$ transforming under an
internal $O(N)$ symmetry group and a scalar $\sigma$ in $6-\epsilon$ dimensions described by the
action (\ref{cubicact}).
Dimensional analysis implies that the interactions are relevant for $d < 6$, so we expect that there
should exist a nontrivial infrared fixed point.  We are interested in the case where $d =
6-\epsilon$.  For small $\epsilon$ and sufficiently large $N$, this fixed point indeed exists, and the coupling constants at
that fixed point have been computed to $\epsilon^{3}$ order by
\cite{Fei:2014yja, Fei:2014xta, Fei:2015kta}.  The answer they obtained  at leading $\eps$-order was:
\begin{equation}
g_{1\star} = \sqrt{\frac{6\epsilon(4\pi)^3}{(N-44)\zeta(N)^2+1}}\zeta (N), \qquad g_{2\star} = \sqrt{\frac{6\epsilon(4\pi)^3}{(N-44)\zeta (N)^2+1}}(1+6\zeta(N))\,,
\label{g1g2star}
\end{equation}
where $\zeta(N)$ is the solution to the cubic equation
\begin{equation}
840\zeta^3-(N-464)\zeta^2+84\zeta+5=0\,,
\end{equation}
which asymptotically tends to $\zeta =N/(840)+\ldots$ at large $N$.\footnote{The other roots correspond to fixed points with 
unstable directions (in the RG sense) that are not related to the $O(N)$ theory with
$(\phi^i\phi^i)^2$ interaction.} Such a solution exists 
for $N> 1038$ \cite{Fei:2014yja}. 

The solution for the fixed point couplings (\ref{g1g2star}) is valid for finite $N$, but its explicit form is somewhat cumbersome. 
Expanding in powers of $1/N$, one gets:
\begin{align}
g_{1\star} &= \sqrt{\frac{6\epsilon(4\pi)^3}{N}}\left(1+\frac{22}{N} + \frac{726}{N^2} +
\dots\right)\,, \label{eq:scalar-g1}\\
g_{2\star} &= 6\sqrt{\frac{6\epsilon(4\pi)^3}{N}}\left(1+\frac{162}{N} + \frac{68766}{N^2} +
\dots\right)\,. \label{eq:scalar-g2}
\end{align}
Our goal is to compute the two-point function of the stress-energy tensor and of a conserved
spin-$1$ current at order $\epsilon$, and in particular compare with
the large $N$ results (\ref{largeNCJ}), (\ref{largeNCT}) obtained in \cite{Petkou:1994ad}. 

\begin{figure}[h!]
   \centering
\includegraphics[width=16cm]{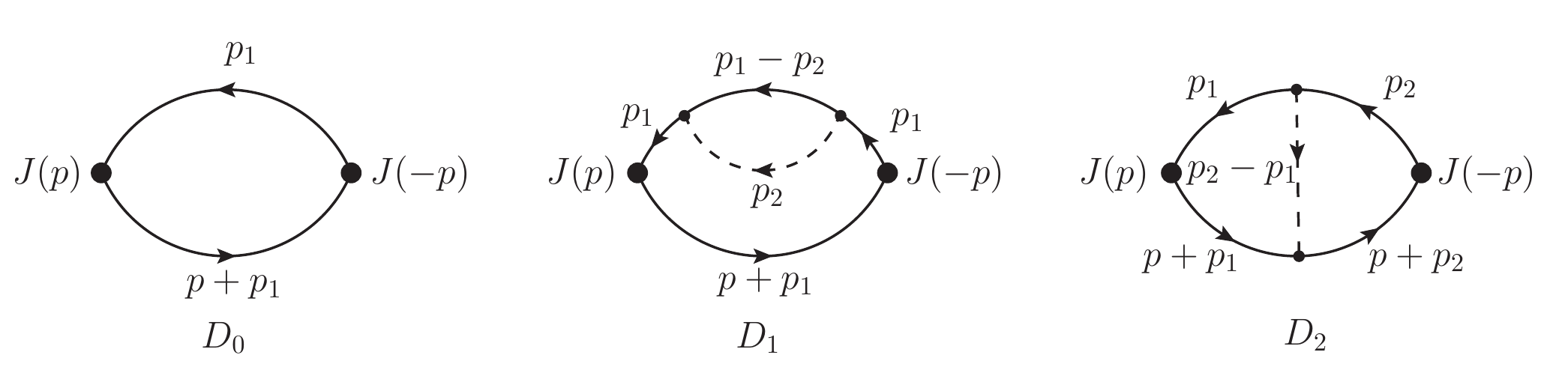}
\caption{Diagrams for $C_{J}$ up to order $\epsilon$.}
\label{CJepsdiags}
\end{figure}

The spin-$1$ current corresponding to the global $O(N)$ symmetry of the model is given by
\begin{equation}
J^a_\mu(x) = \phi^it^a_{ij}\partial_\mu\phi^j\,. 
\end{equation}
Here, the matrices $t^a$ are the generators of the internal $O(N)$ symmetry group.  Since the two
point function of this current is proportional to $\delta_{ab}$, we may as well pick a convenient
generator.  We will choose:
\begin{equation}
J(x)= z^\mu J_\mu(x) = z^\mu (\phi^1\partial_\mu\phi^2 - \phi^2\partial_\mu\phi^1)\,.
\end{equation}
To the first non-trivial order in the $\epsilon$-expansion, we find
\begin{align}
\vev{J(p)J(-p)} = D_{0}+D_1+D_2+\mathcal{O}(\eps^2)\,  ,
\end{align}
where the necessary diagrams are shown in Fig. \ref{CJepsdiags}.
The solid lines here denote the $\phi$ propagators, the dotted line the $\sigma$
propagators, and the arrows here simply denote the flow of momentum. 
 The explicit integrands for $D_0$, $D_1$, $D_2$ and the result of the integrations are given in Appendix \ref{ape}. 
After Fourier transforming to position space and dividing by the free field contribution $D_0$, \footnote{It 
is important to divide by $D_0$ and take the $\epsilon\rightarrow 0$ {\it after} performing the Fourier transform. This 
is because the leading
order behavior of the $\Gamma$ functions arising from the Fourier transform (which are
regularized by expanding in $d=6-\epsilon$) are proportional to $\epsilon/2$ for the second-order
diagrams $D_1$ and $D_2$, but to $\epsilon$ for the one-loop diagram $D_0$. Effectively, this results in an ``enhancement" 
of $D_1$ and $D_2$ by a factor of $2$ relative to $D_0$.}
we obtain the result
\begin{align}
\frac{C^{\textrm{O(N)}}_{J}}{C_{J,\textrm{free}}^{\textrm{O(N)}}} = 1+\frac{D_1+D_2}{D_0}=1+ \left(-\frac{5}{1152\pi^3}+\mathcal{O}(\eps)\right)g_{1\star}^2 = 1+
\epsilon\left(-\frac{5}{3N}-\frac{220}{3N^2} 
+ \mathcal{O}\left(\frac{1}{N^3}\right)\right)+\mathcal{O}(\eps^2)\,, \label{O6modelCJ1}
\end{align}
where in the second step we have substituted the large $N$ expansion (\ref{eq:scalar-g1}) of the critical coupling. One 
may check that this precisely agrees with the $6-\eps$ expansion (\ref{Petkoucjeps}) of the large $N$ result (\ref{largeNCJ}) obtained in \cite{Petkou:1994ad}.

Let us now move to the calculation of $C_T$. The stress-energy tensor may be split into its $\phi$ and $\sigma$ contributions, 
$T=z^\mu z^\nu T_{\mu\nu}= T_{\phi} + T_{\sigma}$, where
\begin{align}
T_{\phi} =& z^\mu z^\nu \Big(\partial_\mu \phi^i \partial_\nu \phi^i - \frac{1}{4}
\frac{d-2}{d-1}\partial_\mu \partial_\nu (\phi^i\phi^i)\Big)
\ ,\nonumber \\
T_\sigma= &  z^\mu z^\nu
\Big( \partial_\mu \sigma \partial_\nu \sigma - \frac{1}{4} \frac{d-2}{d-1}\partial_\mu
\partial_\nu(\sigma^2) \Big)
\,.
\label{tsigma}
\end{align}
Here we have dropped terms proportional to $\delta_{\mu\nu}$ (including terms involving the interactions), since we work with 
the projected stress tensor along the null vector $z^{\mu}$. 
\begin{figure}[h!]
   \centering
\includegraphics[width=15cm]{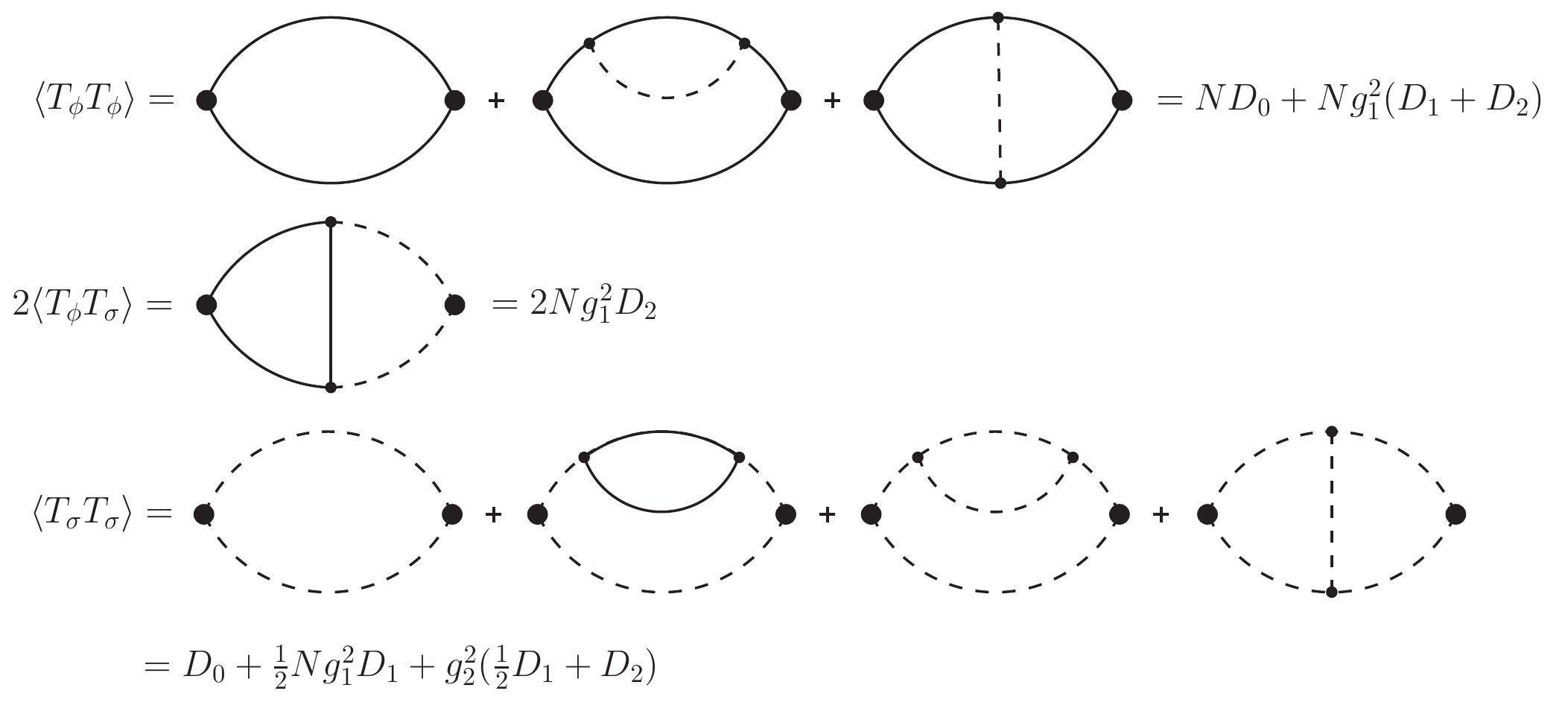}
\caption{Diagrams for $C_{T}$ up to order $\eps$.}
\label{CTscalareps}
\end{figure}

We may write $\vev{T(p) T (-p)} =
\vev{T_\phi (p) T_\phi (-p)} + \vev{T_{\sigma} (p) T_{\sigma} (-p) } + 2\vev{T_{\phi} (p) T_{\sigma} (-p)}$, 
and the diagrams contributing to each term are shown in figure \ref{CTscalareps}. The explicit integrands and 
results are given in Appendix \ref{ape}.
Putting everything together, the final result is:
\begin{align}
\frac{C^{\textrm{O(N)}}_{T}}{C_{T,\textrm{free}}^{\textrm{O(N)}}} &= 1 + \frac{1}{N} + \left(\frac{\frac{1}{2}D_1+D_2}{N D_0}\right)(3 Ng_{1\star}^2 + g_{2\star}^2) \notag\\
&= 1+\frac{1}{N}+\left(-\frac{7}{4608\pi^3}+\mathcal{O}(\eps) \right)\frac{3 N g_{1\star}^2 + g_{2\star}^2}{N} \notag \\
&= 1+ \frac{1}{N}+ \epsilon\left(-\frac{7}{4N} - \frac{98}{N^2} - \frac{10192}{N^3} +
\mathcal{O}\left(\frac{1}{N^4}\right)\right) + \mathcal{O}(\eps^2)\,. \label{O6modelCT1}
\end{align}
Again, we find that this agrees with the $6-\epsilon$ expansion (\ref{Petkoueps}) of Petkou's result (\ref{largeNCT}).

\subsection{$1/N$ expansion}

The $1/N$ expansion of the $O(N)$ model can be developed using the Hubbard-Stratonovich transformation, as reviewed 
in section \ref{Double-trace}. After introducing the Hubbard-Stratonovich auxiliary field 
and dropping the term quadratic in $\sigma$ in the IR limit, we effectively have the following
action, expressed in terms of bare fields: 
\begin{equation}
S_{\textrm{crit scal}}=\frac{1}{2}\int d^{d}x \Big( (\partial \phi_{0}^i)^2+\frac{1}{\sqrt{N}}\sigma_{0} \phi_{0}^i \phi_{0}^i \Big)\,.
\end{equation}
The propagator of the $\phi_{0}^i$ field reads 
\begin{equation}
\langle \phi_{0}^{i}(p) \phi_{0}^{j}(-p) \rangle_{0} =\delta^{ij}/p^{2}.
\end{equation}
After integrating over the fundamental fields $\phi_{0}^i$, the auxiliary field $\sigma_{0}$ develops a non-local kinetic term with an effective propagator 
\begin{align}
\left<\sigma_{0}(p)\sigma_{0}(-p) \right>_{0}&=\tilde{C}_{\sigma0}/(p^2)^{\frac{d}{2}-2+\Delta}\,,
\label{sigsig-ON}
\end{align}
where 
 \begin{align}
\tilde{C}_{\sigma0 } &\equiv 2^{d+1}(4\pi)^{\frac{d-3}{2}}\Gamma\Big( \frac{d-1}{2} \Big)\sin \big(\frac{\pi d}{2}\big)\,, \label{csigmatildeON}
\end{align}
and we have already introduced a regulator $\Delta$ \cite{Vasiliev:1975mq, Vasiliev:1981yc, Vasiliev:1981dg, Derkachov:1997ch}, 
as described in section \ref{Double-trace}. This regulator essentially works analogously to $\epsilon$ in dimensional regularization, 
but there are some subtleties, which we will discuss in this section. 
 
In order to cancel the divergences as $\Delta\rightarrow 0$ we have to renormalize the bare fields $\phi_{0}$ and $\sigma_{0}$: 
\begin{align}
\phi = Z_{\phi}^{1/2} \phi_{0}, \quad  \sigma = Z_{\sigma}^{1/2}\sigma_{0}\,,
\end{align}
where $Z_{\phi}$ and $Z_{\sigma}$ have only poles in $\Delta$ (using a ``minimal subtraction" scheme), and read
\begin{align}
Z_{\phi} = 1+\frac{1}{N}\frac{Z_{\phi 1}}{\Delta}+\mathcal{O}(1/N^{2}), \qquad 
Z_{\sigma} =1+\frac{1}{N}\frac{Z_{\sigma 1}}{\Delta}+\mathcal{O}(1/N^{2})\,.
\end{align}
The full propagators of the renormalized fields in momentum space read
\begin{align}
\langle \phi^{i}(p) \phi^{j}(-p)\rangle = \delta^{ij} \frac{\tilde{C}_{\phi}}{(p^{2})^{\frac{d}{2}-\Delta_{\phi}}}, \qquad 
\langle \sigma(p) \sigma(-p)\rangle =  \frac{\tilde{C}_{\sigma}}{(p^{2})^{\frac{d}{2}-\Delta_{\sigma}}}\,,
\end{align}
where we introduced anomalous dimensions $\Delta_{\phi}$ and $\Delta_{\sigma}$ and two point constants $\tilde{C}_{\phi}$ and $\tilde{C}_{\sigma}$ 
in the momentum space. All of them can be represented as series in $1/N$:
\begin{align}
\Delta_{\phi} = \frac{d}{2}-1+\eta^{\textrm{O(N)}}\,, \qquad \Delta_{\sigma}=2-\eta^{\textrm{O(N)}}-\kappa^{\textrm{O(N)}}\,,
\label{dphidsigma}
\end{align}
where $\eta^{\textrm{O(N)}}=\eta^{\textrm{O(N)}}_{1}/N+\eta^{\textrm{O(N)}}_{2}/N^{2}+\mathcal{O}(1/N^{3})$, $\kappa^{\textrm{O(N)}}=\kappa^{\textrm{O(N)}}_{1}/N+\kappa^{\textrm{O(N)}}_{2}/N^{2}+\mathcal{O}(1/N^{3})$ and 
\begin{align}
\tilde{C}_{\phi} = 1 +\frac{\tilde{C}_{\phi1}}{N}+\frac{\tilde{C}_{\phi 2}}{N^{2}}+\mathcal{O}(1/N^{3})\,, \qquad 
\tilde{C}_{\sigma} = \tilde{C}_{\sigma 0} +\frac{\tilde{C}_{\sigma 1}}{N}+\frac{\tilde{C}_{\sigma 2}}{N^{2}}+\mathcal{O}(1/N^{3}) \,.\label{cphicsigmabos}
\end{align}

Recalling that we may drop all terms proportional to $\delta_{\mu\nu}$ since $z^{\mu}$ is null, the
stress-energy tensor and the $O(N)$ current are:
\begin{align}
T (x) &= z^{\mu}z^{\nu} \left (\partial_{\mu}\phi^{i}_{0}\partial_{\nu}\phi^{i}_{0}-
\frac{1}{4}\frac{d-2}{d-1}\partial_{\mu}\partial_{\nu}(\phi_{0}^{i}\phi_{0}^{i}) \right )\,, \notag\\
J^{a}(x) &= z^\mu \phi_{0}^{i}(t^{a})^{ij}\partial_{\mu}\phi_{0}^{j} \, .
\end{align}
In momentum space: 
\begin{align}
T(p)&= \frac{1}{2}\int \frac{d^{d}p_{1}}{(2\pi)^{d}} (2p_{1z}(p_{1z}+p_{z})+cp_{z}^{2}  )\phi_{0}^{i}(p+p_{1})\phi_{0}^{i}(-p_{1})\,, \notag\\
J^{a}(p)&=\frac{1}{2}\int \frac{d^{d}p_{1}}{(2\pi)^{d}} i(2p_{1z}+p_{z} )\phi_{0}^{i}(-p_{1})(t^{a})^{ij}\phi_{0}^{j}(p+p_{1})\,,
\end{align}
where $c\equiv \frac{d-2}{2(d-1)}$.
\begin{figure}[h!]
   \centering
\includegraphics[width=15cm]{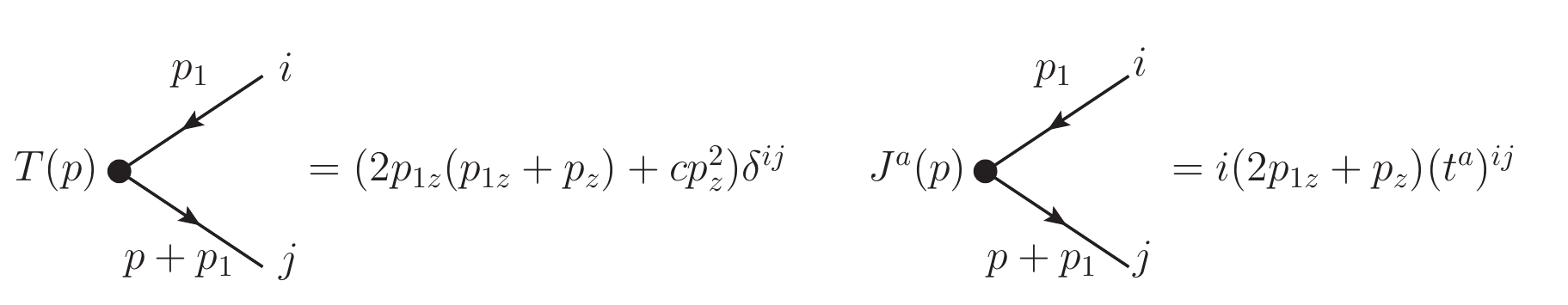}
\caption{Momentum space Feynman rules for $T(p)$ and $J^{a}(p)$.}
\label{TJONdiag}
\end{figure}

\begin{figure}[h!]
   \centering
\includegraphics[width=4cm]{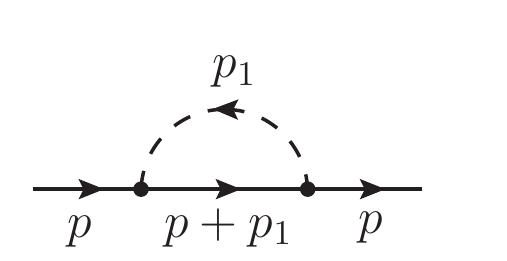}
\caption{One loop correction to the $\langle \phi^{i}(p) \phi^{j}(-p)\rangle$ propagator.}
\label{phiphidiagON}
\end{figure}

For the Ward identity calculation performed below, we will first need to find $\tilde{C}_{\phi 1}$, $\eta_{1}$ and $Z_{\phi 1}$. 
To compute them we have to consider the one loop diagram for the renormalization of the $\langle \phi \phi \rangle$-propagator,  
see figure \ref{phiphidiagON}.  

\noindent Computing this diagram, we find the result (\ref{etaoneon}), and
\begin{align}
\tilde{C}_{\phi 1}=-\frac{1}{2}(3d^{2}-12d +8)\frac{ \sin (\frac{ \pi d}{2} ) \Gamma (d -2)}{\pi  \Gamma (\frac{d}{2} +1)^2}\,.\label{eta1}
\end{align}

As discussed in section \ref{Double-trace}, in order to cancel $1/\Delta$ poles in correlation functions involving $T$ and $J$, 
one may introduce ``$Z_{T}$" and ``$Z_{J}$" factors as 
\begin{align}
T_{\mu\nu}^{\textrm{ren}} = Z_{T} T_{\mu\nu}, \quad J_{\mu}^{\textrm{ren}, a} = Z_{J}J_{\mu}^{a}\,,
\end{align}
which admit the following decomposition:
\begin{align}
Z_{T} = 1+ \frac{1}{N}\Big(\frac{Z_{T1}}{\Delta}+Z'_{T1}\Big)+\mathcal{O}(1/N^{2}), \quad 
Z_{J} = 1+ \frac{1}{N}\Big(\frac{Z_{J1}}{\Delta}+Z'_{J1}\Big)+\mathcal{O}(1/N^{2})\,.\label{zTzJ}
\end{align}
The explicit form of these factors can be obtained from Ward identities. Let us consider $Z_{T}$
first. For this, we can 
examine the three point function $\langle T^{\textrm{ren}}_{\mu\nu} \phi^{i}\phi^{j} \rangle $. 
Its structure is fixed by conformal symmetry and current conservation to be \cite{Osborn:1993cr}
\begin{align}
\langle T^{\textrm{ren}}_{\mu\nu}(x_{1}) \phi^{i}(x_{2})\phi^{j}(x_{3}) \rangle =\frac{-C_{T\phi \phi}}{(x_{12}^{2}x_{13}^{2})^{\frac{d}{2} -1} 
(x_{23}^{2})^{\Delta_{\phi}-\frac{d}{2} +1}} \Big((X_{23})_{\mu}(X_{23})_{\nu}-\frac{1}{d}\delta_{\mu\nu}(X_{23})^{2}\Big)\delta^{ij}\,, \label{Tphiphi}
\end{align}
where 
\begin{align}
(X_{23})_{\nu} = \frac{(x_{12})_{\nu}}{x_{12}^{2}} - \frac{(x_{13})_{\nu}}{x_{13}^{2}} \,.
\end{align}
The structure constant $C_{T\phi \phi}$ is not arbitrary and is related to $C_{\phi}$ by the Ward identity.  To show this, 
we note that for the infinitesimal scaling transformation $\varepsilon_{\nu}=\varepsilon x_{\nu}$:
\begin{align}
\langle \delta_{\varepsilon} \phi^{i}(x_{2}) \phi^{j}(x_{3})\rangle =- \varepsilon\int d^{d}\Omega\, r^{d-2}r_{\mu}r_{\nu} \langle T^{\textrm{ren}}_{\mu\nu}(x_{1}) \phi^{i}(x_{2})\phi^{j}(x_{3})\rangle,
\end{align}
where $r= |x_{1}-x_{2}|$ and $\delta_{\varepsilon} \phi^{i}(x) = \varepsilon(\Delta_{\phi}+x_{\mu}\partial_{\mu}) \phi^{i}(x)$. Perfoming the integral in the limit $r\to 0$ we find 
\begin{align}
C_{T \phi \phi } = \frac{1}{S_{d}} \frac{d\Delta_{\phi}}{d-1}C_{\phi}\,, \label{TphiphiCphi}
\end{align}
where $S_{d}\equiv 2\pi^{d/2}/\Gamma(d/2)$ and $C_{\phi}$ is the two-point function constant in
coordinate space; it is related to $\tilde{C}_{\phi}$ in momentum space (\ref{cphicsigmabos}) through
the Fourier transform\footnote{Notice that it is important that we define $\tilde{C}_{\sigma 0}$ in
(\ref{csigmatildeON}) in momentum space. Thus, $C_{\sigma}$ in the coordinate space will depend on $\Delta$. This dependence will affect the loop calculations in coordinate space.}. 
Taking the Fourier transform of (\ref{Tphiphi}) and using (\ref{TphiphiCphi}) we find\footnote{Here
we fix some field, say $\phi =\phi^{1}$, and do not write the $O(N)$-index explicitly.} 
\begin{align}
\langle T^{\textrm{ren}}(0) \phi(p)\phi(-p) \rangle= (d-2\Delta_{\phi})\tilde{C}_{\phi} \frac{p_{z}^{2}}{(p^{2})^{\frac{d}{2} -\Delta_{\phi}+1}}\,, \label{TphiphiWard}
\end{align}
where we took the stress-energy tensor at zero momentum for simplicity. 
Now, to fix $Z_{T}$ we compute (\ref{TphiphiWard}) using a direct Feynman diagram calculation:
\begin{equation}
\begin{aligned}
\langle T^{\textrm{ren}}(0) \phi(p)\phi(-p) \rangle = Z_{T} Z_{\phi} \langle T(0)
\phi_{0}(p)\phi_{0}(-p) \rangle\, .
\label{TphiphiTphiphi}
\end{aligned}
\end{equation}
To $1/N$ order we have four diagrams
\begin{align}
&\langle T(0) \phi_{0}(p)\phi_{0}(-p) \rangle  = D_{0}+D_{1}+D_{2}+D_{3} +\mathcal{O}(1/N^{2})\,,
\end{align}
which are shown in figure \ref{Tphiphidiagbos} and  given explicitly in  Appendix \ref{ape}.
\begin{figure}[h!]
   \centering
\includegraphics[width=16cm]{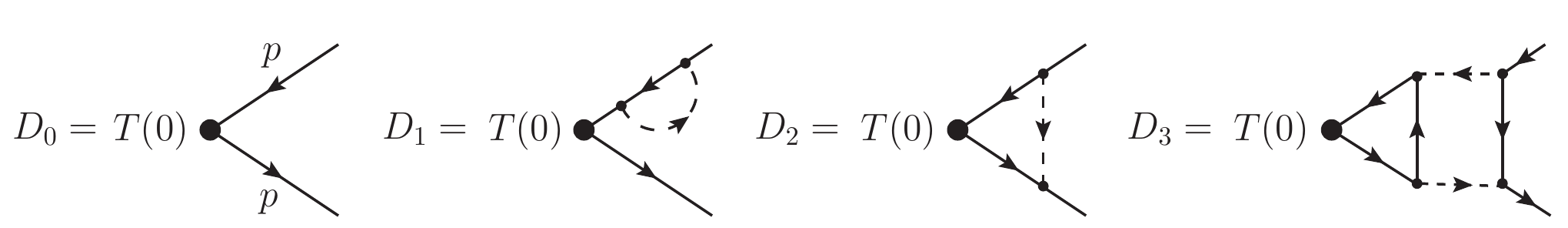}
\caption{Diagrams contributing to $ \langle T(0) \phi_{0}(p)\phi_{0}(-p) \rangle$ up to order $1/N$.}
\label{Tphiphidiagbos}
\end{figure}

\noindent Computing these diagrams and using  (\ref{TphiphiWard}) and  (\ref{TphiphiTphiphi}), we find 
\begin{align}
Z_{T1} =\frac{2\eta_{1}^{\textrm{O(N)}}}{d+2},\qquad Z'_{T1}= \frac{8\eta_{1}^{\textrm{O(N)}}}{(d +2) (d -4)} \,, \label{zTON1}
\end{align}
where $\eta_{1}^{\textrm{O(N)}}$ is given in (\ref{etaoneon}). These renormalization constants will
be of  great importance for the $C_{T}$ calculation. 

\begin{figure}[h!]
   \centering
\includegraphics[width=13.5cm]{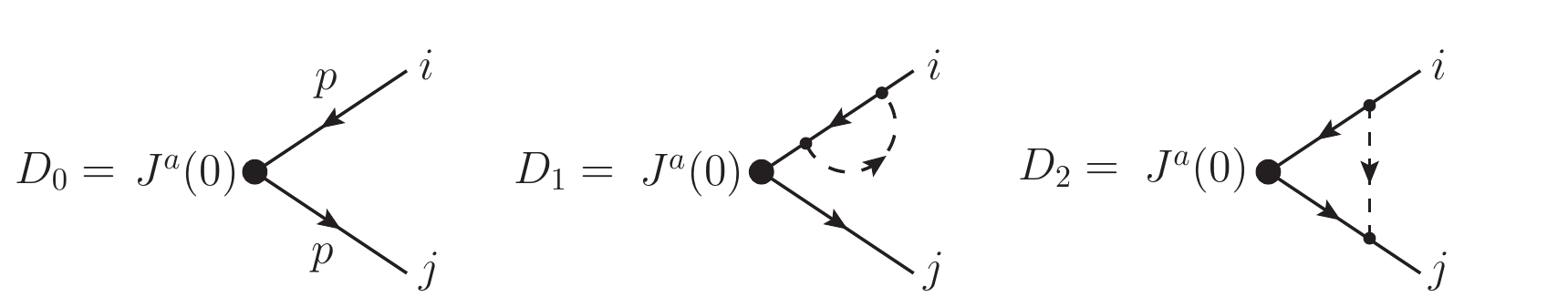}
\caption{Diagrams contributing to $ \langle J^{a}(0) \phi^{i}_{0}(p)\phi^{j}_{0}(-p) \rangle$ up to order $1/N$.}
\label{Jphiphidiag}
\end{figure}

To find $Z_{J}$, we again consider the three-point function $\langle J^{a}_{\nu} \phi^{i}\phi^{j}\rangle$, which is fixed by conformal invariance 
and current conservation \cite{Osborn:1993cr}
\begin{align}
\langle J^{\textrm{ren}, a}_{\mu}(x_{1}) \phi^{i}(x_{2})\phi^{j}(x_{3})\rangle = \frac{C_{J\phi\phi}}{(x_{12}^{2}x_{13}^{2})^{\frac{d}{2} -1}(x_{23}^{2})^{\Delta_{\phi}-\frac{d}{2} +1}}(X_{23})_{\mu} (t^{a})^{ij} \label{Jphiphi}\, ,
\end{align}
and again the structure constant $C_{J\phi \phi}$ is exactly related to $C_{\phi}$ by the Ward identity. 
To show this, we perform an infinitesimal $O(N)$ rotation of fields $\delta_{\varepsilon}\phi^{i} =
\varepsilon (t^{a})^{ik}\phi^{k}$, and we get
\begin{align}
\langle \delta_{\varepsilon} \phi^{i}(x_{2}) \phi^{j}(x_{3})\rangle= \varepsilon\int d^{d}\Omega r^{d-2}r_{\mu}\langle J_{\mu}^{\textrm{ren}, a}(x_{1})\phi^{i}(x_{2})\phi^{j}(x_{3})\rangle\,,
 \end{align}
where $r=|x_{1}-x_{2}|$. Using (\ref{Jphiphi}) and performing the integral in the limit $r\to 0$ we find
\begin{align}
C_{J \phi \phi} = \frac{1}{S_{d}}C_{\phi}\,. \label{JphiphiCphi}
\end{align}
Taking the Fourier transform of (\ref{Jphiphi}) and using (\ref{JphiphiCphi}),  we get
\begin{align}
\langle J^{\textrm{ren},a}(0) \phi^{i}(p)\phi^{j}(-p) \rangle= i(d-2\Delta_{\phi})\tilde{C}_{\phi} \frac{p_{z}}{(p^{2})^{\frac{d}{2} -\Delta_{\phi}+1}}(t^{a})^{ij}\,, \label{JphiphiWard}
\end{align}
where again we took the current at zero momentum to simplify the calculation. 
Now to fix $Z_{J}$, we can compute (\ref{JphiphiWard}) by a direct perturbative calculation
\begin{align}
\langle J^{\textrm{ren},a}(0) \phi^{i}(p)\phi^{j}(-p) \rangle = Z_{J} Z_{\phi} \langle J^{a}(0) \phi^{i}_{0}(p)\phi^{j}_{0}(-p) \rangle\,, \label{JphiphiJphiphi}
\end{align}
and to $1/N$ order we have three diagrams
\begin{align}
\langle J^{a}(0) \phi^{i}_{0}(p)\phi^{j}_{0}(-p) \rangle  = D_{0}+D_{1}+D_{2} +\mathcal{O}(1/N^{2})\,,
\end{align}
which are shown in figure \ref{Jphiphidiag}.
Computing these diagrams and using (\ref{JphiphiWard}) and (\ref{JphiphiJphiphi}),  we find 
\begin{align}
Z_{J} =1+\mathcal{O}(1/N^{2}) \,. \label{zJ}
\end{align}
Therefore $Z_{J}$ is trivial to order $1/N$ and will not affect the $C_{J1}$ calculation.

\subsection{Calculation of $C^{\textrm{O(N)}}_{J1}$ and $C^{\textrm{O(N)}}_{T1}$}
\label{cjlargen}
\begin{figure}[h!]
   \centering
\includegraphics[width=17cm]{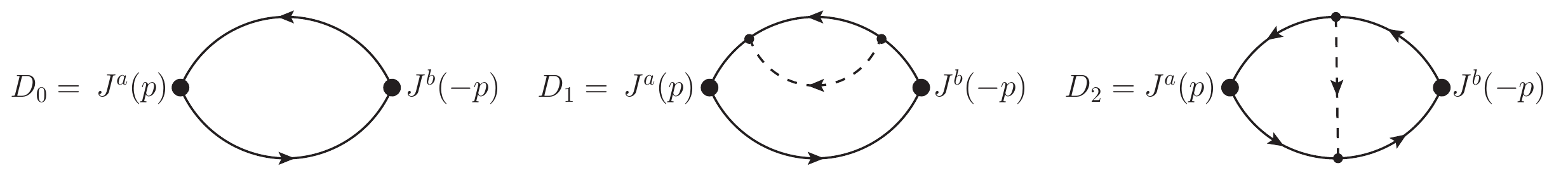}
\caption{The diagrams contributing to $\langle J^{a}(p) J^{b}(-p)\rangle$ up to order $1/N$.}
\label{CJdiagsbosons}
\end{figure}

\begin{figure}[h!]
   \centering
\includegraphics[width=7cm]{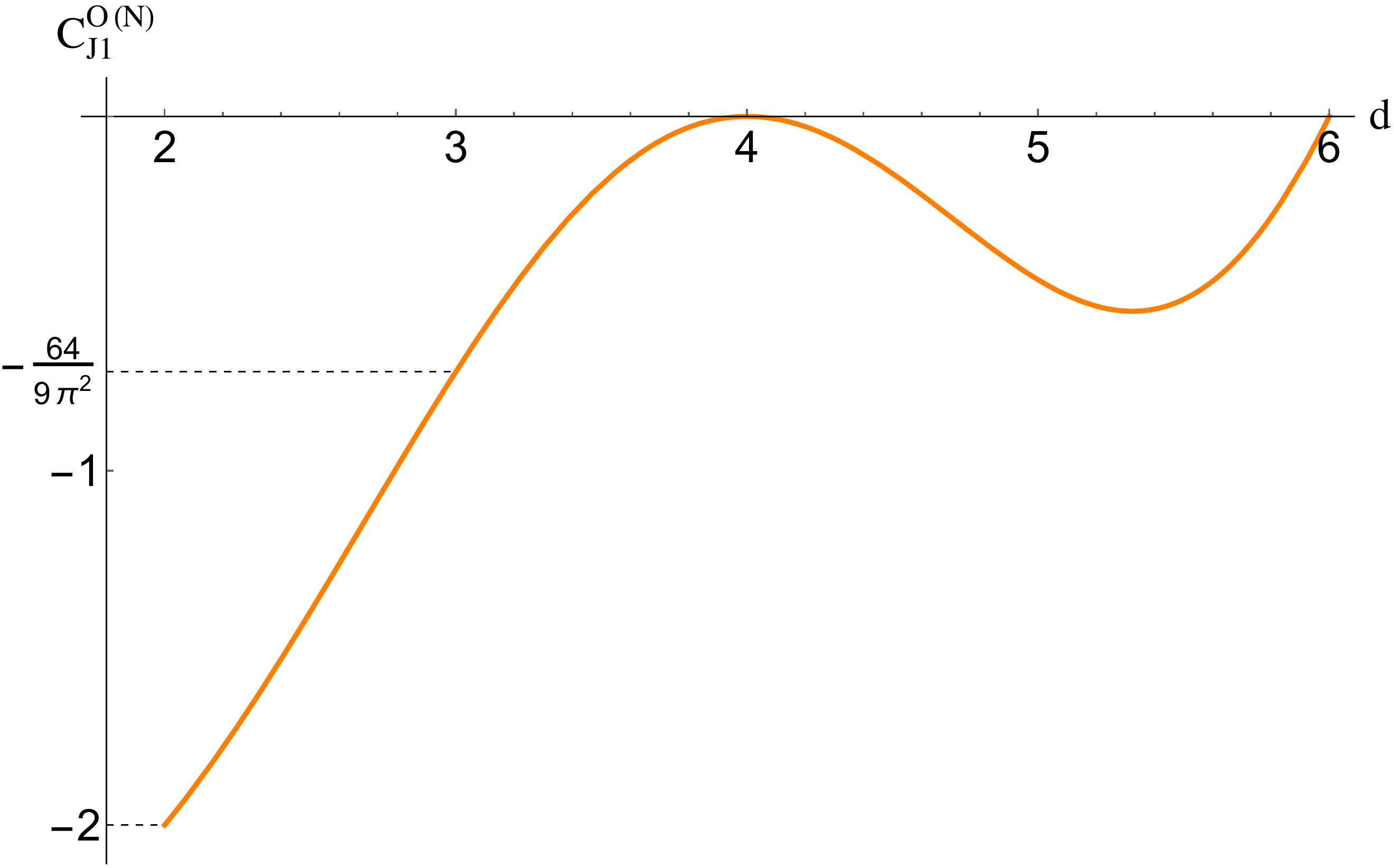}
\caption{Plot of $C^{\textrm{O(N)}}_{J1}$, which is negative throughout the range $2<d<6$.}
\label{CJgraphbos}
\end{figure}

There are three diagrams contributing to the $1/N$ correction to $C_J$, depicted in figure \ref{CJdiagsbosons}. 
The current two-point function up to order $1/N$ is then
\begin{align}
\langle J^{a}(p)J^{b}(-p)\rangle = D_{0}+D_{1}+D_{2}+\mathcal{O}(1/N^{2})\,.
\end{align}
The sum of $D_1$ and $D_2$ corresponds to the contribution 
denoted $I_{\langle JJOO\rangle}$ in section \ref{Double-trace}. The explicit integrands and results for each diagram are given in Appendix \ref{ape}. 
To compute these diagrams,
we use standard techniques to perform tensor reductions and partial fraction decompositions of the integrand, 
which are discussed in Appendix \ref{apa}.  This results in a sum of simpler scalar integrals which involve either the product 
of two elementary one-loop integrals of the form
\begin{equation}
\begin{aligned}
\int \frac{d^{d}p_{1}}{(2\pi)^{d}}\frac{1}{p_{1}^{2\alpha}(p+p_{1})^{2\beta}}=
\frac{\Gamma(\frac{d}{2}-\alpha)\Gamma(\frac{d}{2}-\beta)\Gamma(\alpha+\beta-\frac{d}{2})}
{(4\pi)^{d/2}\Gamma(\alpha)\Gamma(\beta)\Gamma(d-\alpha-\beta)} (p^2)^{d/2-\alpha-\beta}\equiv l(\alpha,\beta)(p^2)^{d/2-\alpha-\beta}\,,
\label{intl}
\end{aligned}
\end{equation}
or the two-loop ``kite" diagram with the topology of $D_2$ and general power of the middle line
\begin{equation}
K(a) = \int \frac{d^dp_1 d^d p_2}{(2\pi)^{2d}}\frac{1}{p_{1}^{2}(p+p_{1})^{2}p_{2}^{2}(p+p_{2})^{2}(p_{1}-p_{2})^{2a}}\,.
\label{kite-Ka}
\end{equation}
The result for this integral as a function of $d$ and $a$ can be obtained, 
for instance, by using the Gegenbauer polynomial technique \cite{Chetyrkin:1980pr, Kotikov:1995cw}.
Putting all contributions together, the final result is 
\begin{align}
\langle J^{a}(p)J^{b}(-p)\rangle = \frac{\pi ^{\frac{d}{2} } \Gamma (2-\frac{d}{2} ) }{2^{d -3} \Gamma (d )}C^{\textrm{O(N)}}_{J0}\bigg(1-\frac{1}{N}\frac{8(d-1)}{d(d-2)}\eta^{\textrm{O(N)}}_{1}+\mathcal{O}(1/N^{2})\bigg)\frac{p_{z}^{2}}{(p^{2})^{2-\frac{d}{2}}}\,,
\end{align}
where $\eta^{\textrm{O(N)}}_{1}$ is given in (\ref{etaoneon}) and
\begin{align}
C^{\textrm{O(N)}}_{J0} = -\frac{\tr(t^{a}t^{b})}{(d-2)S_{d}^{2}}\,.
\end{align}
Using that in this case $Z_{J}=1+\mathcal{O}(1/N^{2})$, we find 
\begin{align}
C^{\textrm{O(N)}}_{J1} =-\frac{8(d-1)}{d(d-2)}\eta_{1}^{\textrm{O(N)}}\,. \label{Cj1bos}
\end{align}
This agrees with the result of \cite{Petkou:1994ad}, who derived it using the conformal bootstrap
technique. We can verify that $C_{J,1}$ is negative throughout the range $2<d<6$, as shown in figure \ref{CJgraphbos}. 
The value in $d=3$ is given in eq.~(\ref{CTCJON-3d}), and from (\ref{Cj1bos}) one can also get
\begin{align}
C^{\textrm{O(N)}}_{J1}|_{d=2+\eps}=-2+\epsilon +\frac{\epsilon ^2}{2}, \quad 
C^{\textrm{O(N)}}_{J1}|_{d=4-\eps}=-\frac{3 \epsilon ^2}{4}-\frac{\epsilon ^3}{8},\quad  C^{\textrm{O(N)}}_{J1}|_{d=6-\eps}=-\frac{5 \epsilon }{3}+\frac{7 \epsilon ^2}{6}\,
\label{Petkoucjeps}
\end{align}
We note that the $d=6-\eps$ expansion precisely agrees with the result (\ref{O6modelCJ1}) that we derived above from the cubic model. 

\begin{figure}[h!]
   \centering
\includegraphics[width=15cm]{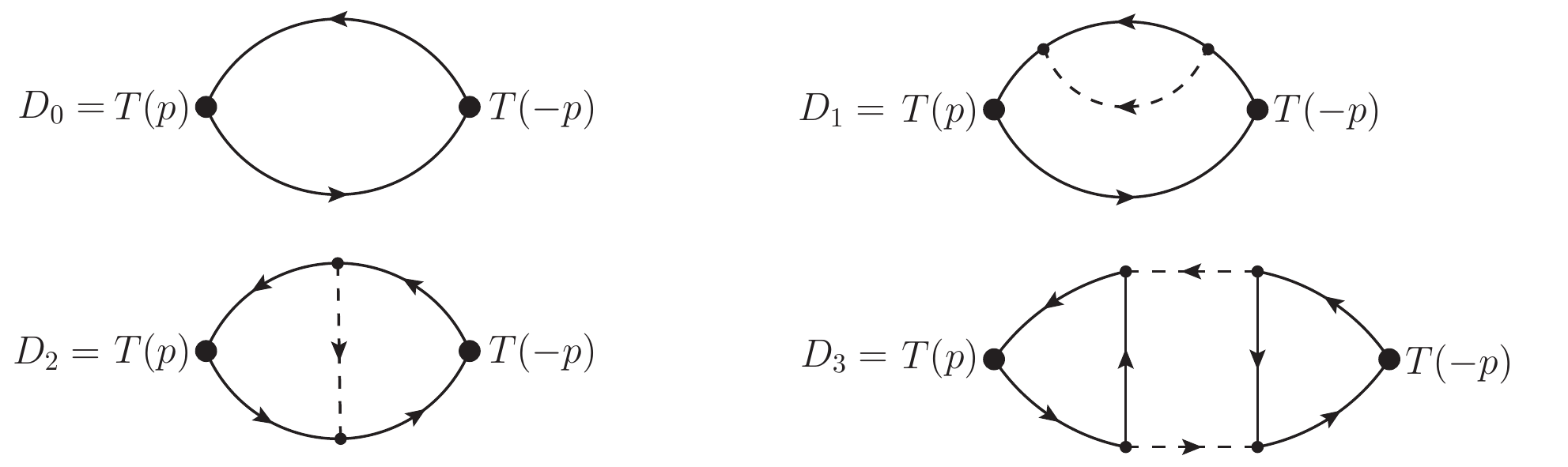}
\caption{Diagrams contributing to $\langle T(p) T(-p)\rangle$ up to order $N^0$. The last one is the three-loop Aslamazov-Larkin diagram.}
\label{CTdiagsbosons}
\end{figure}

Let us now turn to the calculation of $C_T$. There are four diagrams contributing to $\langle TT\rangle$ to order $N^0$ 
\begin{align}
\langle T(p) T(-p)\rangle = D_{0}+D_{1}+D_{2}+D_{3}+\mathcal{O}(1/N)\,,
\end{align}
including the three-loop diagram of Aslamazov-Larkin type \cite{Aslamasov1968238}, which was not
present in the calculation of $C_J$, as shown in figure \ref{CTdiagsbosons}.
After tensor reductions, one obtains a large sum of scalar integrals that, in addition to (\ref{intl}) and (\ref{kite-Ka}), involve 
three-loop ladder scalar integrals with various powers of the propagator lines. The evaluation of this type of integrals is discussed 
in detail in Appendix \ref{apb}, and the results for the individual diagrams are listed in Appendix \ref{ape}. After a very laborious computation, we obtain 
\begin{align}
&\langle T(p) T(-p)\rangle= \frac{\pi ^{\frac{d}{2} } \Gamma (2-\frac{d}{2} )}{2^{d-2} \Gamma (d +2)}\times \notag\\
&~~ \times C_{T0}^{\textrm{O(N)}} \bigg(1-\frac{1}{N}\Big(\frac{1}{\Delta}\frac{4\eta^{\textrm{O(N)}}_{1}}{(d+2)}+\eta_{1}^{\textrm{O(N)}}\Big(\frac{4 \mathcal{C}_{\textrm{O(N)}}(d)}{d +2}+\frac{2 \left(d^3+10 d^2-48 d+32\right)}{(d-4) (d-2) d (d+2)}\Big)\Big)+\mathcal{O}(1/N^{2})\bigg)\frac{p_{z}^{4}}{(p^{2})^{2-\frac{d}{2}}}\,, 
\label{TTbare}
\end{align}
where $\mathcal{C}_{\textrm{O(N)}}(d) =\psi(3-\frac{d}{2})+\psi(d-1)-\psi(1)-\psi(\frac{d}{2})$ and 
$\eta^{\textrm{O(N)}}_{1}$ is given in (\ref{etaoneon}), and 
\begin{align}
C_{T0}^{\textrm{O(N)}}= \frac{Nd}{(d-1)S_{d}^{2}}\,.
\label{ctzero}
\end{align}
As we have already discussed, the $1/\Delta$-pole is present, but there is no $\log (p^{2}/\mu^{2})$ term, 
as expected since the stress-energy tensor is exactly conserved and cannot develop an anomalous dimension. 
In order to get an expression free of the $1/\Delta$ poles, we have to use ``renormalized" stress-energy  tensor 
$T^{\textrm{ren}}_{\mu\nu}=Z_{T}T_{\mu\nu}$, where $Z_{T}$ was derived above and given in (\ref{zTON1}). 
Therefore, we obtain 
\begin{align}
&\langle T^{\textrm{ren}}(p) T^{\textrm{ren}}(-p)\rangle = Z_{T}^{2}\langle T(p) T(-p)\rangle \notag\\
&~~~~~= \frac{\pi ^{\frac{d}{2} } \Gamma (2-\frac{d}{2} )}{2^{d-2} \Gamma (d +2)}C_{T0}^{\textrm{O(N)}}\bigg(1-\frac{\eta_{1}^{\textrm{O(N)}}}{N}\Big(\frac{4 \mathcal{C}_{\textrm{O(N)}}(d)}{d +2}+\frac{2 \left(d^2+6 d-8\right)}{(d-2) d (d+2)}\Big)+\mathcal{O}(1/N^{2})\bigg)\frac{p_{z}^{4}}{(p^{2})^{2-\frac{d}{2}}}\,.
\label{TTren}
\end{align}
Note that, as desired, the $1/\Delta$ pole was cancelled. This is a non-trivial consistency check of our procedure, since the $Z_T$ factor 
was obtained above from an independent Ward identity calculation. From (\ref{TTren}), we thus find 
\begin{align}
C^{\textrm{O(N)}}_{T1}=-2\eta^{\textrm{O(N)}}_{1}\left(\frac{2 \mathcal{C}_{\textrm{O(N)}}(d)}{d +2}+\frac{d^2+6 d-8}{(d-2) d (d+2)}\right)\,, \label{Ct1bos}
\end{align}
which exactly agrees with the result of \cite{Petkou:1994ad}. We note that we may also write this result in a simpler form as 
\begin{align}
C^{\textrm{O(N)}}_{T1}=-2\eta^{\textrm{O(N)}}_{1}\left(\frac{2 \Psi_{\textrm{O(N)}}(d)}{d +2}+\frac{ d+4}{d (d+2)}\right)\,, \label{Ct1bosAlt}
\end{align}
where  $\Psi_{\textrm{O(N)}}(d)\equiv \psi(3-\frac{d}{2} )+\psi(d -1)-\psi(1)-\psi(\frac{d}{2} -1)$. 

\begin{figure}[h!]
   \centering
\includegraphics[width=7cm]{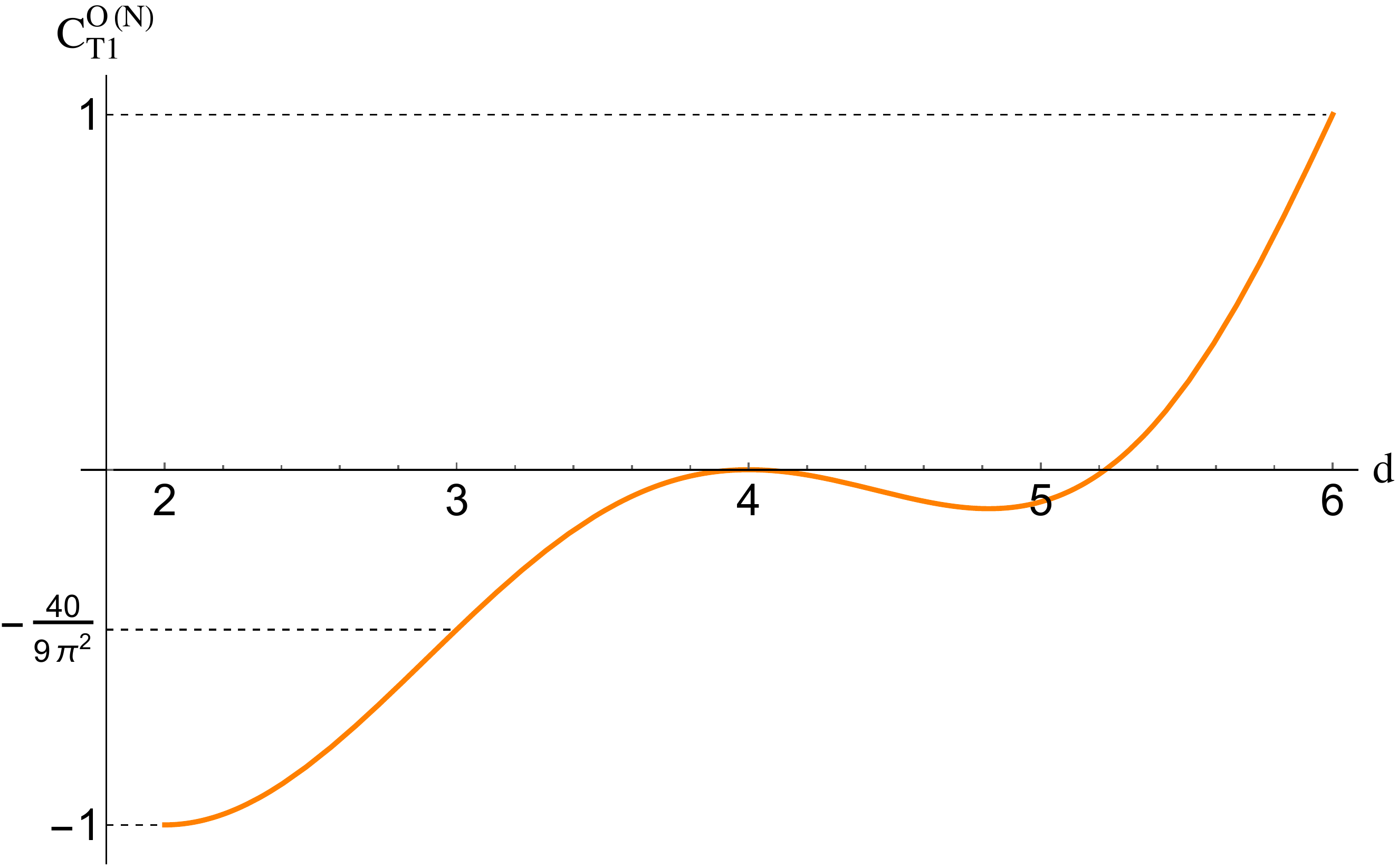}
\caption{Plot of $C_{T1}$.}
\label{CTgraphbos}
\end{figure}

A plot of $C^{\textrm{O(N)}}_{T1}$ in $2<d<6$ is given 
in figure \ref{CTgraphbos}. 
The value in $d=3$ was already given in (\ref{CTCJON-3d}). From  (\ref{Ct1bos}), one can also get
\begin{align}
C^{\textrm{O(N)}}_{T1}|_{d=2+\eps}=-1 +\frac{3\epsilon ^2}{4}, \quad 
C^{\textrm{O(N)}}_{T1}|_{d=4-\eps}=-\frac{5 \epsilon ^2}{12}-\frac{7 \epsilon ^3}{36},\quad  C^{\textrm{O(N)}}_{T1}|_{d=6-\eps}=1-\frac{7 \epsilon }{4}+\frac{23 \epsilon ^2}{288}\,.
\label{Petkoueps}
\end{align}
We note that the result for $C^{\textrm{O(N)}}_{T1}$ expanded in $d=6-\eps$ precisely agrees with the 
the calculation in the cubic model, see (\ref{O6modelCT1}).  This constitutes a new perturbative check of the formula (\ref{Ct1bos}) 
for $C^{\textrm{O(N)}}_{T1}$. 
Note that the leading term in $d=6-\epsilon$ is just the contribution of the free scalar field $\sigma$ in the cubic model. As discussed in 
the Introduction, for all even $d$, the critical $O(N)$ model is expected to reduce to a free theory of $N$ ordinary conformal scalars, 
plus a $\Delta=2$ scalar with kinetic term $\sim \sigma (\partial^2)^{\frac d 2-2}\sigma$, see eq.~(\ref{sigsig-ON}). From (\ref{Ct1bos}) it follows that
\begin{equation}
C_{T1}^{\textrm{O(N)}}|_{{\rm even }\ d} = \frac{(-1)^{{\frac d 2}+1}(d-4)(d-2)!}{({\frac d 2}+1)!({\frac d 2}-1)!} 
= (-1)^{\frac{d}{2}+1}\left[\begin{pmatrix}d-4\\ \frac{d}{2}-3\end{pmatrix}-\begin{pmatrix}d-4\\ \frac{d}{2}-5\end{pmatrix}\right]
\ .\label{CTON-2n}
\end{equation}
Interestingly, this is an integer for all even dimensions \cite{Stergiou:2015roa}.\footnote{In fact, we note that (\ref{CTON-2n}) appears to be equal (for $d>4$)
to $(-1)^{d/2+1}$ times the dimension of the irreducible representation of $Sp(d-4)$ labelled by the Young tableaux $[\underbrace{1,\ldots,1}_{d/2-3},0,\ldots,0]$.} 
The formula (\ref{CTON-2n}) is the ratio of the $C_T$ of a free $(d-4)$-derivative scalar to that of a canonical scalar. 
This means that
\begin{equation}
C_{T}^{(d-4)-\textrm{deriv.\ scalar}}|_{{\rm even }\ d} = \frac{(-1)^{{\frac d 2}+1}d(d-4)(d-2)!}{(d-1) ({\frac d 2}+1)!({\frac d 2}-1)! S_d^2} 
\ .\label{dminusfour}
\end{equation}
It would be interesting to check this result
via an explicit calculation using the action for a higher derivative scalar.

\subsection{Pad\' e approximations}

For any quantity $f(d)$ 
known in the $\epsilon=4-d$ and $\epsilon=d-2$ expansions up to a given order, we can construct a Pad\'e approximant
\begin{equation}
\textrm{Pad\'e}_{[m,n]}(d) = \frac{A_0 + A_1 d + A_2 d^2 + \ldots + A_m d^m}{1 + B_1 d + B_2 d^2 + \ldots + B_n d^n}\,,
\label{Pade}
\end{equation}
where the coefficients $A_i, B_i$ are fixed by requiring that the expansion of (\ref{Pade}) agrees with 
the known terms in $f (4-\epsilon)$ and $f (2+\epsilon)$ obtained by perturbation theory.  For the $O(N)$ model the $4-\epsilon$ expansion can be developed
for any integer $N$ using the weakly coupled Wilson-Fisher IR fixed point \cite{Wilson:1971dc}. The $2+\epsilon$ expansion can be developed using standard
perturbation theory only for $N>2$, because this is when the $O(N)$
non-linear $\sigma$ model has a weakly coupled UV fixed point \cite{Brezin:1975sq, Bardeen:1976zh,Moshe:2003xn}.

\begin{figure}[h!]
   \centering
\includegraphics[width=10cm]{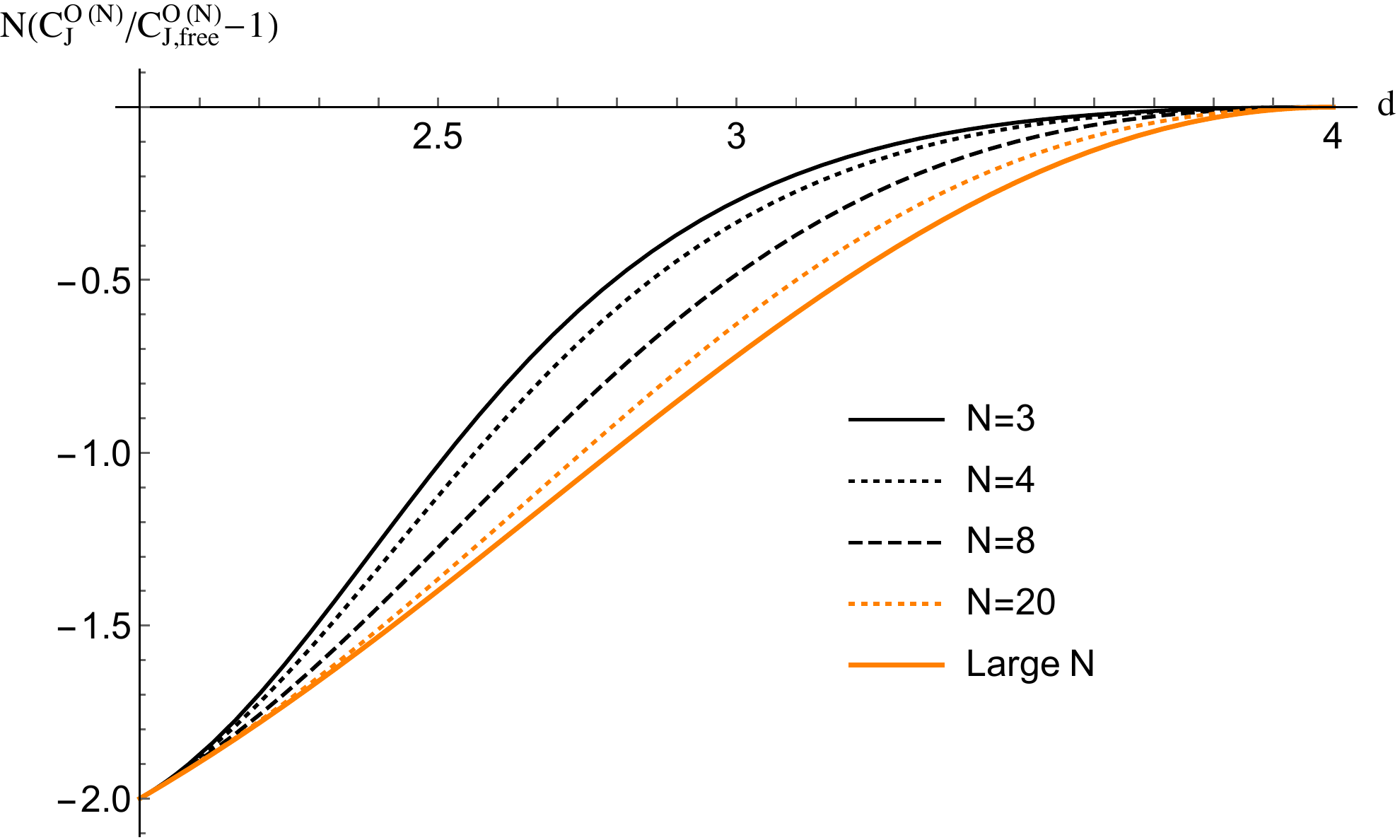}
\caption{Plot of $N(C^{\textrm{O(N)}}_{J}/C^{\textrm{O(N)}}_{J,\textrm{free}}-1)$ for $\textrm{Pad\' e}_{[2,2]}$.}
\label{PadeCJBos}
\end{figure}

For $C^{\textrm{O(N)}}_{J}/C_{J,\textrm{free}}^{\textrm{O(N)}}$, the $\epsilon$ expansions read
(the $\epsilon/N$ correction in $d=2+\epsilon$ was guessed on the basis of the large $N$ results and plausible assumptions, 
and the $d=4-\epsilon$ expansion can be found in \cite{Jack:1983sk,Petkou:1994ad}):
\begin{equation}
C^{\textrm{O(N)}}_{J}/C_{J,\textrm{free}}^{\textrm{O(N)}}(d )=
\begin{cases}\frac{N-2}{N}+\frac{\epsilon }{N}+\mathcal{O}(\epsilon^{2}) &\mbox{in }\quad d=2+\epsilon\,, \\
1-\frac{3 (N+2) \epsilon ^2}{4 (N+8)^2}+\mathcal{O}(\epsilon^{3}) &\mbox{in }\quad d=4-\eps\,.
\end{cases}
\end{equation}
In this case we find that only the approximant $\textrm{Pad\' e}_{[2,2]}$ is well-behaved, being free of poles 
and in good agreement at large $N$ with the result (\ref{Cj1bos}) in $2<d<4$. We plot $\textrm{Pad\' e}_{[2,2]}$ for different 
values of $N$ in figure \ref{PadeCJBos}, and list a few of its numerical values in $d=3$ in table \ref{tablePadeNBosCJ}.

\begin{figure}[h!]
   \centering
\includegraphics[width=9cm]{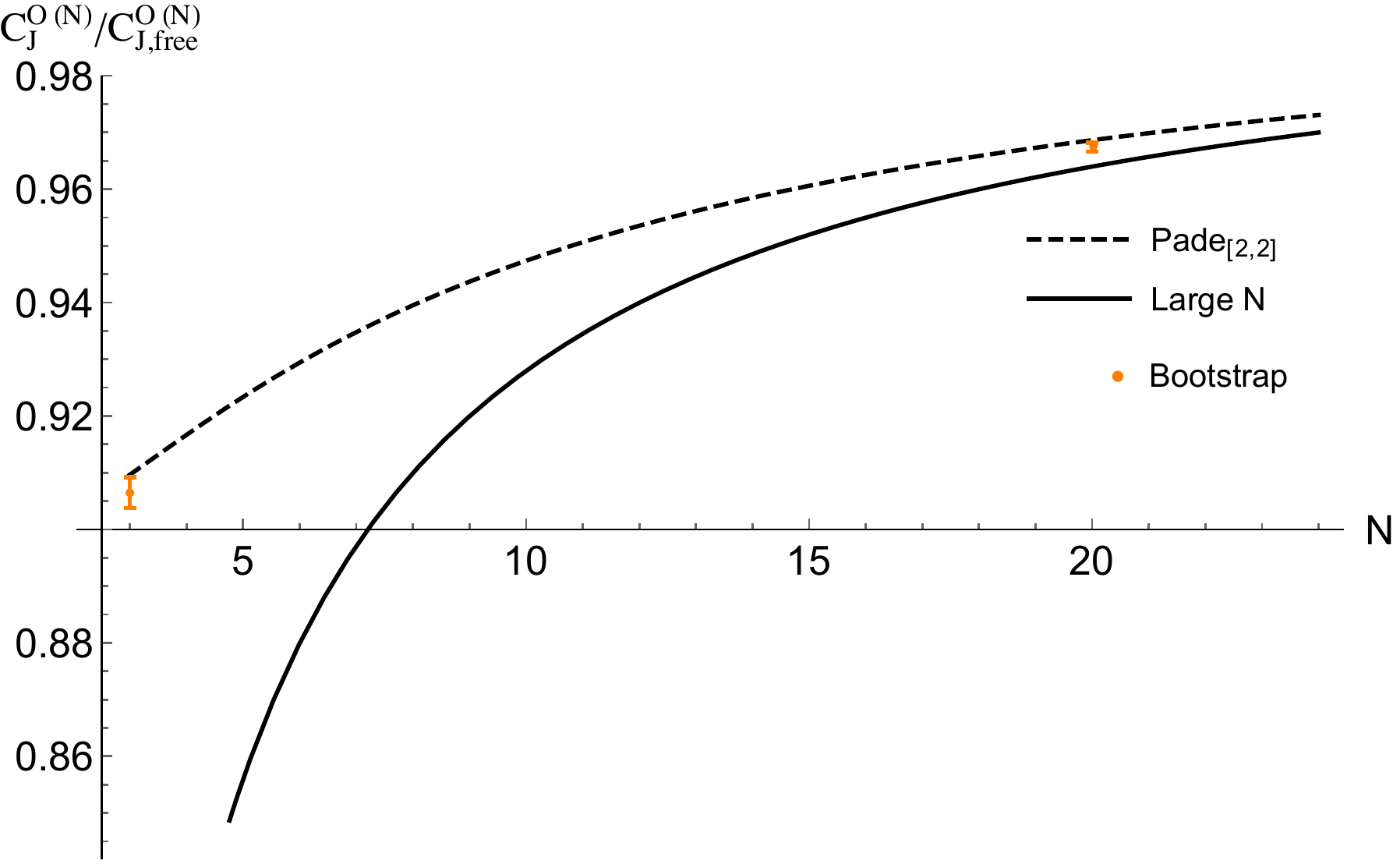}
\caption{Plot of $C^{\textrm{O(N)}}_{J}/C^{\textrm{O(N)}}_{J,\textrm{free}}$ in $d=3$}
\label{CJForN}
\end{figure}

\begin{table}[h]
\centering
\begin{tabular}{ccccccccc}
\hline
\multicolumn{1}{|c|}{$N$}         
& \multicolumn{1}{c|}{3} & \multicolumn{1}{c|}{4} & \multicolumn{1}{c|}{5} & \multicolumn{1}{c|}{8}& \multicolumn{1}{c|}{12} & \multicolumn{1}{c|}{20}& \multicolumn{1}{c|}{50} \\ \hline
\multicolumn{1}{|c|}{$\textrm{Pad\' e}_{[2,2]}$ }     
& \multicolumn{1}{c|}{0.9096} & \multicolumn{1}{c|}{0.9167} & \multicolumn{1}{c|}{0.9234} & \multicolumn{1}{c|}{0.9395} & \multicolumn{1}{c|}{0.9535}& \multicolumn{1}{c|}{0.9686}  & \multicolumn{1}{c|}{0.9860}\\ \hline
\multicolumn{1}{|c|}{$1-\frac{64}{9\pi^{2}N}$ }     
& \multicolumn{1}{c|}{0.7598} & \multicolumn{1}{c|}{0.8199} & \multicolumn{1}{c|}{0.8559} & \multicolumn{1}{c|}{0.9099} & \multicolumn{1}{c|}{0.9400}& \multicolumn{1}{c|}{0.9640}  & \multicolumn{1}{c|}{0.9856}\\ \hline
\end{tabular}
\caption{List of $\textrm{Pad\'e}_{[2,2]}$ extrapolations for $C^{\textrm{O(N)}}_{J}/C^{\textrm{O(N)}}_{J,\textrm{free}}$ for $d=3$. 
The second line corresponds to the large $N$ result (\ref{Cj1bos}) in $d=3$. 
}
\label{tablePadeNBosCJ}
\end{table}
We observe that the results we find are close to the $C_J$ values obtained using the conformal bootstrap \cite{Kos:2015mba}. The quoted bootstrap value 
$C^{\textrm{O(3)}}_{J}/C^{\textrm{O(3)}}_{J,\textrm{free}}= 0.9065(27)$ should be compared with our $\textrm{Pad\' e}_{[2,2]}$ result $0.9096$,
and the bootstrap value $C^{\textrm{O(20)}}_{J}/C^{\textrm{O(20)}}_{J,\textrm{free}}= 0.9674(8)$ with our $\textrm{Pad\' e}_{[2,2]}$ result $0.9686$.

For the $C^{\textrm{O(N)}}_{T}/C_{T,\textrm{free}}^{\textrm{O(N)}}$  we use the following $\epsilon$-expansions: 
\begin{equation}
C^{\textrm{O(N)}}_{T}/C_{T,\textrm{free}}^{\textrm{O(N)}}(d )=
\begin{cases}
1-\frac{1}{N}+\frac{3 (N-1) \epsilon ^2}{4 N (N-2)}+\mathcal{O}(\epsilon^{3}) &\mbox{in }\quad d=2+\epsilon\,, \\
1-\frac{5 (N+2) \epsilon^2}{12 (N+8)^2}+\mathcal{O}(\epsilon^{3}) &\mbox{in }\quad d=4-\eps\,.
\end{cases}
\end{equation}
The leading correction in $d=4-\eps$ can be found in \cite{Jack:1983sk,Cappelli:1990yc, Petkou:1995vu}. To determine the $2+\eps$ expansion we used the fact that there is a $R_{abcd}^2$
correction to the central charge in the $d=2$ sigma model with general target space curvature
\cite{Callan:1986jb,Metsaev:1987zx}. After specializing 
to the case of $N-1$ dimensional sphere, we find that this term $\sim (N-1)(N-2) g^2$. 
The $O(N)$ sigma model has a UV fixed point in $d=2+\eps$ for $N> 2$ \cite{Brezin:1975sq, Bardeen:1976zh,Moshe:2003xn}.
Setting the sigma model coupling $g$ to 
its fixed point value $\sim \frac{\epsilon}{N-2}$, and
using the large $N$ result to normalize the correction, we find the result above.

\begin{figure}[h!]
   \centering
\includegraphics[width=10cm]{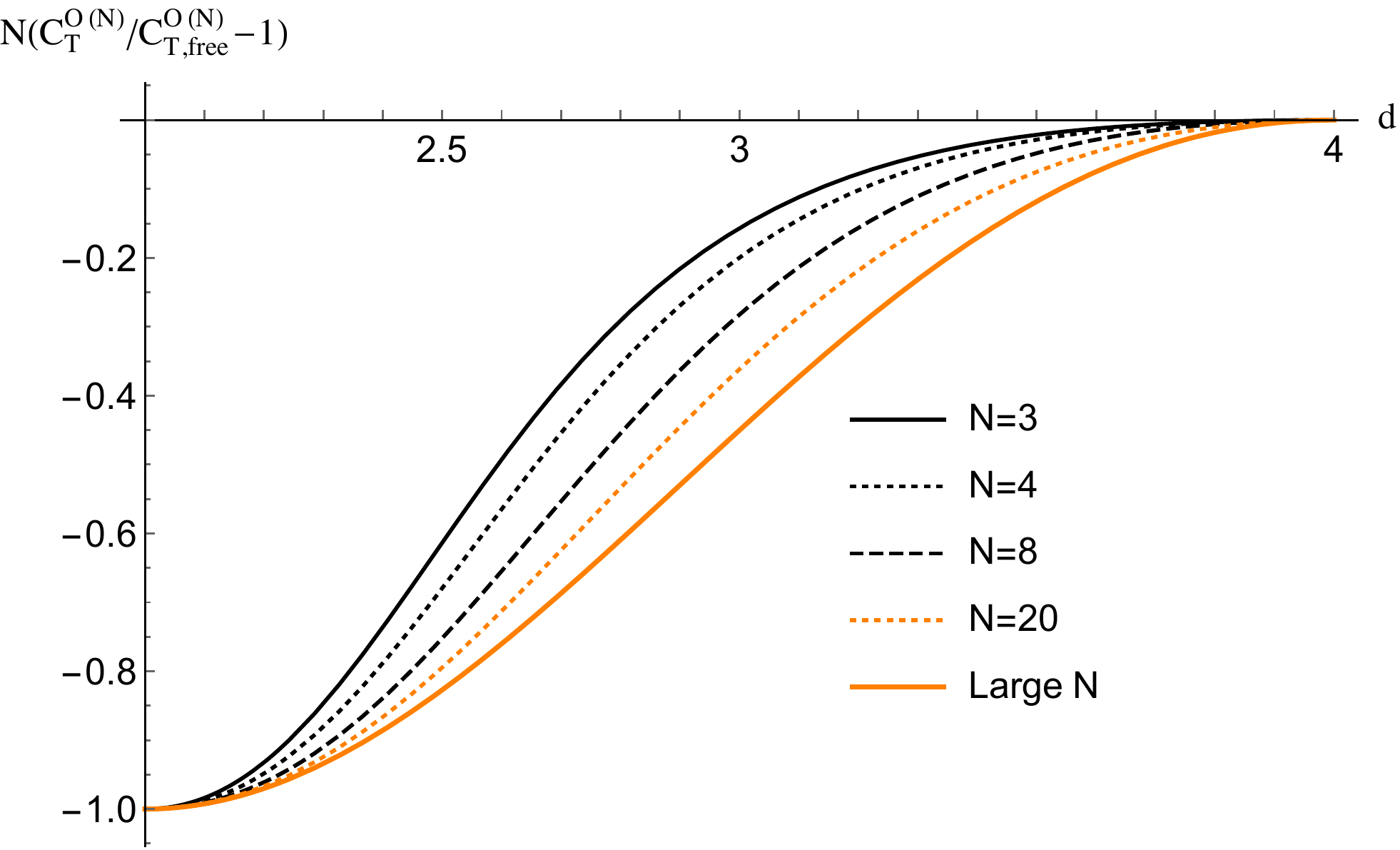}
\caption{Plot of $N(C^{\textrm{O(N)}}_{T}/C^{\textrm{O(N)}}_{T,\textrm{free}}-1)$ for $\textrm{Pad\' e}_{[3,2]}$.}
\label{PadeCTBos}
\end{figure}

The best approximant we find is $\textrm{Pad\' e}_{[3,2]}$; it does not have  poles 
and approaches the large $N$ result (\ref{Ct1bos}) quite well. We plot  $\textrm{Pad\' e}_{[3,2]}$ for different $N$ in figure \ref{PadeCTBos}. 
Also, we give  the values of  
$C^{\textrm{O(N)}}_{T}/C^{\textrm{O(N)}}_{T,\textrm{free}}$ for different $N$ in $d=3$ in table \ref{tablePadeBosCT}.

\begin{figure}[h!]
   \centering
\includegraphics[width=10cm]{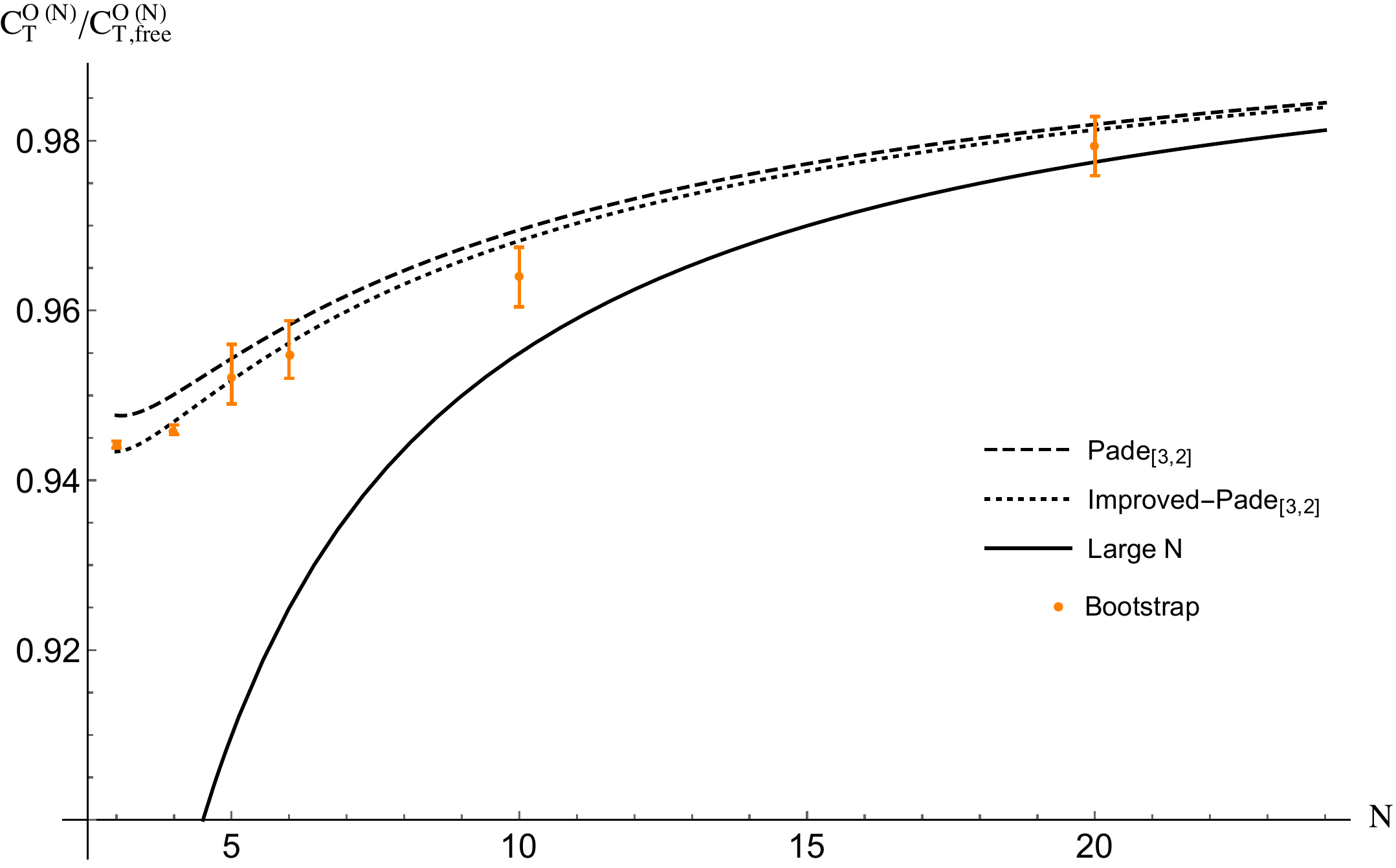}
\caption{Plot of $C^{\textrm{O(N)}}_{T}/C^{\textrm{O(N)}}_{T,\textrm{free}}$ in $d=3$}
\label{CTForN}
\end{figure}

\begin{table}[h]
\centering
\begin{tabular}{ccccccccc}
\hline
\multicolumn{1}{|c|}{$N$}         
& \multicolumn{1}{c|}{3} & \multicolumn{1}{c|}{4} & \multicolumn{1}{c|}{5} & \multicolumn{1}{c|}{8}& \multicolumn{1}{c|}{12} & \multicolumn{1}{c|}{20}& \multicolumn{1}{c|}{50} \\ \hline
\multicolumn{1}{|c|}{$\textrm{Pad\' e}_{[3,2]}$ }     
& \multicolumn{1}{c|}{0.9477} & \multicolumn{1}{c|}{0.9501} & \multicolumn{1}{c|}{0.9543} & \multicolumn{1}{c|}{0.9647} & \multicolumn{1}{c|}{0.9732}& \multicolumn{1}{c|}{0.9819}  & \multicolumn{1}{c|}{0.9919}\\ \hline
\multicolumn{1}{|c|}{$1-\frac{40}{9\pi^{2}N}$ }     
& \multicolumn{1}{c|}{0.8499} & \multicolumn{1}{c|}{0.8874} & \multicolumn{1}{c|}{0.9099} & \multicolumn{1}{c|}{0.9437} & \multicolumn{1}{c|}{0.9625}& \multicolumn{1}{c|}{0.9775}  & \multicolumn{1}{c|}{0.9910}\\ \hline
\end{tabular}
\caption{List of $\textrm{Pad\'e}_{[3,2]}$ extrapolations for $C^{\textrm{O(N)}}_{T}/C^{\textrm{O(N)}}_{T,\textrm{free}}$ in $d=3$. 
The second line is the large $N$ result (\ref{Ct1bos}) in $d=3$.
}
\label{tablePadeBosCT}
\end{table}
The results we find are close to the $C_T$ values obtained using the conformal bootstrap \cite{Kos:2013tga}. The quoted bootstrap values (see Table 3 in  \cite{Kos:2013tga}) are in good agreement with our $\textrm{Pad\' e}_{[3,2]}$. This is shown in figure \ref{CTForN}, where we also include 
the result of an ``improved" Pad\' e$_{[3,2]}$ approximant obtained by imposing 
exact agreement with the large $N$ result (\ref{Ct1bos}) in $2<d<4$.  Explicitly, this may be defined as 
\begin{align}
\textrm{Improved-Pad\' e}(d,N)=\textrm{Pad\' e}(d,N)+\frac{1}{N}\Big(C_{T1}-\lim_{N\to \infty}\big(N(\textrm{Pad\' e}(d,N)-1)\big)\Big)\, ,
\end{align} 
which by construction exactly approaches the large $N$ result when $N$ goes to infinity.  
From figure \ref{CTForN}, we see that it fits the bootstrap data even better than the regular Pad\' e.

\section{Gross-Neveu Model}
\label{section:Fermions}

\subsection{$1/N$ expansion}

\begin{figure}[h!]
   \centering
\includegraphics[width=12.5cm]{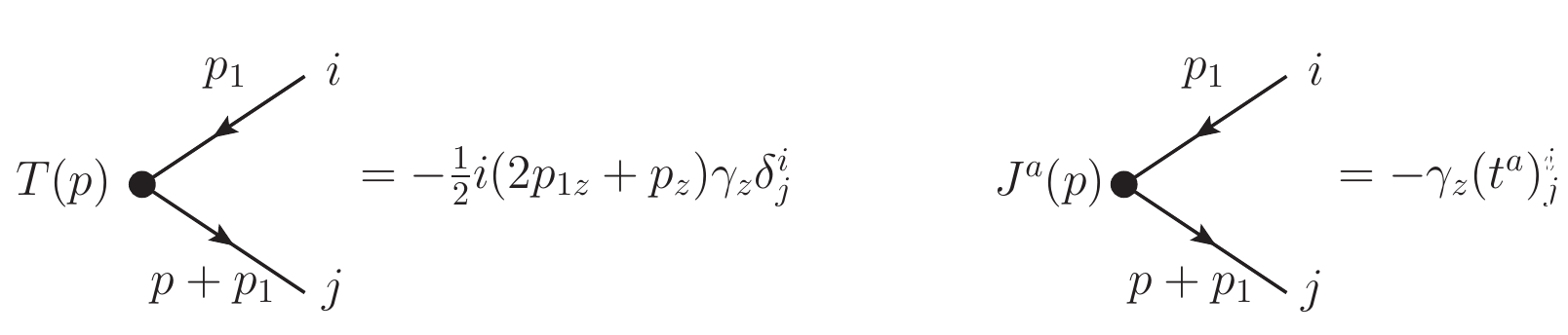}
\caption{Momentum space Feynman rules for $T(p)$ and  $J^{a}(p)$.}
\label{TJdiag}
\end{figure}

The Hubbard-Stratonovich analysis reviewed in Section \ref{Double-trace} can be also applied to the Gross-Neveu model. 
Introducing the auxiliary field $\sigma$, and dropping the quadratic term in the critical limit, we have the action
\begin{equation}
S_{\textrm{crit ferm}}= \int d^{d}x \Big(-\bar{\psi}_{0i}\slashed{\partial}\psi_{0}^{i}+\frac{1}{\sqrt{N}}\sigma_{0} \bar{\psi}_{0i} \psi^{i}_{0} \Big)\,,
\end{equation}
where $i=1,\ldots,\tilde{N}$ and $N = \tilde{N}{\rm Tr}{\bf 1}$.
The propagator of the $\psi_{0}^i$ field reads 
\begin{equation}
\langle \psi_{0}^{i}(p) \bar{\psi}_{0j}(-p) \rangle_{0} =\delta^{i}_{j}\frac{i\slashed{p}}{p^{2}}.
\end{equation}
The $\sigma$ effective propagator obtained after integrating over the fundamental fields $\psi_{0}^i$ reads
\begin{align}
\left<\sigma_{0}(p)\sigma_{0}(-p) \right>_{0}&=\tilde{C}_{\sigma0}/(p^2)^{\frac{d}{2}-1+\Delta}\,,
\label{sigma-eff-GN}
\end{align}
where 
 \begin{align}
\tilde{C}_{\sigma0 } &\equiv -2^{d+1}(4\pi)^{\frac{d-3}{2}}\Gamma\Big( \frac{d-1}{2} \Big)\sin \big(\frac{\pi d}{2}\big)\, \label{csigmatilde}
\end{align}
and we have introduced the regulator $\Delta$. 
Note that the power of $p^2$ in the propagator is $\frac{d}{2}-1+\Delta$ instead of $\frac{d}{2}-2+\Delta$ found in the scalar case.
In order to cancel the divergences as $\Delta\rightarrow 0$ we have to renormalize the bare fields $\psi_{0}$ and $\sigma_{0}$: 
\begin{align}
\psi = Z_{\psi}^{1/2} \psi_{0}, \qquad  \sigma = Z_{\sigma}^{1/2}\sigma_{0}\,, \label{zpsizsig}
\end{align}
where 
\begin{align}
Z_{\psi} = 1+\frac{1}{N}\frac{Z_{\psi 1}}{\Delta}+\mathcal{O}(1/N^{2}), \qquad 
Z_{\sigma} =1+\frac{1}{N}\frac{Z_{\sigma 1}}{\Delta}+\mathcal{O}(1/N^{2})\,. \label{zpsizsig2}
\end{align}
The full propagators of the renormalized fields read
\begin{align}
\langle \psi^{i}(p) \bar{\psi}_{j}(-p)\rangle = \delta^{i}_{j}\tilde{C}_{\psi} \frac{i\slashed{p}}{(p^{2})^{\frac{d}{2}-\Delta_{\psi}+\frac{1}{2}}}, \qquad 
\langle \sigma(p) \sigma(-p)\rangle =  \frac{\tilde{C}_{\sigma}}{(p^{2})^{\frac{d}{2}-\Delta_{\sigma}}}\,, \label{fermprops}
\end{align}
where we introduced anomalous dimensions $\Delta_{\psi}$ and $\Delta_{\sigma}$ and two-point
function normalizations $\tilde{C}_{\psi}$ and $\tilde{C}_{\sigma}$ in momentum space. Each of them
may be represented as a series in $1/N$:
\begin{align}
\Delta_{\psi} = \frac{d}{2}-\frac{1}{2}+\eta^{\textrm{GN}}\,, \qquad \Delta_{\sigma}=1-\eta^{\textrm{GN}}-\kappa^{\textrm{GN}}\,, 
\label{dphidsigma}
\end{align}
where $\eta^{\textrm{GN}} = \eta^{\textrm{GN}}_{1}/N+\eta^{\textrm{GN}}_{2}/N^{2}+\mathcal{O}(1/N^{3})$, $\kappa^{\textrm{GN}}=\kappa^{\textrm{GN}}_{1}/N+\kappa^{\textrm{GN}}_{2}/N^{2}+\mathcal{O}(1/N^{3})$  and 
\begin{align}
\tilde{C}_{\psi} = 1 +\frac{\tilde{C}_{\psi1}}{N}+\frac{\tilde{C}_{\psi 2}}{N^{2}}+\mathcal{O}(1/N^{3})\,, \qquad 
\tilde{C}_{\sigma} = \tilde{C}_{\sigma 0} +\frac{\tilde{C}_{\sigma 1}}{N}+\frac{\tilde{C}_{\sigma 2}}{N^{2}}+\mathcal{O}(1/N^{3})\,. \label{cphicsigma}
\end{align}
The stress-energy tensor and the current are 
\begin{equation}
\begin{aligned}
T &=  - \frac{1}{2}(\bar{\psi}_{0 i}\gamma_{\mu}\partial_{\nu}\psi^{i}_{0} - \partial_{\mu}\bar{\psi}_{0i} \gamma_{\nu} \psi^{i}_{0}) z^{\mu}z^{\nu}\,,\\
J^{a} &= -z^\mu \bar{\psi}_{0i}(t^{a})^{i}_{j}\gamma_{\mu}\psi_{0}^{j} \, 
\label{TJ-GN}
\end{aligned}
\end{equation}
and in momentum space
\begin{align}
T(p)&= -\frac{1}{2}\int \frac{d^{d}p_{1}}{(2\pi)^{d}} \bar{\psi}_{0i}(-p_{1})i\gamma_{z}(2p_{1z}+p_{z})\psi_{0}^{i}(p+p_{1})\,, \notag\\
J^{a}(p)&=-\int \frac{d^{d}p_{1}}{(2\pi)^{d}} \bar{\psi}_{0 i}(-p_{1})(t^{a})^{i}_{j}\gamma_{z}\psi_{0}^{j}(p+p_{1})\,.
\end{align}
The diagrammatic representation is shown in figure \ref{TJdiag}.

\noindent As in the scalar case, we define
\begin{align}
T_{\mu\nu}^{\textrm{ren}} = Z_{T} T_{\mu\nu}, \qquad  J_{\mu}^{\textrm{ren}, a} = Z_{J}J_{\mu}^{a}\,,
\end{align}
where 
\begin{align}
Z_{T} = 1+ \frac{1}{N}\Big(\frac{Z_{T1}}{\Delta}+Z'_{T1}\Big)+\mathcal{O}(1/N^{2}), \quad Z_{J} = 1+ \frac{1}{N}\Big(\frac{Z_{J1}}{\Delta}+Z'_{J1}\Big)+\mathcal{O}(1/N^{2})\,.
\label{zTferm}
\end{align}
By a direct calculation presented in Appendices \ref{apc} and \ref{apd}, we show that Ward identities fix
\begin{align}
Z_{T1} =\frac{2\eta_{1}^{\textrm{GN}}}{d +2},\qquad Z'_{T1}=\frac{8\eta_{1}^{\textrm{GN}}}{(d +2) (d -2)} \,,  \label{zt1zt1p}
\end{align}
where  $\eta^{\textrm{GN}}_{1}$ is defined in (\ref{dphidsigma}) and reads 
\begin{align}
\eta_{1}^{\textrm{GN}}=\frac{ \Gamma (d -1)(\frac{d}{2} -1)^2}{\Gamma (2-\frac{d}{2} ) \Gamma (\frac{d}{2} +1) \Gamma (\frac{d}{2} )^2}\,. \label{etaferm}
\end{align}
For the spin 1 current, we find $Z_{J} = 1+ \mathcal{O}(1/N^{2})$, 
which means that it does not affect the $C_{J1}$ calculation.

\subsection{Calculation of $C^{\textrm{GN}}_{J1}$ and $C^{\textrm{GN}}_{T1}$}

\label{GNcjct}

\begin{figure}[h!]
   \centering
\includegraphics[width=17cm]{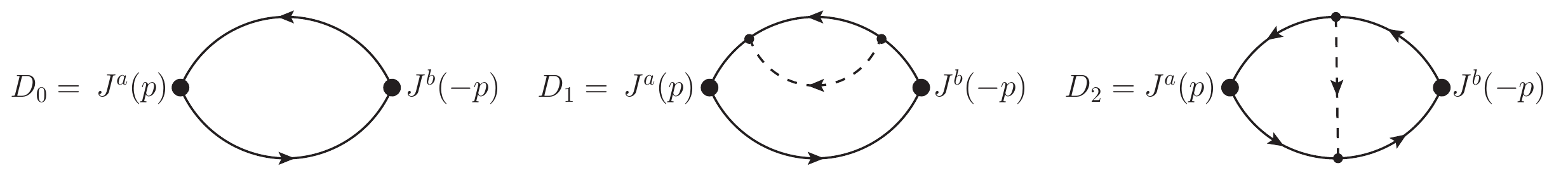}
\caption{Diagrams contributing to $\langle J^a(p) J^b(-p) \rangle$ up to order $1/N$.}
\label{CJdiagsfermion}
\end{figure}

There are again three diagrams contributing to $C_J/C_{J0}$ up to order $1/N$, given in figure \ref{CJdiagsfermion}.
They are identical to the ones for the critical scalar, except the solid lines are fermionic instead of scalar.

To compute the diagrams we use the same methods as for the case of the $O(N)$ model (see Appendices \ref{apa}, \ref{apb}). 
We  find that the $1/\Delta$ divergence is canceled in the combination $D_1+D_2$, yielding the result 
(see Appendix \ref{ape} for the integrands and results for each diagram):
\begin{align}
\langle J^{a}(p)J^{b}(-p)\rangle&= D_{0}+D_{1}+D_{2}+\mathcal{O}(1/N^{2}) \notag\\
&=\frac{\pi ^{\frac{d}{2} } \Gamma (2-\frac{d}{2} ) }{2^{d -3} \Gamma (d )}C_{J0}^{\textrm{GN}}\bigg(1-\frac{1}{N}\frac{8(d-1)}{d(d-2)}\eta_{1}^{\textrm{GN}}+\mathcal{O}(1/N^{2})\bigg)\frac{p_{z}^{2}}{(p^{2})^{2-\frac{d}{2}}}\,,
\end{align}
where $\eta_1^{\textrm{GN}}$ is given in (\ref{etaferm}) and 
\begin{align}
C_{J0}^{\textrm{GN}}=-\tr(t^{a}t^{b}){\rm Tr}{\bf 1}\frac{1}{S_{d}^{2}}\,.
\end{align}
Therefore, we find the final result
\begin{equation}
C_{J1}^{\textrm{GN}} = -\frac{8(d-1)}{d(d-2)}\eta_{1}^{\textrm{GN}} \,.
\label{Cj1GN}
\end{equation}
We see that $C_{J1}^{\textrm{GN}}$ for the critical fermion is always negative in the range $2<d<4$, thus a ``$C_J$-theorem" 
inequality $C_J^{\rm UV}>C_J^{\rm IR}$ does not hold for the flow from the UV fixed point to the free fermions in the IR. 

\begin{figure}[h!]
   \centering
\includegraphics[width=7cm]{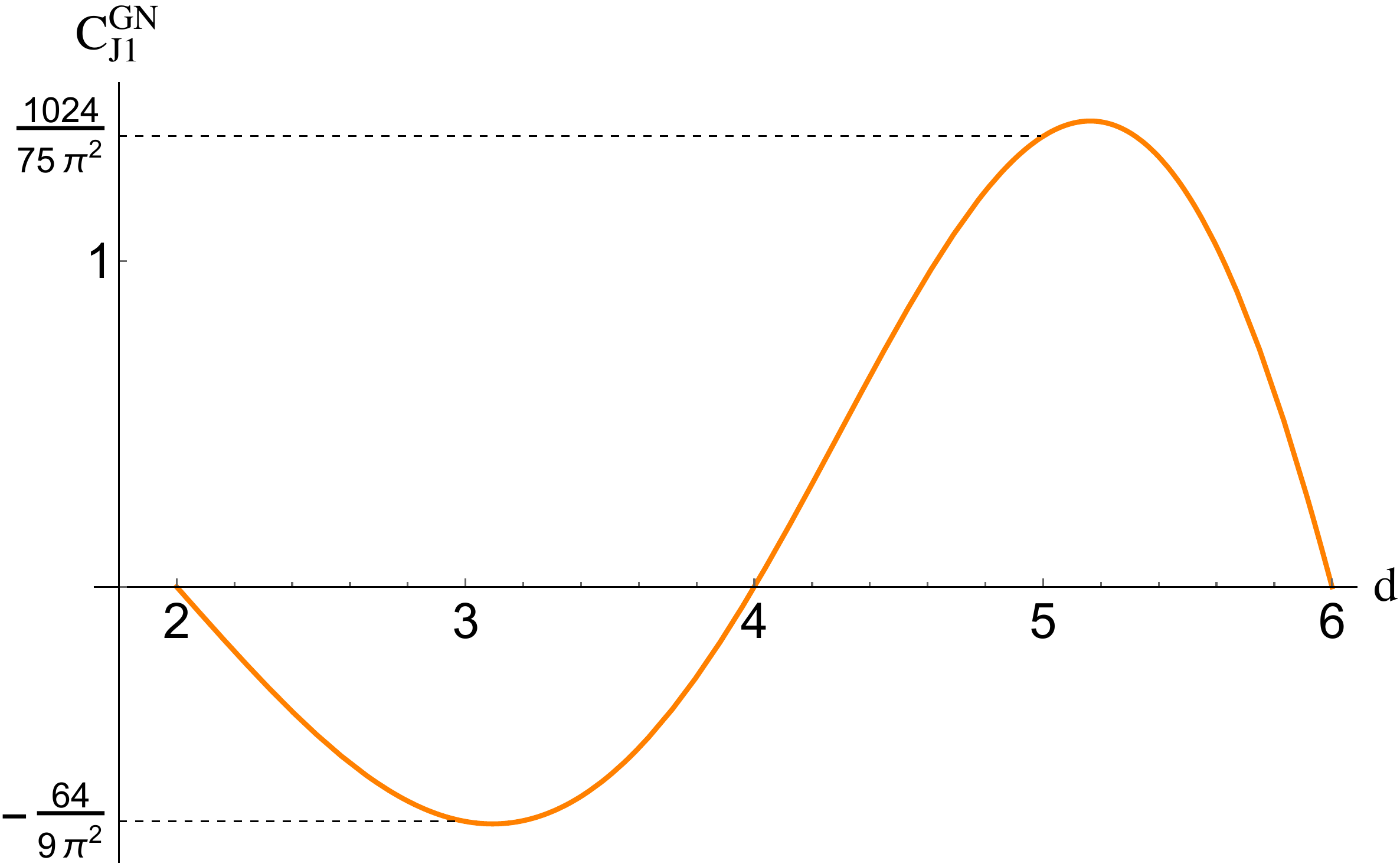}
\caption{Plot of $C^{\textrm{GN}}_{J1}$, which is negative throughout the range $2<d<4$.}
\label{CJgraphferm}
\end{figure}

In $d=3$, we obtain the value reported in eq.~(\ref{CTCJGN-3d}). In $d=2+\eps$ and $d=4-\eps$ dimensions, we find
\begin{align}
C^{\textrm{GN}}_{J,1}|_{d=2+\eps}=-\eps+\frac{\eps^3}{4}+\mathcal{O}(\epsilon^{4})\,,
\qquad C^{\textrm{GN}}_{J,1}|_{d=4-\eps}=-\frac{3\eps}{2}+\frac{\eps^2}{2}+\frac{15\eps^3}{32}+\mathcal{O}(\epsilon^{4})\,.
\label{Cj1GN-eps}
\end{align}
We will show that these values are in precise agreement with our $C_J$ calculations for the GN and GNY models performed in sections \ref{CTCJ-GNY} and 
\ref{CTCJ-GN} below.


\begin{figure}[h!]
   \centering
\includegraphics[width=18cm]{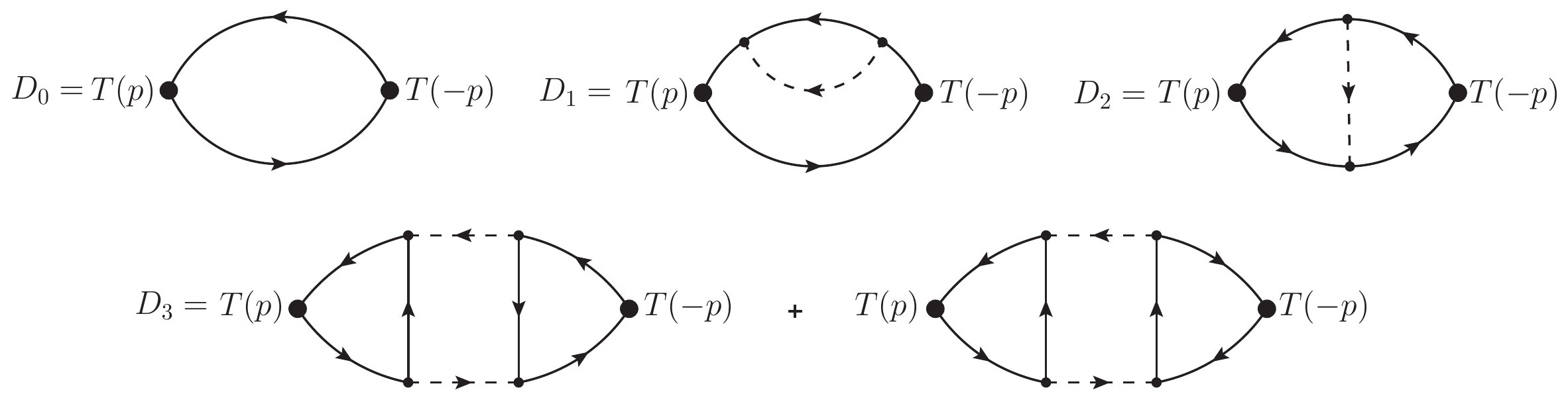}
\caption{Diagrams contributing to $\langle T(p) T(-p)\rangle$ up to order $N^0$.}
\label{CTdiagsfermions}
\end{figure}

The diagrams contributing to the stress tensor two-point function
\begin{align}
\langle T(p) T(-p)\rangle = D_{0}+D_{1}+D_{2}+D_{3}+\mathcal{O}(1/N^{2})\, ,
\end{align}
are shown in figure \ref{CTdiagsfermions} (see Appendix \ref{ape} for the results).
After a very laborious computation, the details of which are discussed in the Appendices, we obtain the final result
\begin{align}
&\langle T(p) T(-p)\rangle= \frac{\pi ^{\frac{d}{2} } \Gamma (2-\frac{d}{2} )}{2^{d-2} \Gamma (d +2)}\times \notag\\
&~~~~ \times C_{T0}^{\textrm{GN}}\bigg(1-\frac{1}{N}\Big(\frac{1}{\Delta}\frac{4\eta_{1}^{\textrm{GN}}}{(d+2)}+\eta_{1}^{\textrm{GN}}\Big(\frac{4 \mathcal{C}_{\textrm{GN}}(d)}{d +2}+\frac{4 \left(5 d^2-8 d+4\right)}{(d-2) (d-1) d (d+2)}\Big)\Big)+\mathcal{O}(1/N^{2})\bigg)\frac{p_{z}^{4}}{(p^{2})^{2-\frac{d}{2}}}\,, 
\end{align}
where $\mathcal{C}_{\textrm{GN}}(d)\equiv \psi(2-\frac{d}{2})+\psi(d-1)-\psi(1)-\psi(\frac{d}{2})$,
$\eta_{1}^{\textrm{GN}}$ is given in (\ref{etaferm}) and
\begin{align}
C_{T0}^{\textrm{GN}} = \frac{Nd}{2S_{d}^{2}}\,.
\label{ctzero-GN}
\end{align}
As we already discussed, we see that $1/\Delta$-pole is present, but the $\log (p^{2}/\mu^{2})$ term cancels out; this means that, as 
expected, the stress tensor does not have an anomalous dimension, because it is exactly conserved. 
In order to get a finite expression we have to use the renormalized stress-energy  tensor $T^{\textrm{ren}}_{\mu\nu}=Z_{T}T_{\mu\nu}$, 
where $Z_{T}$ is given in (\ref{zTferm}) and (\ref{zt1zt1p}). Therefore, we obtain 
\begin{align}
&\langle T^{\textrm{ren}}(p) T^{\textrm{ren}}(-p)\rangle = Z_{T}^{2}\langle T(p) T(-p)\rangle \notag\\
&~~~=\frac{\pi ^{\frac{d}{2} } \Gamma (2-\frac{d}{2} )}{2^{d-2} \Gamma (d +2)}C_{T0}^{\textrm{GN}} \bigg(1-\frac{\eta_{1}^{\textrm{GN}}}{N}\Big(\frac{4 \mathcal{C}_{\textrm{GN}}(d)}{d +2}+\frac{4 (d-2)}{(d-1) d (d+2)}\Big)+\mathcal{O}(1/N^{2})\bigg)\frac{p_{z}^{4}}{(p^{2})^{2-\frac{d}{2}}}\,.
\label{TTren-GN}
\end{align}
As in the scalar case discussed earlier, it is a non-trivial test of our procedure that the $Z_T$ factor fixed by Ward identities has precisely the 
correct pole to cancel the $1/\Delta$ divergence in $\langle TT\rangle$. From (\ref{TTren-GN}), we then find one of our main results 
\begin{align}
C^{\textrm{GN}}_{T1}=-4\eta^{\textrm{GN}}_{1}\left(\frac{ \mathcal{C}_{\textrm{GN}}(d)}{d +2}+\frac{ d-2}{(d-1) d (d+2)}\right)\,.\label{Ct1ferm}
\end{align}

\begin{figure}[h!]
\begin{minipage}[h]{0.53\linewidth}
\center{\includegraphics[width=0.9\linewidth]{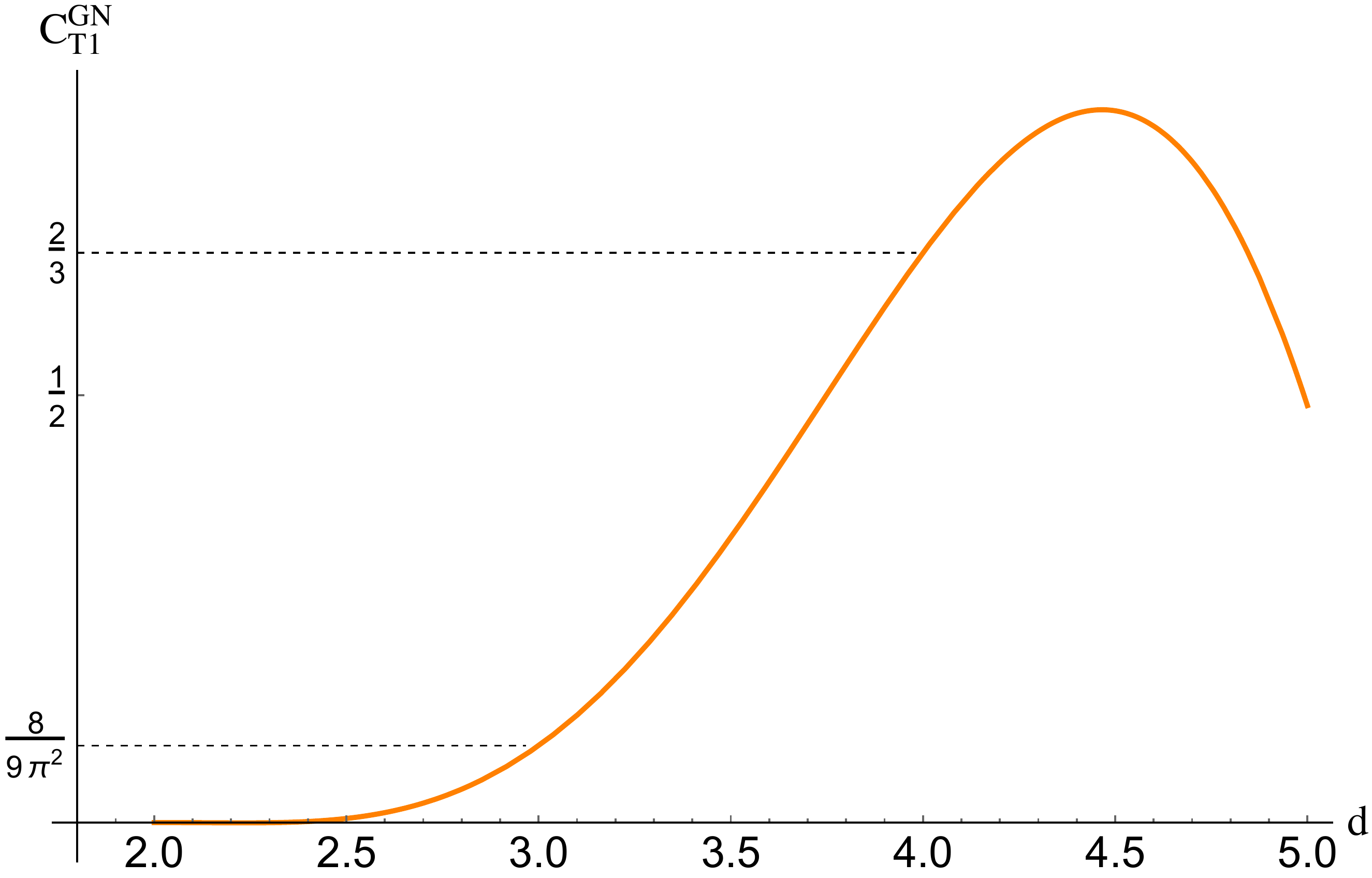} }
\end{minipage}
\hfill
\begin{minipage}[h]{0.53\linewidth}
\center{\includegraphics[width=0.8\linewidth]{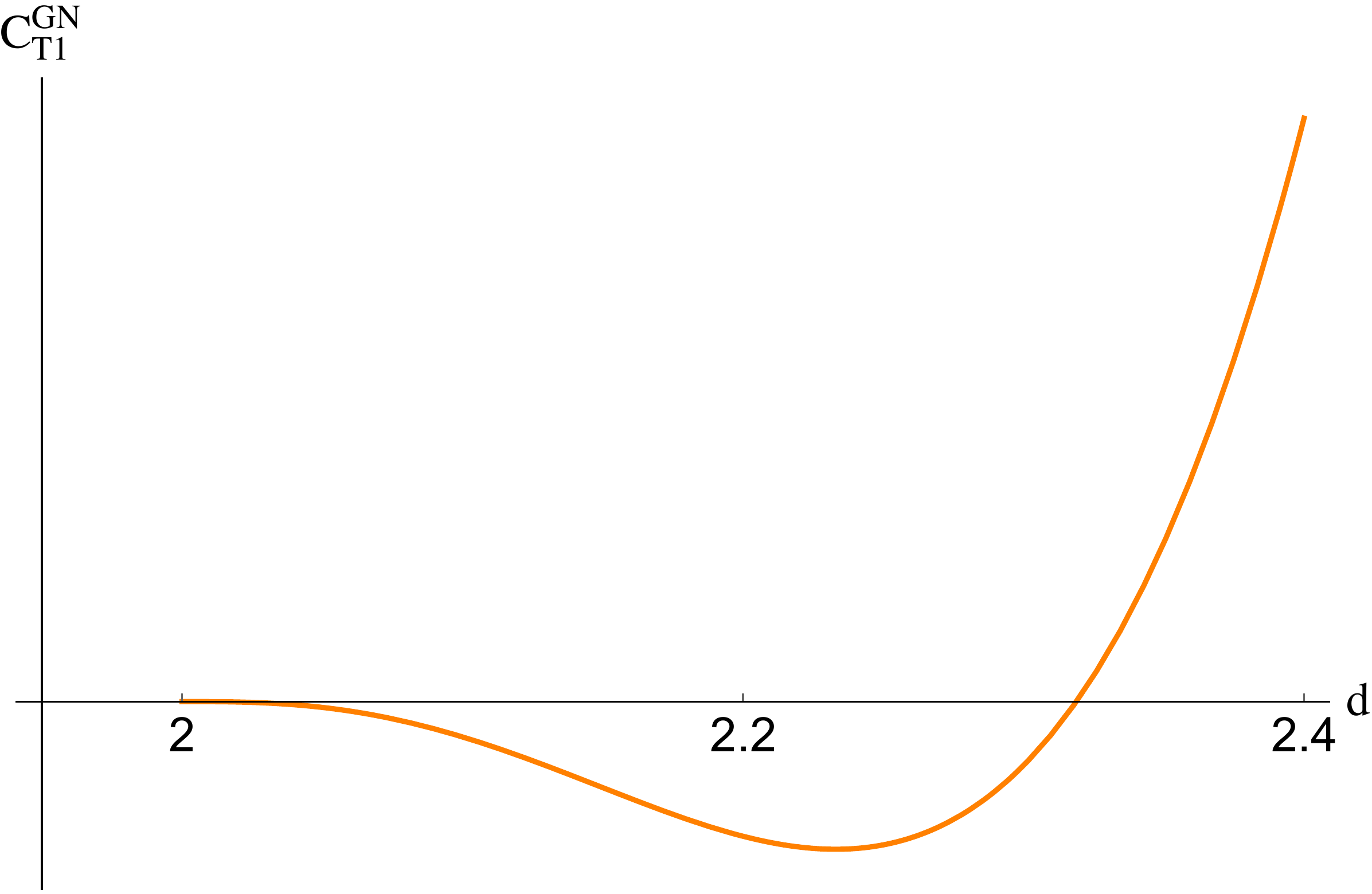} }
\end{minipage}
\caption{Plot of $C_{T1}^{\textrm{GN}}$.}
\label{CTgraphferm}
\end{figure}

In $d=3$, we get the result quoted in eq.~(\ref{CTCJGN-3d}). It is interesting that $C^{\textrm{GN}}_{T1}>0$ in $d=3$. This means that the 
``$C_T$-theorem" inequality $C_T^{\textrm{UV}} > C_T^{\textrm{IR}}$ applies to the large $N$ Gross-Neveu model in
$d=3$. However, as plot \ref{CTgraphferm} shows, this inequality is violated for $2< d \lesssim 2.3$.

In $d=2+\eps$ and $d=4-\eps$, we find
\begin{align}
C^{\textrm{GN}}_{T1}|_{d=2+\eps}=-\frac{\epsilon ^3}{8}+\mathcal{O}(\epsilon^{4})\,,
\qquad C^{\textrm{GN}}_{T1}|_{d=4-\eps}=\frac{2}{3}-\frac{11 \epsilon }{18}-\frac{17 \epsilon ^2}{54}+\mathcal{O}(\epsilon^{3}) \,. \label{ctfermresults}
\end{align}
As we show below, these precisely agree with the results obtained using the $\epsilon$ expansion in the GN and GNY models, respectively. 

It is also interesting to look at general even dimensions $d$. In this case, the GN model is expected to be equivalent 
to a theory of $\tilde N$ free fermions plus a higher derivative scalar with local kinetic term $\sim \sigma (\partial^2)^{\frac {d} {2} -1}\sigma$ (see 
the form of the induced propagator (\ref{sigma-eff-GN})). The contribution to $C_T$ of such a free scalar can be obtained from 
(\ref{Ct1ferm}), which has a finite non-zero limit for all even $d>2$
\begin{equation}
C^{\textrm{GN}}_{T1}|_{{\rm even}\ d} = \frac{(-1)^{\frac d 2} (d-2)(d-2)!}{({\frac d 2}+1)!({\frac d 2}-1)!}\,,
\label{CTGN-2n}
\end{equation}  
From this, after multiplying by the overall free fermion factor (\ref{ctzero-GN}), 
one may read off the $C_T$ coefficient of the $(d-2)$-derivative scalar for all even $d$:
\begin{equation}
C^{(d-2)-\textrm{deriv.\ scalar}}_{T}|_{{\rm even}\ d}=
\frac{(-1)^{\frac d 2} d (d-2)(d-2)!}{2 ({\frac d 2}+1)!({\frac d 2}-1)! S_d^2}\ .
\label{dminustwo}
\end{equation}
Its ratio to $C_T$ of a canonical scalar is
\begin{equation}
\frac{(-1)^{\frac d 2} (d - 1) (d-2)(d-2)!}{2 ({\frac d 2}+1)!({\frac d 2}-1)!}=(-1)^{\frac{d}{2}} \begin{pmatrix}d-1\\\frac{d}{2}-2\end{pmatrix}\, .
\label{CTGN-2nratio}
\end{equation}
Interestingly, this is an integer; in  $d=6,8,10,\ldots$ we find $-5, 21, -84,\ldots$\footnote{These correspond to $\pm$ the dimensions
of the rank-$(d/2-2)$ totally antisymmetric representations of $SO(d-1)$.}
It would be interesting to check the formula (\ref{CTGN-2nratio}) by a direct calculation using the stress-energy tensor of
the free $(d-2)$-derivative scalar.

\subsection{Gross-Neveu-Yukawa model and $4-\epsilon$ expansions of $C_J$ and $C_T$}
\label{CTCJ-GNY}

In this section we consider the Gross-Neveu-Yukawa (GNY) model \cite{Hasenfratz:1991it,ZinnJustin:1991yn}.  It is a theory of $\tilde{N}$
Dirac fermions $\psi^i$ transforming under an internal $U(\tilde{N})$ symmetry group and a scalar
field $\sigma$ in $d = 4-\epsilon$ dimensions described by the action (\ref{GNYact}).
As above, we define $N = \tilde{N}{\rm Tr}{\bf 1}$, where ${\bf 1}$ is the identity matrix for the Dirac
representation. The model has a weakly coupled fixed point in $d=4-\eps$, with the coupling constants given by,
to leading order in $\eps$ \cite{Moshe:2003xn}, 
\begin{align}
g_{1\star} &= \sqrt{\frac{16\pi^2\epsilon}{N+6}} \label{eq:fermion-g1}\,,\\
g_{2\star} &= 16\pi^2\epsilon\frac{24N}{(N+6)\left((N-6) + \sqrt{N^2+132N+36}\right)}\,.
\label{eq:fermion-g2}
\end{align}
As before, we will compute $C_J$ and $C_T$ up to two-loop level. We have not found such
a calculation in the literature, so our results appear to be new.


\begin{figure}[h!]
   \centering
\includegraphics[width=16cm]{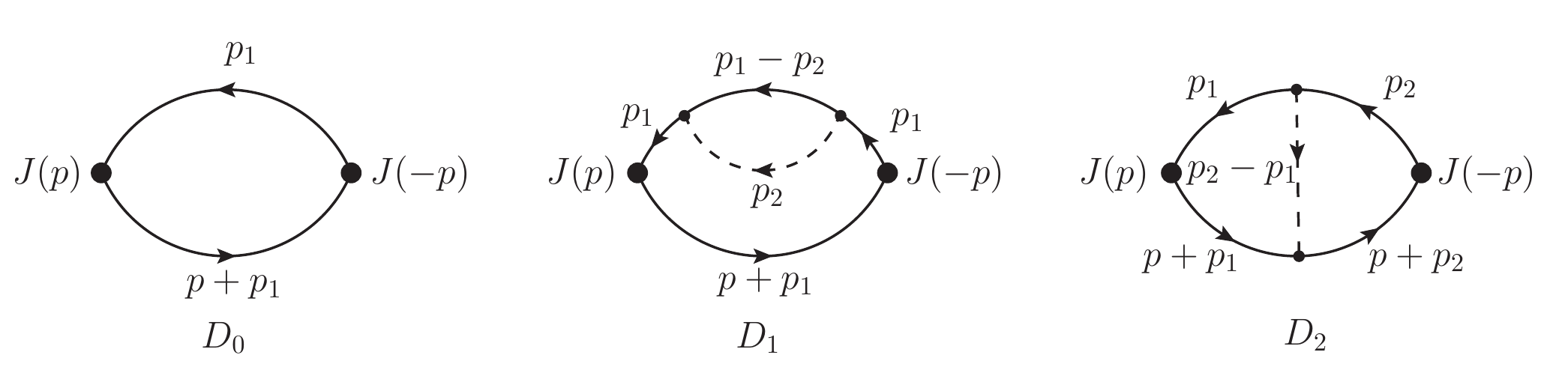}
\caption{Diagrams for $C_{J}$ to $1/N$ order}
\label{CJepsdiags2}
\end{figure}

For simplicity, we will consider the two-point function of the $U(1)$ current 
\begin{equation}
J= z^{\mu} \bar{\psi}_i \gamma_{\mu}\psi^i \label{Jfermion}
\,,
\end{equation}
which, in the notation used above in eq.~(\ref{TJ-GN}), 
just corresponds to a particular choice of generator of $U(N)$ (the one proportional to the identity).
The diagrams contributing to
\begin{align}
\vev{J(p)J(-p)} = D_{0}+D_1+D_2+\mathcal{O}(1/N^{2})\,.
\end{align}
are shown in figure \ref{CJepsdiags2} (see Appendix \ref{ape} for the integrands and results).
The arrows are fermionic arrows, and we have defined our momenta in such a way
that the flow of momentum coincides with the fermionic arrows.  
As before, the dashed line denotes the $\sigma$ field.  

After evaluating the integrals, Fourier
transforming to position space, substituting the fixed-point values
(\ref{eq:fermion-g1}) and (\ref{eq:fermion-g2}) of the coupling constants, and extracting the $C_J$
coefficient from each term according to (\ref{eq:cj-def}), we obtain:
\begin{equation}
C_J^{\textrm{GNY}} = \frac{1}{S_d^2}\left(N - \frac{3N\epsilon}{2(N+6)} + 
\mathcal{O}(\eps^2)\right)\,,
\end{equation}
where $S_d=2\pi^{d/2}/\Gamma(d/2)$ is the volume of the $(d-1)$-dimensional sphere (evaluated here in $d = 4-\epsilon$). 
Normalizing by the free field contribution, we find 
\begin{align}
\frac{ C_{J}^{\textrm{GNY}}}{C_{J}^{\textrm{free}}} =1-\frac{3 \epsilon }{2 (N+6)}+\mathcal{O}(\epsilon^{2})\,,
\end{align}
which precisely agrees, to leading order at large $N$, with the result (\ref{Cj1GN}) expanded in $d=4-\eps$, see eq.~(\ref{Cj1GN-eps}). 


To study $C_T$ we write $T=T_\psi+ T_\sigma$, where
\begin{equation}
T_\psi = - \frac{1}{2}\left (\bar{\psi}_i\gamma_\mu \partial_{\nu}\psi^i 
- \partial_{\mu}\bar{\psi}_i\gamma_{\nu }\psi^i  \right ) z^\mu z^\nu\ , \label{eq:fermion-t-pp-f}
\end{equation}
and $T_\sigma$ is given in (\ref{tsigma}).
We have $\vev{TT} = \vev{T_\psi T_\psi} + 2\vev{T_\psi T_\sigma} + \vev{T_\sigma T_\sigma}$.

At leading order, $\vev{T_\psi T_\sigma} = 0$, while $\vev{T_\sigma T_\sigma}$ and  
$\vev{T_\psi T_\psi}$ are given by the free field one-loop integrals.
At the next to leading order we have four diagrams, which we call $D_1$, $D_2$, $D_3$, and $D_4$;
they are shown in figure \ref{CTGNYeps} (see Appendix \ref{ape} for the explicit results).

\begin{figure}[h!]
   \centering
\includegraphics[width=11cm]{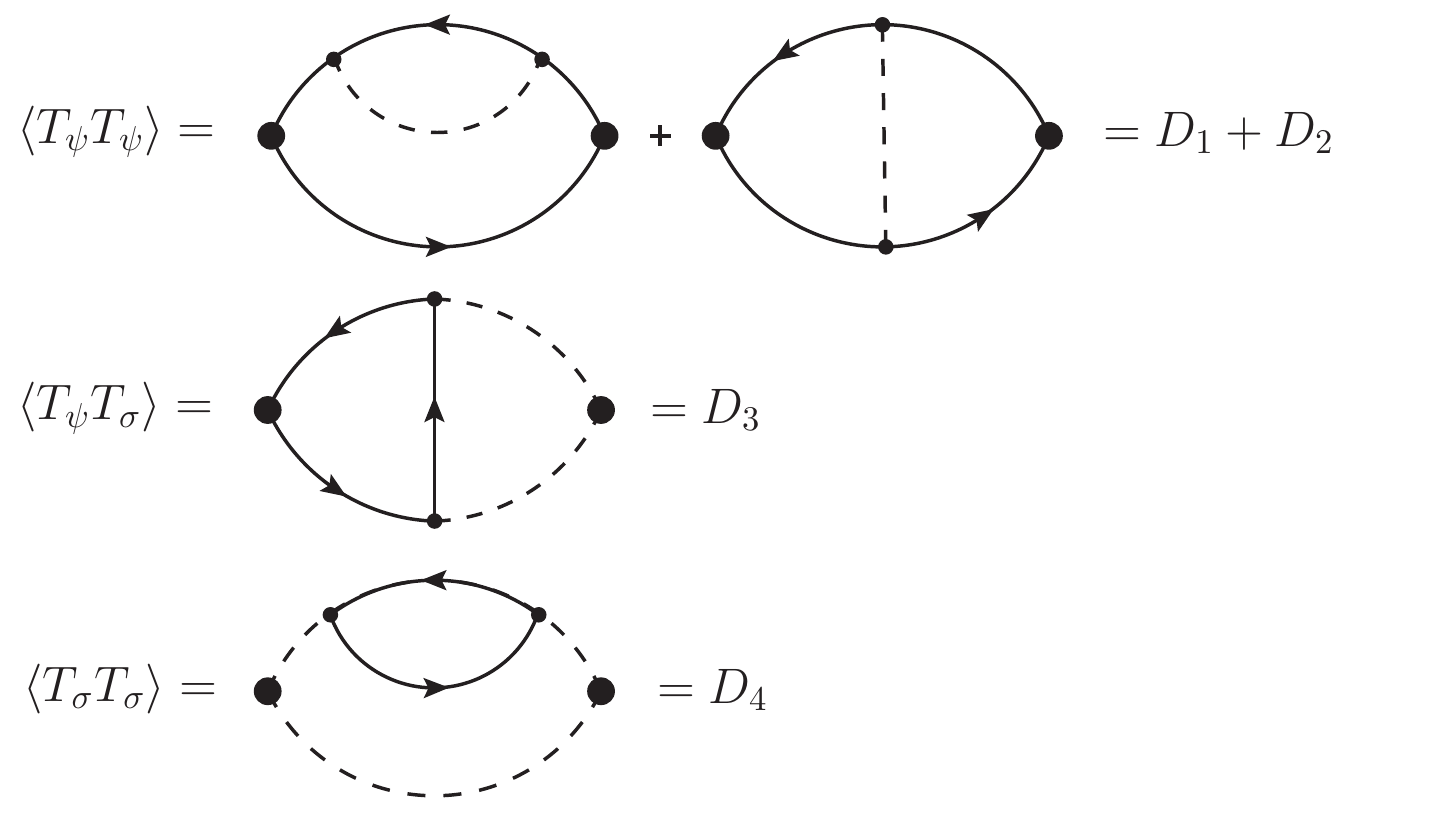}
\caption{Diagrams for $C_{T}$ in GNY model}
\label{CTGNYeps}
\end{figure}

After evaluating the integrals, Fourier transforming to position space, and plugging in the expression
(\ref{eq:fermion-g1}) for the coupling constant $g_1$ at the fixed point, we get
\begin{equation}
C_T^{\textrm{GNY}} = \frac{d}{S_d^2}\left(\frac{N}{2} + \frac{1}{d-1} - \frac{5N\epsilon}{12(N+6)}\right)\,,
\end{equation}
To compare to the large $N$ calculation in the previous section, 
we should normalize this result by the contribution of $\tilde N$ free Dirac fermions. Using (\ref{ctzero-GN}), we find
\begin{align}
\frac {C_{T}^{\textrm{GNY}}}{N C_{T0}^{\textrm{GN}}} =1+\frac{2}{3N}-\frac{11N-24}{18 N (N+6)} \epsilon  +\mathcal{O}(\epsilon^{2})\,.
\end{align}
Comparing with (\ref{ctfermresults}), we again find precise agreement with our large $N$ result (\ref{Ct1ferm}) expanded in 
$d=4-\eps$. 

\subsection{$2+\epsilon$ expansion of $C_J$ and $C_T$}
\label{CTCJ-GN}

In this section, we will consider the Gross-Neveu model (\ref{GNact}) in $d=2+\epsilon$.   
The beta function and the critical value of $g$ at the UV fixed point are \cite{Moshe:2003xn}
\begin{align}
\beta &= \eps g - (N-2)\frac{g^2}{2\pi} + (N-2)\frac{g^3}{4\pi^2}+(N-2)(N-7)\frac{g^4}{32\pi^3}+\mathcal{O}(g^5)\,, \notag \\
g_* &= \frac{2\pi}{N-2}\eps + \frac{2\pi}{(N-2)^2}\eps^2+\frac{(N+1)\pi}{2(N-2)^3}\eps^3 +\mathcal{O}(\eps^4)\,, \label{gstarGN}
\end{align}
where $N = \tilde{N}{\rm Tr}{\bf 1}$.
From the beta function we can also deduce the relation between the bare and renormalized couplings (here $\mu$ denotes the renormalization scale):
\begin{equation}
g_0 = \mu^{-\eps}\left(g+\frac{N-2}{2\pi}\frac{g^2}{\eps}-\frac{N-2}{8\pi^2}\frac{g^3}{\eps}+\frac{(N-2)^2}{4\pi^2}\frac{g^3}{\eps^2}+\mathcal{O}(g^4) \right) \,.\label{g0GN}
\end{equation}
The UV fixed point of this model is related to the IR fixed point of the GNY model. One can check
this by comparing the anomalous dimensions of the $\psi$ and $\sigma$ fields as in \cite{Moshe:2003xn}. In this 
section we will derive $C_J$  and $C_T$ for the critical fermionic theory at to next-to-leading order.

\begin{figure}[h!]
   \centering
\includegraphics[width=6.5cm]{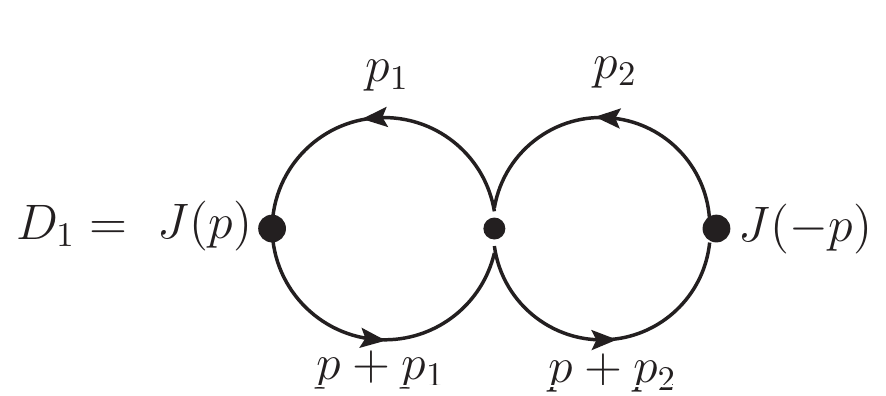}
\caption{Two-loop diagram contributing to $C_J$ to order $g$}
\label{GNCJ12loop}
\end{figure}

\begin{figure}[h!]
   \centering
\includegraphics[width=16cm]{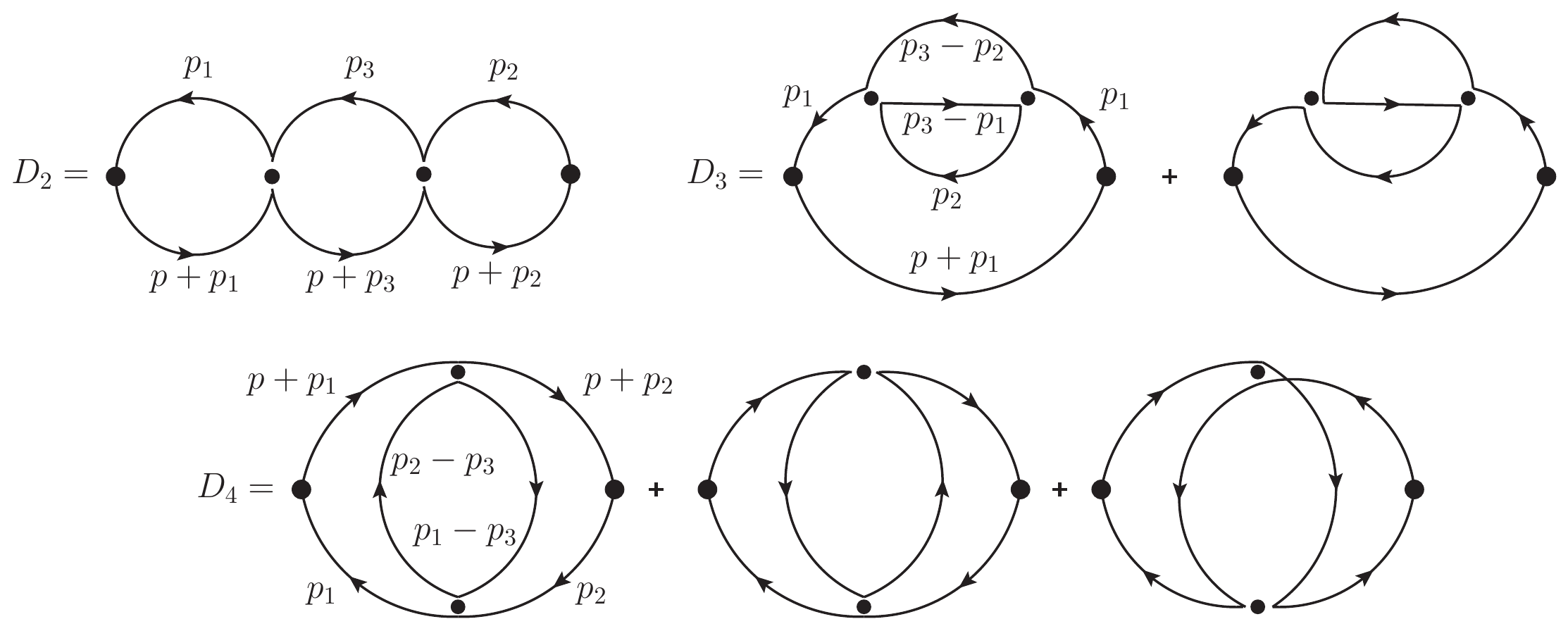}
\caption{Diagrams contributing to $C_J$ to order $g^{2}$.}
\label{GNCJdiag3loop}
\end{figure}

To extract $C_J$, we may calculate the two-point function of the $U(1)$ current (\ref{Jfermion}).
The leading order contribution to $C_J$ is the same diagram $D_0$ as in the GNY model, and the contribution of order $g$ is depicted in figure 
\ref{GNCJ12loop}. 
The diagrams contributing to $g^{2}$ order are shown in \ref{GNCJdiag3loop}. \footnote{We did not
draw some of the diagrams with the $D_{4}$ topology 
because they cancel each other after using the formula $\Tr(\slashed{A}\,\slashed{B}\slashed{C}\slashed{D})=\Tr(\slashed{A}\,\slashed{D}\slashed{C}\slashed{B})$, 
but diagrams with such a topology do appear in the $\langle TT\rangle$ computation.  
Also, the  second diagram for $D_{4}$ in the figure has a partner with different orientation of the fermion line, but one can show that these diagrams are equal, therefore we have a factor of $2$ for the integral of this diagram in formula (\ref{GNCJD4}).} There are three different topologies, and multiple ways of directing the fermion lines within each. As before, the arrows are fermionic arrows, and we have defined momenta in such a way
that the flow of momentum coincides with the fermionic arrows.
Notice that each insertion of $J_{\mu}$ carries a $\gamma_{\mu}$, and we have omitted the diagrams that are zero due to having an odd number of $\gamma$'s in the trace.

The explicit results for the diagrams $D_{0}, \dots,D_{4}$ are collected in Appendix \ref{ape}. After plugging in the critical 
coupling from (\ref{gstarGN}) and normalizing by the free field contribution, we find
\begin{align}
C_{J}^{\textrm{GN}}/C^{\textrm{GN}}_{J,\textrm{free}} &= \frac{D_{0} + D_{1} + D_{2} + 2D_{3} + D_{4}}{D_{0}} \notag \\
&=1 - \frac{\eps}{N-2} - \frac{\eps^2 }{2(N-2)^2}+ \mathcal{O}(\eps^3)\,.
\end{align}
This agrees with our large-$N$ formula (\ref{Cj1GN}) for $C_{J1}^{\textrm{GN}}$ of the critical fermionic theory, 
expanded in $d=2+\eps$ to $\mathcal{O}(\eps^2)$.


The calculation of $C_T$ proceeds similarly to the computation for $C_J$ in the previous section. 
All the diagrams have identical topologies, with the difference that instead of $J$ we insert
the stress-energy tensor (\ref{eq:fermion-t-pp-f}). The two-loop diagram $D_1$ with the same
topology as the one in figure \ref{GNCJ12loop} actually vanishes; see eq.~(\ref{D1-GN}).  
\begin{figure}[h!]
   \centering
\includegraphics[width=16cm]{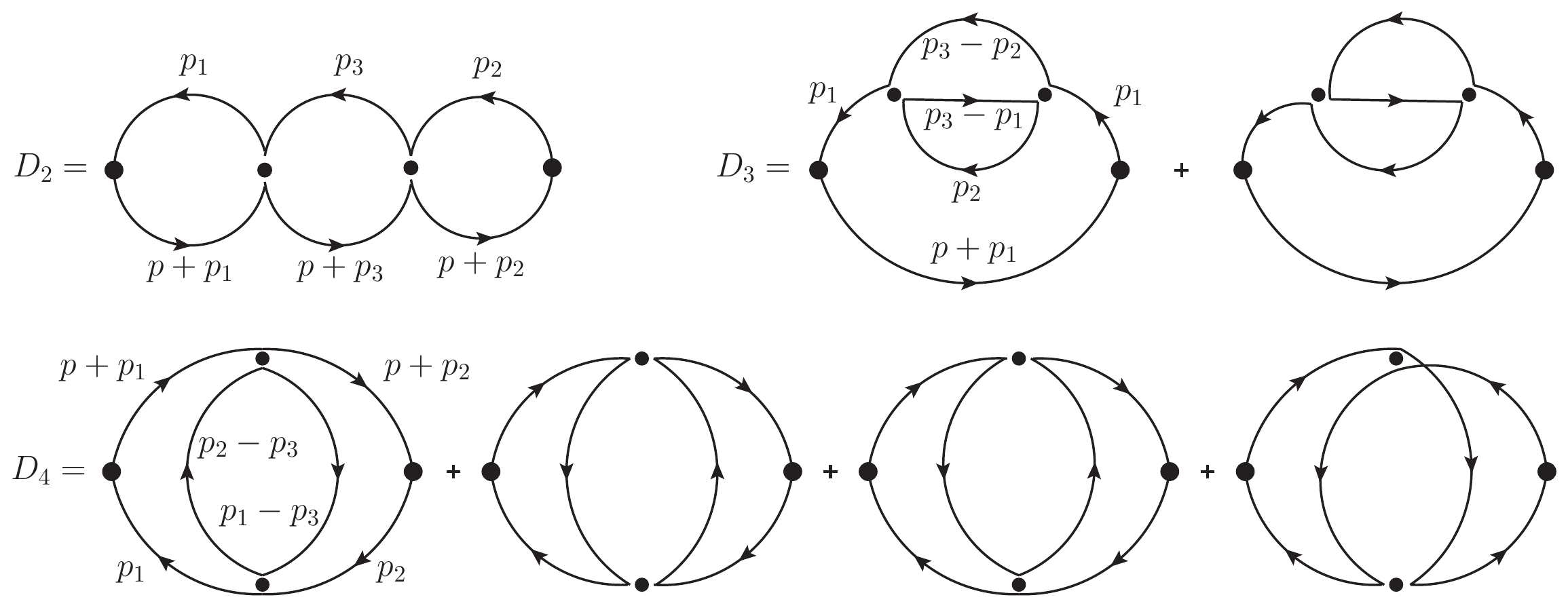}
\caption{Diagrams contributing to $C_{T}$  to order $g^{2}$.}
\label{GNCTdiag3loop}
\end{figure}
Computing the three loop diagrams in 
figure \ref{GNCTdiag3loop} (see Appendix \ref{ape}) and normalizing 
by the free field contribution, we find the following contribution to $C^{\textrm{GN}}_{T}/C^{\textrm{GN}}_{T,\textrm{free}}$:
\begin{align}
\frac{D_{0} + D_{2} + 2D_{3} + D_{4}}{D_{0}}= 1+ g^2 \left (\frac{3(N-1)}{8(2\pi)^{2}} \eps +  \mathcal{O}(\eps^2) \right )
\,. \label{geps2res}
\end{align}
Note that this $\mathcal{O}(g^2)$ term vanishes in $d=2$. Therefore, for $g=g_*$ the leading correction is of order $\eps^3$; 
this is consistent with the vanishing of the $\mathcal{O}(\eps^2)$ term in our large-$N$ result (\ref{ctfermresults}) for $C_{T1}$. 

In order to determine the coefficient of the $\mathcal{O}(\eps^3)$ correction to $C^{\textrm{GN}}_{T}/C^{\textrm{GN}}_{T,\textrm{free}}$ at the critical point, we also need the
$g^{3}$ term, which comes from four-loop Feynman diagrams.  
We will not perform this calculation directly, but rather use a shortcut involving the conformal perturbation theory in $d=2$. 
The GN-model involves the  free Dirac fermions
perturbed by a marginal operator $O= \frac{1}{2}(\bar{\psi}^{i}\psi_{i})^{2}$ with the scaling dimension $\Delta_{O}=2+\mathcal{O}(g)$
\begin{equation}
S = S_{\textrm{free ferm}} +g\int d^{2}x \,O(x)\,. \label{pertaction}
\end{equation}
The Zamolodchikov $c$-function is defined as follows \cite{Zamolodchikov:1986gt, Zamolodchikov:1987ti}: 
\begin{align}
c(g)= C(g) +4\beta(g) H(g) -6 \beta^{2}(g)G(g)\,,
\end{align}
where 
\begin{align}
&C(g) = 2w^{4}\langle T_{ww}(x) T_{ww}(0)\rangle|_{x^{2}=x_{0}^{2}}\,, \notag\\
&H(g)= w^{2}x^{2} \langle T_{ww}(x)O(0)\rangle |_{x^{2}=x_{0}^{2}}\,, \notag\\
&G(g)= x^{4} \langle O(x)O(0)\rangle |_{x^{2}=x_{0}^{2}} \ .\label{defchg}
\end{align}
Here $w=x^{1}+ix^{2}$,  $T_{ww}=T_{11}-T_{22}-2iT_{12}$, and  
\begin{align}
\beta(g) &=  - (N-2)\frac{g^2}{2\pi} +\mathcal{O}(g^{3})\,. \label{betGN}
\end{align}
We notice that 
\begin{align}
C_{T}\propto C(g)\,.
\end{align}
Therefore, to find the $g^{3}$ term in $C_{T}$ we have to find 
the central charge $c(g)$ to order $g^{3}$ and the function $H(g)$ to order $g$. The term $\beta^{2}G$ obviously does not contribute to this order. 
Thus, we have 
\begin{align}
C_{T}(g)\propto c(g)-4\beta(g)H(g)+\mathcal{O}(g^{4})\,.
\end{align}
Let us find $H(g)$ to order $g$.
Using (\ref{pertaction}) we get
\begin{align}
H(g)= gw^{2}x^{2} \int d^{2}y \langle T_{ww}(x)O(0)O(y)\rangle |_{x^{2}=x_{0}^{2}} +\mathcal{O}(g^{2})\,.
\end{align}
To compute this integral it is convenient to use dimensional regularization. We have 
\begin{align}
\langle T_{\mu\nu}(x)O(0)O(y) \rangle =\frac{-C_{TO O}}{(x^{2}(x-y)^{2})^{\frac{d}{2}-1} (y^{2})^{\Delta_{O}-\frac{d}{2}+1}} \Big(X_{\mu}X_{\nu}-\frac{1}{d}\delta_{\mu\nu}X^{2}\Big)\,,
\end{align}
where  $X_{\nu}\equiv x_{\nu}/x^{2}-(x-y)_{\nu}/(x-y)^{2}$. Therefore, we find 
\begin{align}
\int d^{d}y\langle T_{\mu\nu}(x)O(0)O(y) \rangle = -\frac{2C_{TOO}(d-\Delta_{O})}{d(d-2\Delta_{O})}\frac{\pi^{\frac{d}{2}}}{\Gamma(\frac{d}{2}+1)} \frac{1}{(x^{2})^{\Delta_{O}}}\Big(\delta_{\mu\nu}-d \frac{x_{\mu}x_{\nu}}{x^{2}}\Big).
\end{align}
In $d=2$ we obtain
 \begin{align}
H(g)&= gw^{2}x^{2}C_{TOO}  \frac{\pi (2-\Delta_{O})}{1-\Delta_{O}}\frac{\bar{w}^{2}}{(x^{2})^{\Delta_{O}+1}} +\mathcal{O}(g^{2})\notag\\
&=gC_{TOO}\frac{\pi (2-\Delta_{O})}{1-\Delta_{O}}\frac{1}{(x^{2})^{\Delta_{O}-2}} +\mathcal{O}(g^{2})\,.
\end{align}
Since the operator $O$ is marginal, $\Delta_{O}=2+\mathcal{O}(g)$, we have $H(g)=\mathcal{O}(g^{2})$. 
This implies
\begin{align}
C_{T}(g)\propto c(g)+\mathcal{O}(g^{4})\,.
\end{align}
So we can write 
\begin{align}
C_{T}^{d=2}/C_{T}^{\textrm{free}} = 1+ (c(g)-c_{\textrm{free}})/c_{\textrm{free}}= 1+\delta \tilde{F}/\tilde{F}_{\textrm{free}}\,,\label{cttf}
\end{align}
where $\tilde F=-\sin (\pi d/2) F$, and $F$ is the free energy on the $d$-dimensional sphere  \cite{Giombi:2014xxa}.
For a CFT in $d=2$ we have $\tilde F= \pi c/6$; therefore,
$c(g) = 6\tilde{F}(g)/\pi$.
For the free fermion free-energy in $d=2$ we have \cite{Giombi:2014xxa}
\begin{align}
\tilde{F}_{\textrm{free}} = \frac{\pi N}{12}
\end{align}
corresponding to the standard value for the free fermion central charge, $c_{\textrm{free}}=N/2$.
For the change of the free-energy we find \cite{Fei:2015oha,FGKT-to-appear}
\begin{align}
 \delta \tilde{F}=\pi^{3}\int_{0}^{g} G(g) \beta(g)dg\,,
\end{align}
where from (\ref{defchg}) we have $G(g)=N(N-1)/(2(2\pi)^{4})+\mathcal{O}(g)$ and
$\beta(g)=-(N-2)g^{2}/(2\pi)+\mathcal{O}(g^{3})$. Therefore, we obtain for  (\ref{cttf})
\begin{align}
C_{T}^{d=2}/C_{T}^{\textrm{free}} = 1-\frac{\pi (N-1)(N-2)}{(2\pi)^{4}}g^{3}\,.
\end{align}
Thus, in $d=2$ the leading correction is of order $g^3$.
In $d=2+\epsilon$ this term, evaluated at the fixed point  $g^{*}=2\pi\epsilon /(N-2)$, gives a correction of order $\eps^3$.
 Adding this correction to the one coming from the order $g_*^{2}\epsilon$ term (\ref{geps2res})
 we finally find  
\begin{align}
C^{\textrm{GN}}_{T}/C^{\textrm{GN}}_{T,\textrm{free}} &=1-\frac{(N-1)}{8(N-2)^{2}}\eps^3 +\mathcal{O}(\eps^{4})\,.
\label{cteps}\end{align}
In the large $N$ limit this agrees with
(\ref{ctfermresults}), providing a check of our large $N$ calculation. The negative sign of the correction in (\ref{cteps}) means that in $2+\eps$ dimensions the $C_T$ theorem is violated for
the GN model with all $N>2$.

\subsection{Pad\' e approximations}
\label{padeGN}

\begin{figure}[h!]
   \centering
\includegraphics[width=10cm]{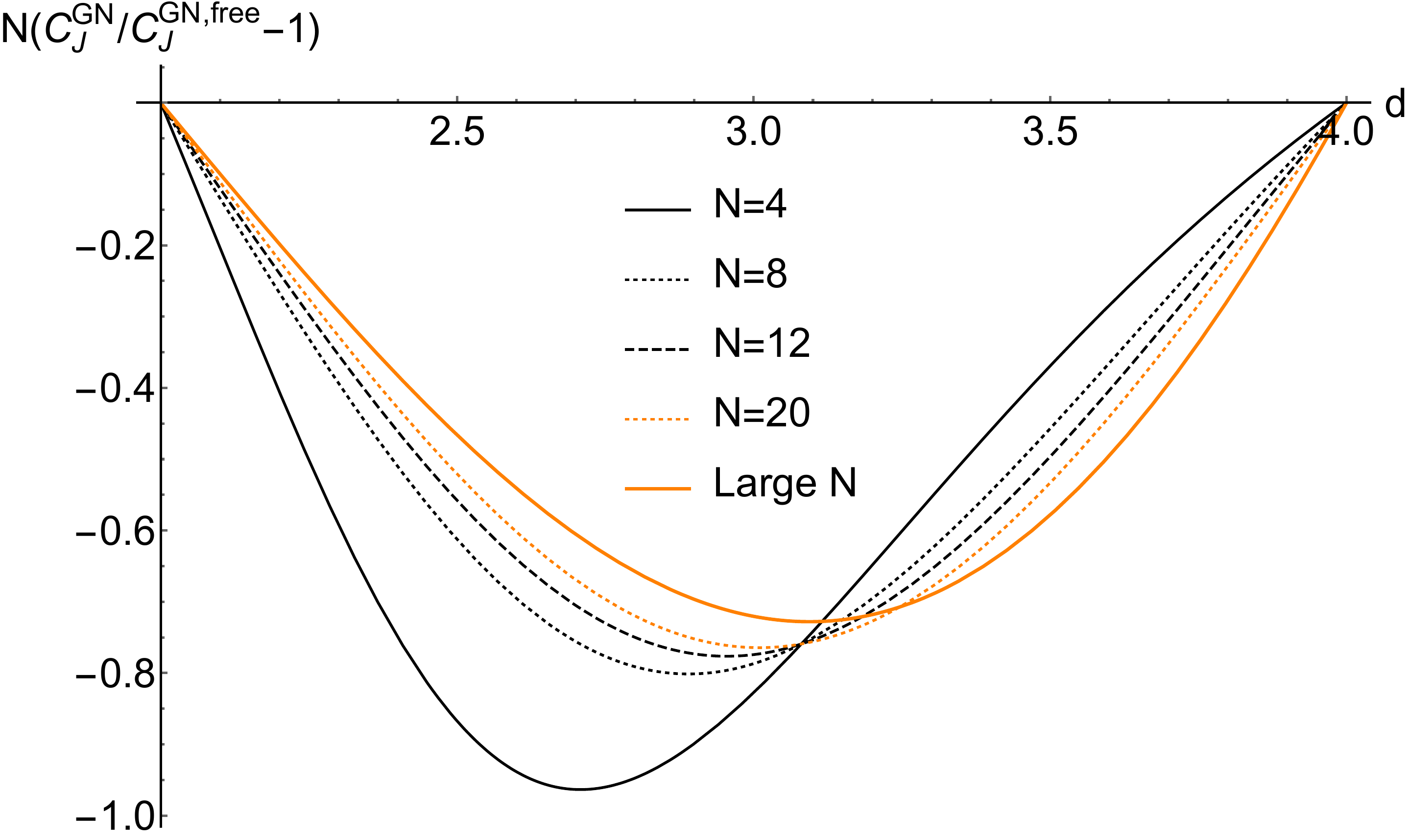}
\caption{Plot of $N(C^{\textrm{GN}}_{J}/C^{\textrm{GN}}_{J,\textrm{free}}-1)$ for $\textrm{Pad\' e}_{[2,2]}$.}
\label{PadeCJFerm}
\end{figure}

We have the following $\epsilon$ expansions for $C^{\textrm{GN}}_{J}/C^{\textrm{GN}}_{J,\textrm{free}}$:
\begin{equation}
C^{\textrm{GN}}_{J}/C^{\textrm{GN}}_{J,\textrm{free}}(d )=
\begin{cases}1 - \frac{\eps}{N-2} - \frac{\eps^2 }{2(N-2)^2}+\mathcal{O}(\epsilon^{3}) &\mbox{in }\quad d=2+\epsilon\,, \\
1-\frac{3 \epsilon }{2 (N+6)}+\mathcal{O}(\epsilon^{2}) &\mbox{in }\quad d=4-\eps\,.
\end{cases}
\end{equation}
In this case we find that only the approximant $\textrm{Pad\' e}_{[2,2]}$ has no poles; it  
approaches the targe $N$ result well. We plot  $\textrm{Pad\' e}_{[2,2]}$ for different $N$ in figure \ref{PadeCJFerm}. We also give the $d=3$ values of 
$C^{\textrm{GN}}_{J}/C^{\textrm{GN}}_{J,\textrm{free}}$ 
 for different $N$ in table \ref{tablePadeFermCJ}. Note that $N$ should be a multiple of 4, since when using the GNY description, we 
take $N=\tilde N {\rm tr}{\bf 1}=4\tilde N$.

\begin{table}[h]
\centering
\begin{tabular}{cccccccc}
\hline
\multicolumn{1}{|c|}{$N=\tilde{N}\Tr (\mathbbm{1})$}         & \multicolumn{1}{c|}{4} & \multicolumn{1}{c|}{8} & \multicolumn{1}{c|}{12} & \multicolumn{1}{c|}{16}& \multicolumn{1}{c|}{20} & \multicolumn{1}{c|}{24}& \multicolumn{1}{c|}{100} \\ \hline
\multicolumn{1}{|c|}{$\textrm{Pad\' e}_{[2,2]}$ }      & \multicolumn{1}{c|}{0.7931} & \multicolumn{1}{c|}{0.9016} & \multicolumn{1}{c|}{0.9355} & \multicolumn{1}{c|}{0.9520} & \multicolumn{1}{c|}{0.9618}& \multicolumn{1}{c|}{0.9683}  & \multicolumn{1}{c|}{0.9925}\\ \hline
\multicolumn{1}{|c|}{$1-\frac{64}{9\pi^{2}N}$ }      & \multicolumn{1}{c|}{0.8199} & \multicolumn{1}{c|}{0.9099} & \multicolumn{1}{c|}{0.9400} & \multicolumn{1}{c|}{0.9550} & \multicolumn{1}{c|}{0.9640}& \multicolumn{1}{c|}{0.9700}  & \multicolumn{1}{c|}{0.9928}\\ \hline
\end{tabular}
\caption{List of $\textrm{Pad\'e}_{[2,2]}$ extrapolations for $C^{\textrm{GN}}_{J}/C^{\textrm{GN}}_{J,\textrm{free}}$ in $d=3$. 
The second line is the large $N$ result (\ref{Cj1GN}) in $d=3$.
}
\label{tablePadeFermCJ}
\end{table}

\begin{figure}[h!]
   \centering
\includegraphics[width=10cm]{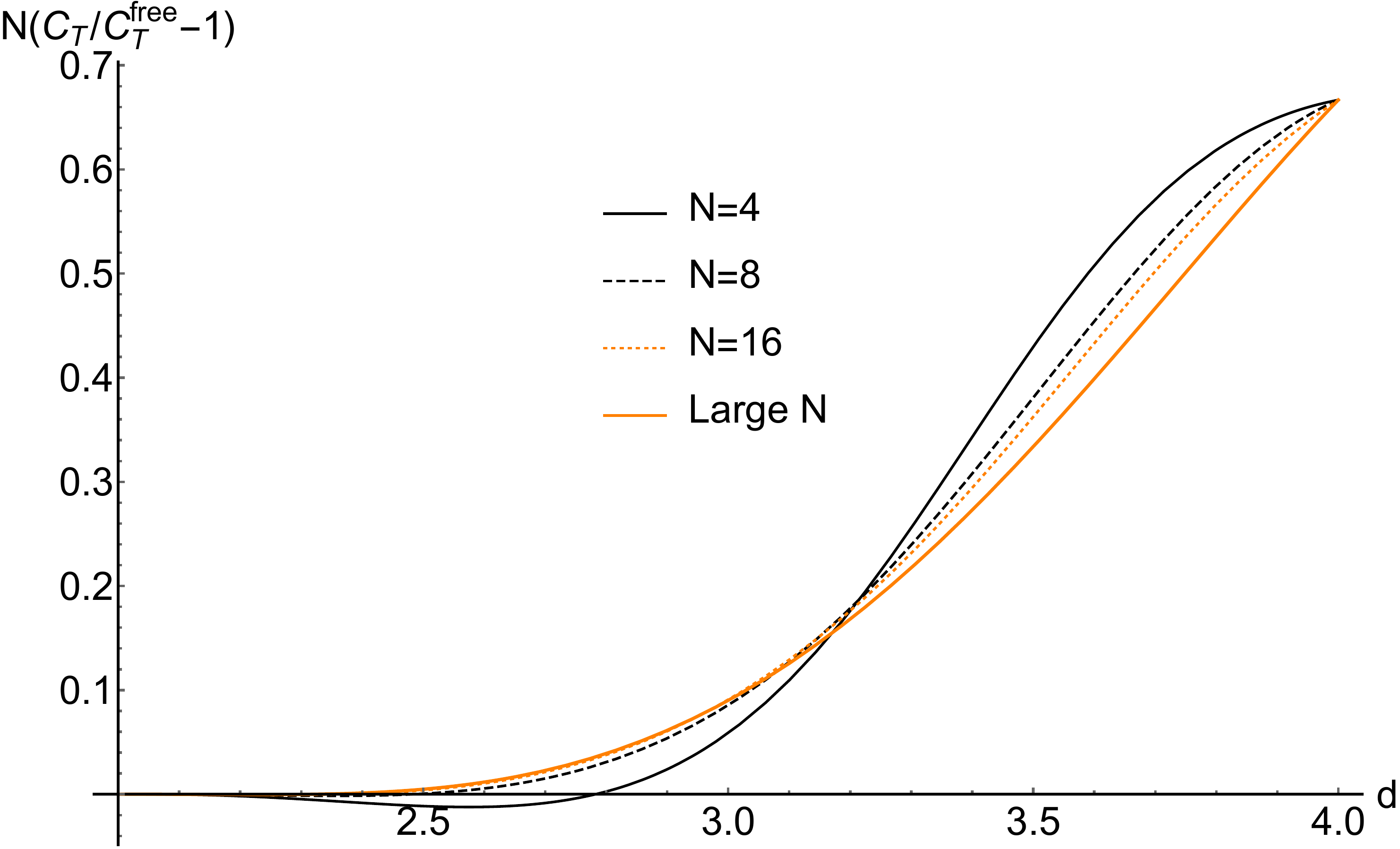}
\caption{Plot of $N(C^{\textrm{GN}}_{T}/C^{\textrm{GN}}_{T,\textrm{free}}(d )-1)$ for $\textrm{Pad\' e}_{\textrm{aver}}\equiv (\textrm{Pad\' e}_{[1,4]}+\textrm{Pad\' e}_{[4,1]})/2$.}
\label{PadeCTFerm}
\end{figure}

We have the following $\epsilon$-expansions for $C^{\textrm{GN}}_{T}/C^{\textrm{GN}}_{T,\textrm{free}}(d)$
\begin{equation}
C^{\textrm{GN}}_{T}/C^{\textrm{GN}}_{T,\textrm{free}}(d )=
\begin{cases} 1-\frac{(N-1)}{8(N-2)^{2}}\eps^3+\mathcal{O}(\epsilon^{4})  &\mbox{in }\quad d=2+\epsilon\,, \\
1+\frac{2}{3N}-\frac{11N-24}{18 N (N+6)} \epsilon+\mathcal{O}(\epsilon^{2}) &\mbox{in }\quad d=4-\eps\,,
\end{cases}
\end{equation}
In this case we find that all two-sided Pad\'e approximants have poles. One reason for this behavior
is the non-monotonicity of the function 
we are trying to approximate. To make our approximation  better, we apply instead the Pad\'e procedure to the following combination 
\begin{align}
f(d)\equiv\Big(\frac{N}{2} \left(C^{\textrm{GN}}_{T}/C^{\textrm{GN}}_{T,\textrm{free}}(d)-1\right) -\frac{1}{d-1}\Big)/\Big(\frac{N}{2}+\frac{1}{d-1}\Big)\,.
\end{align}
This combination is natural from the point of view of the GNY model. It corresponds to writing $C_T^{GNY} = C_{T0}^{GNY}\left(1+f(d)\right)$, 
where $C_{T0}^{GNY} = (\frac{Nd}{2}+\frac{d}{d-1})/S_d^2$ is the contribution of the $\tilde N$ free fermions and the single free scalar. 
We find that $f(d)$ is now a monotonic function at large $N$, and has the $\epsilon$ expansions 
\begin{equation}
f(d)=
\begin{cases} -\frac{2}{N+2}+\frac{2 N \eps}{(N+2)^2}-\frac{2  N^2 \eps^{2}}{(N+2)^3}+\frac{(15 N^5-69 N^4+58 N^3+4 N^2+8 N)\eps^3 }{8 (N-2)^2 (N+2)^4}+\mathcal{O}(\epsilon^{4})  &\mbox{in}\quad d=2+\epsilon\,, \\-\frac{5N \epsilon }{2 (N+6) (3 N+2)}+\mathcal{O}(\epsilon^{2})
 &\mbox{in}\quad d=4-\eps\,,
\end{cases}
\end{equation}
Applying Pad\'e approximation to this function we find that $\textrm{Pad\' e}_{[1,4]}$ and $\textrm{Pad\' e}_{[4,1]}$
do not have poles for $N\geqslant 4$ and are in a good agreement with the large $N$ result. Now we may return to the function $N(C^{\textrm{GN}}_{T}/C^{\textrm{GN}}_{T,\textrm{free}}(d )-1)$.  We plot $\textrm{Pad\' e}_{\textrm{aver}}\equiv (\textrm{Pad\' e}_{[1,4]}+\textrm{Pad\' e}_{[4,1]})/2$ for different $N$ in figure \ref{PadeCTFerm}.  
We also give the $d=3$ values of  $C^{\textrm{GN}}_{T}/C^{\textrm{GN}}_{T,\textrm{free}}$ for different $N$ in table \ref{tablePadeFermCT}.
These values differ from one by only around a percent even for small $N$. We also note that the convergence to the large $N$ limit appears 
to be very fast.
\begin{table}[h]
\centering
\begin{tabular}{cccccccc}
\hline
\multicolumn{1}{|c|}{$N=\tilde{N}\Tr (\mathbbm{1})$}         & \multicolumn{1}{c|}{4} & \multicolumn{1}{c|}{8} & \multicolumn{1}{c|}{12} & \multicolumn{1}{c|}{16}& \multicolumn{1}{c|}{20} & \multicolumn{1}{c|}{24}& \multicolumn{1}{c|}{100} \\ \hline
\multicolumn{1}{|c|}{$\textrm{Pad\' e}_{\textrm{aver}}$ }      & \multicolumn{1}{c|}{1.0147} & \multicolumn{1}{c|}{1.0107} & \multicolumn{1}{c|}{1.0076} & \multicolumn{1}{c|}{1.0057} & \multicolumn{1}{c|}{1.0045}& \multicolumn{1}{c|}{1.0037}  & \multicolumn{1}{c|}{1.0008}\\ \hline
\multicolumn{1}{|c|}{$1+\frac{8}{9 \pi^2 N}$ }      & \multicolumn{1}{c|}{1.0225} & \multicolumn{1}{c|}{1.0113} & \multicolumn{1}{c|}{1.0075} & \multicolumn{1}{c|}{1.0056} & \multicolumn{1}{c|}{1.0045}& \multicolumn{1}{c|}{1.0038}  & \multicolumn{1}{c|}{1.0009}\\ \hline
\end{tabular}
\caption{List of $\textrm{Pad\' e}_{\textrm{aver}}$ extrapolations for $C^{\textrm{GN}}_{T}/C^{\textrm{GN}}_{T,\textrm{free}}$ in $d=3$. 
The second line corresponds to the large $N$ result (\ref{Ct1ferm}) in $d=3$. 
}
\label{tablePadeFermCT}
\end{table}

\bigskip
{\bf Note Added:} After the first version of this paper appeared, we were informed by H. Osborn and A. Stergiou that, via a direct calculation, they obtained values of $C_T$
for the higher-derivative scalar fields that agree with (\ref{dminusfour}) and (\ref{dminustwo}). The latter agreement provides additional evidence in favor of our results for the
Gross-Neveu model.

\section*{Acknowledgments}

We thank Z. Komargodski, F. Kos, H. Osborn, A. Petkou, S. Pufu, K. Sen and A. Stergiou for useful discussions and communications.
The work of SG was supported in part by the US NSF under Grant No.~PHY-1318681.
The work of IRK and GT was supported in part by the US NSF under Grant No.~PHY-1314198.
The work of KD was supported in part by the US NSF Graduate Research Fellowship under Grant No.~DGE 1148900.

\appendix

\section{Tensor reduction} \label{apa}
In this appendix we describe the standard tensor reduction for Feynman integrals in general
dimension (see for example  \cite{Ghinculov:1997pd}).  We use this type of reduction because it
doesn't change the dimension of the integrals, but unfortunately it sometimes adds new denominators
to  the integrals\footnote{This is why in our Aslamazov-Larkin (ladder) type diagrams we have
$a_{9}$  index (see (\ref{genladddiag})).  In order to bring this index to zero we apply a
complicated recursion relation, discussed in Appendix \ref{apb}. }. On the other hand, there is
another type of tensor reduction called Davydychev recursion relations \cite{Davydychev:1991va,
Davydychev:1992xr}. This method does not  add  new denominators to  Feynman integrals, but it
changes their dimension.  This type of reduction was applied in the papers
\cite{Huh:2013vga,Huh:2014eea} for a very similar computations in $d=3$.

Let us first briefly review the main logic. Suppose we are trying to evaluate a $m$-loop Feynman integral with loop momenta $p_{i}$, where $i=1,...,m$, a 
single external momentum $p$, and  $n$ uncontracted Euclidean indices:
\begin{equation}
I^{\mu_{1}...\mu_{n}}(p)=\int_{p_{1},...,p_{m}}  \frac{p_{i_{1}}^{\mu_1}\ldots p_{i_{n}}^{\mu_n}\text{(Numer)}}{\text{(Denom)}}\,,
\label{originalI}
\end{equation}
where $ \int_{p}\equiv \int \frac{d^d p}{(2\pi)^d}$ and  the $\textrm{(Numer)}$ denotes some function of $(p_{i}\cdot p)$, $(p_{i}\cdot p_{j})$ and $p^{2}$.
We would like to convert this into a sum of scalar integrals only. First, we define the components of the loop momenta transverse to the external momentum as:
\begin{align}
p_{i\perp}^{\mu} \equiv p^{\mu}_{i}-\frac{p_{i}\cdot p}{p^2}p^{\mu}\,.
\end{align}
Using this formula in (\ref{originalI}), we get that the original integral $I^{\mu_{1}...\mu_{n}}(p)$ is equal to a sum of integrals of the following form:
\begin{equation}
I_{\perp}^{\mu_1 \cdots \mu_{k}}(p)=\int_{p_{1},...,p_{m}} \frac{p_{j_{1}\perp}^{\mu_1}\ldots p_{j_{k}\perp}^{\mu_{k}}(\text{Numer})}{\text{(Denom)}}\,.
\end{equation}
Now we notice that the tensor $I_{\perp}^{\mu_1 \cdots \mu_{k}}$ is transverse with respect to all its indices:
\begin{equation}
p_{\mu_{l}}I_{\perp}^{\mu_1 \cdots \mu_{l}\cdots \mu_{k}}(p)=0,\quad \textrm{for all}\quad l=1,...,k\,.
\end{equation}
 At the same time $I_{\perp}^{\mu_1 \cdots \mu_{k}}$ can be expressed only from  the external
momentum $p^{\mu}$  and the Kronecker delta  $\delta^{\mu \nu}$. Notice that if $k$ is odd, then the
integral is zero, because for instance there must be a term
$p^{\mu_{1}}\delta^{\mu_{2}\mu_{3}}...\delta^{\mu_{k-1}\mu_{k}}$, and  $I_{\perp}^{\mu_1 \cdots
\mu_{k}}$ cannot be made transverse to $p^{\mu_{1}}$. Therefore, we can focus only on even $k$.

In this paper we are dealing with  the cases of $k=2$ and $k=4$. Let us start with the case of $k=2$, so we have 
\begin{equation}
I_{\perp}^{\mu_1  \mu_{2}}(p)=\int_{p_{1},...,p_{m}} \frac{p_{j_{1}\perp}^{\mu_1} p_{j_{2}\perp}^{\mu_{2}}(\text{Numer})}{\text{(Denom)}} =(\delta^{\mu_{1}\mu_{2}}-p^{\mu_{1}}p^{\mu_{2}}/p^{2})I(p) \,, \label{ik2}
\end{equation}
where $I(p)$ is  some scalar function and $j_{1}, j_{2}$ can be $1,...,m$. Now if we contract (\ref{ik2}) with $\delta_{\mu_1\mu_2}$ we can easily find
 \begin{equation}
I(p)= \frac{1}{d-1}\int_{p_{1},...,p_{m}} \frac{(p_{j_{1}\perp}\cdot p_{j_{2}\perp})(\text{Numer})}{\text{(Denom)}}\,.
\end{equation}
Further reduction to usual scalar integrals can be made by using:
\begin{align}
p_{i\perp}\cdot p_{j\perp} = p_{i}\cdot p_{j} - \frac{1}{p^2}(p_{i}\cdot p)(p_{j}\cdot p),\quad 
(p_{i}\cdot p) = \frac{1}{2}((p+p_{i})^{2}-p^{2}-p_{i}^{2})\,. \label{redformulas}
\end{align}
Now consider the case of $k=4$. We have
\begin{align}
I_{\perp}^{\mu_{1}\mu_{2}\mu_{3}\mu_{4}}(p)=&\int_{p_{1},...,p_{m}} \frac{p_{j_{1}\perp}^{\mu_1}p_{j_{2}\perp}^{\mu_2}p_{j_{3}\perp}^{\mu_3}p_{j_{4}\perp}^{\mu_4}(\text{Numer})}{\text{(Denom)}} \notag\\
=&\Big(\delta^{\mu_1\mu_2}p^{\mu_3}p^{\mu_4} +\delta^{\mu_3\mu_4}p^{\mu_1}p^{\mu_2}-p^2 \delta^{\mu_1\mu_2}\delta^{\mu_3\mu_4} - p^{\mu_1}p^{\mu_2}p^{\mu_3}p^{\mu_4}/p^{2}\Big)I_{1}(p) \notag\\
&+\Big(\delta^{\mu_1\mu_3}p^{\mu_2}p^{\mu_4} +\delta^{\mu_2\mu_4}p^{\mu_1}p^{\mu_3}-p^2 \delta^{\mu_1\mu_3}\delta^{\mu_2\mu_4} -  p^{\mu_1}p^{\mu_2}p^{\mu_3}p^{\mu_4}/p^{2}\Big)I_{2}(p)\notag\\
&+\Big(\delta^{\mu_1\mu_4}p^{\mu_2}p^{\mu_3} +\delta^{\mu_2\mu_3}p^{\mu_1}p^{\mu_4}-p^2 \delta^{\mu_1\mu_4}\delta^{\mu_2\mu_3} - p^{\mu_1}p^{\mu_2}p^{\mu_3}p^{\mu_4}/p^{2}\Big)I_{3}(p)\,, \label{TR}
\end{align}
where  $j_{1},..., j_{4}$ can be $1,...,m$ and  $I_{1}$, $I_{2}$, $I_{3}$ are some scalar functions. The particular combination of tensor structures in front of them are  fixed by the fact that they should vanish when contracted with $p_{\mu_1}$(or $p_{\mu_2}$) and  $p_{\mu_3}$(or $p_{\mu_4}$). These are the only three structures with four Euclidean indices, constructed from $p^{\mu}$ and $\delta^{\mu\nu}$, and transverse with respect to all indices, so this decomposition is general.

Now, if we contract (\ref{TR}) with $\delta_{\mu_1\mu_2}\delta_{\mu_3\mu_4}$, $\delta_{\mu_1\mu_3}\delta_{\mu_2\mu_4}$, and $\delta_{\mu_1\mu_4}\delta_{\mu_2\mu_3}$, we get three equations, which have the solution 
\begin{align}
I_{1} &= \frac{1}{(d^2-1)(d-2)p^2}\notag\\
&\quad \times \int\limits_{p_{1},...,p_{m}}  \frac{\big((p_{j_{1}\perp} p_{j_{3}\perp})(p_{j_{2}\perp} p_{j_{4}\perp})+(p_{j_{1}\perp} p_{j_{4}\perp})(p_{j_{2}\perp} p_{j_{3}\perp})-d(p_{j_{1}\perp} p_{j_{2}\perp})(p_{j_{3}\perp} p_{j_{4}\perp})\big)\text{(Numer)}}{\text{(Denom)}}\,
\end{align}
and $I_{2}$ and $I_{3}$ can be obtained from $I_{1}$ by replacements $j_{2}\leftrightarrow j_{3}$ and $j_{2}\leftrightarrow j_{4}$ correspondingly. Further reduction can be made by using formulas (\ref{redformulas}) and 
finally, everything  reduces to scalar integrals.

\section{Recursion relations} \label{apb}

\begin{figure}[h!]
   \centering
\includegraphics[width=15cm]{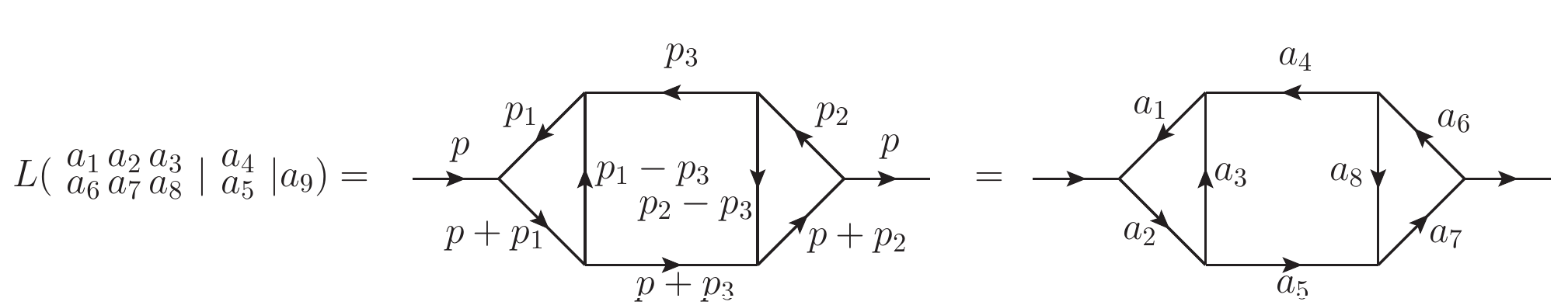}
\caption{Example of a general ladder diagram ( $p_{1}-p_{2}$ and  $a_{9}$ are not included )}
\label{fig1}
\end{figure}

The most difficult part of the calculation is the three-loop ladder (Aslamazov-Larkin) diagram with some non-trivial numerator. After the tensor reduction we are required to compute integrals of the form:
\begin{align}
&L\Big( \begin{array}{c} \\ [-20pt]
a_{1}\, a_{2}\, a_{3} \\[-5pt]
a_{6}\, a_{7}\, a_{8} 
\end{array} 
\Big |\begin{array}{c} \\ [-20pt]
a_{4} \\[-5pt]
a_{5} 
\end{array} \Big | a_9\Big)=\notag\\
&~~=\int_{p_{1},p_{2},p_{3}}\frac{1}{p_{1}^{2a_1}(p+p_{1})^{2a_2}(p_{1}-p_{3})^{2a_3}p_{3}^{2a_4}(p+p_{3})^{2a_5}p_{2}^{2a_6}(p+p_{2})^{2a_7}(p_{2}-p_{3})^{2a_8}(p_{1}-p_{2})^{2a_9}}\,, \label{genladddiag}
\end{align}
where $ \int_{p}\equiv \int \frac{d^d p}{(2\pi)^d}$ and $p$ is the external momentum. 
The indices $a_1$ to $a_8$ correspond to lines shown in Figure \ref{fig1}. Note that $a_9$, which corresponds to the momentum combination $p_{1}-p_{2}$, does not appear in the figure. It is generated by tensor reductions and it can only be a negative integer in our calculation. It is not feasible to evaluate such a large number of diagrams individually. Therefore, we seek to reduce these into a small number of ``master integrals'' through integration by parts relations. However, programs such as FIRE  \cite{Smirnov:2008iw} does not work well when multiple non-integer indices are included. We therefore need to implement our own reduction relations.

We would first like to use some recursion relation to reduce to $a_9=0$. 

\begin{figure}[h!]
   \centering
\includegraphics[width=12.3cm]{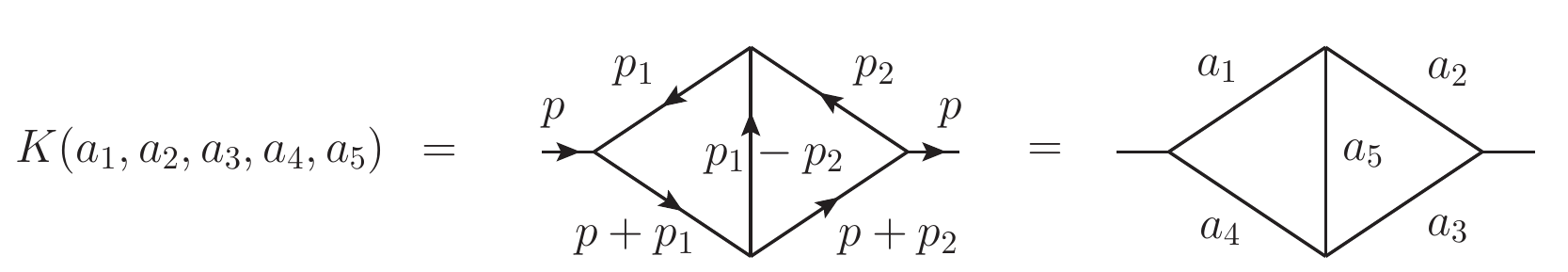}
\caption{Diagrammatic representation of the integral $K(a_{1},a_{2},a_{3},a_{4},a_{5})$.}
\label{KITEdiag}
\end{figure}

\noindent 
The non-trivial general relation to reduce $a_9$ is:
\begin{align}
&L\Big( \begin{array}{c} \\ [-20pt]
a_{1}\, a_{2}\, a_{3} \\[-5pt]
a_{6}\, a_{7}\, a_{8} 
\end{array} 
\Big |\begin{array}{c} \\ [-20pt]
a_{4} \\[-5pt]
a_{5} 
\end{array} \Big | a_9\Big)=  \frac{(d-a_{134689})(d-a_{235789})p^2}{(d-a_{1239}-1)(d-a_{6789}-1)}L\Big( \begin{array}{c} \\ [-20pt]
a_{1}\, a_{2}\, a_{3} \\[-5pt]
a_{6}\, a_{7}\, a_{8} 
\end{array} 
\Big |\begin{array}{c} \\ [-20pt]
a_{4} \\[-5pt]
a_{5} 
\end{array} \Big | a_9+1\Big)\notag\\
&~~~~~~~~~~~~~ -\frac{(d-a_{235789})(5d/2-a_{124567}-2a_{389}-1)}{(d-a_{1239}-1)(d-a_{6789}-1)} L\Big( \begin{array}{c} \\ [-20pt]
a_{1}\, a_{2}\, a_{3} \\[-5pt]
a_{6}\, a_{7}\, a_{8} 
\end{array} 
\Big |\begin{array}{c} \\ [-20pt]
a_{4}-1 \\[-5pt]
a_{5} 
\end{array} \Big | a_9+1\Big)\notag\\
&~~~~~~~~~~~~~  -\frac{(d-a_{134689})(5d/2-a_{124567}-2a_{389}-1)}{(d-a_{1239}-1)(d-a_{6789}-1)} L\Big( \begin{array}{c} \\ [-20pt]
a_{1}\, a_{2}\, a_{3} \\[-5pt]
a_{6}\, a_{7}\, a_{8} 
\end{array} 
\Big |\begin{array}{c} \\ [-20pt]
a_{4} \\[-5pt]
a_{5}-1 
\end{array} \Big | a_9+1\Big)
\notag\\
&~~~~~~~~~~~~~ +\frac{3d/2-a_{123678}-2a_9-1}{d-a_{6789}-1} L\Big( \begin{array}{c} \\ [-20pt]
a_{1}\, a_{2}\, a_{3}-1 \\[-5pt]
a_{6}\, a_{7}\, a_{8} 
\end{array} 
\Big |\begin{array}{c} \\ [-20pt]
a_{4} \\[-5pt]
a_{5}
\end{array} \Big | a_9+1\Big)\notag\\
&~~~~~~~~~~~~~+\frac{3d/2-a_{123678}-2a_9-1}{d-a_{1239}-1} L\Big( \begin{array}{c} \\ [-20pt]
a_{1}\, a_{2}\, a_{3} \\[-5pt]
a_{6}\, a_{7}\, a_{8}-1 
\end{array} 
\Big |\begin{array}{c} \\ [-20pt]
a_{4} \\[-5pt]
a_{5}
\end{array} \Big | a_9+1\Big)\,, \label{relG9}
\end{align}
where $a_{nml...}\equiv a_{n}+a_{m}+a_{l}+\dots$.
The relation (\ref{relG9}) is expected to hold for arbitrary indices.  This relation can be used to reduce all integrals to have $a_9=0$. We will denote  the ladder diagrams with $a_{9}=0$ as 
$L\Big( \begin{array}{c} \\ [-20pt]
a_{1}\, a_{2}\, a_{3} \\[-5pt]
a_{6}\, a_{7}\, a_{8} 
\end{array} 
\Big |\begin{array}{c} \\ [-20pt]
a_{4} \\[-5pt]
a_{5} 
\end{array} \Big)\equiv L\Big( \begin{array}{c} \\ [-20pt]
a_{1}\, a_{2}\, a_{3} \\[-5pt]
a_{6}\, a_{7}\, a_{8} 
\end{array} 
\Big |\begin{array}{c} \\ [-20pt]
a_{4} \\[-5pt]
a_{5} 
\end{array} \Big |0\Big)$.  After the reduction of $a_{9}$ the majority of the integrals can be reduced to two-loop integrals of the form
\begin{align}
K(a_{1},a_{2},a_{3},a_{4},a_{5}) \equiv \int_{p_{1},p_{2}} \frac{1}{p_{1}^{2a_{1}}(p+p_{1})^{2a_{4}}(p_{1}-p_{2})^{2a_{5}}p_{2}^{2a_{2}}(p+p_{2})^{2a_{3}}}\,. \label{Kdef}
\end{align}
This integral is shown in figure \ref{KITEdiag}.  

\bigskip

\noindent There is an extensive literature about different methods for the computation of this type of integrals \cite{Vasiliev:1981dg, Gracey:1992ns, Gracey:1992cp, Gracey:1992ew, Gracey:1990sx, Gracey:2013sz, Chetyrkin:1980pr, Kotikov:1995cw}.
The other diagrams can be reduced to the diagram of type  $L\Big( \begin{array}{c} \\ [-17pt]
1\, 1\, 1 \\[-5pt]
1\, 1\, 1 
\end{array} 
\Big |\begin{array}{c} \\ [-17pt]
\alpha \\[-5pt]
\beta
\end{array} \Big)$, where
\begin{equation}
\alpha = \frac{d}{2}-n+\Delta ,\quad \beta= \frac{d}{2}-m+\Delta \,
\end{equation}
and $n$ and $m$ are some integers. The diagram with $\alpha=\beta=\frac{d}{2}-2+\Delta$ was originally computed in \cite{Vasiliev:1981dg}  and the result reads
\begin{align}
&L\Big( \begin{array}{c} \\ [-17pt]
1\, 1\, 1 \\[-5pt]
1\, 1\, 1 
\end{array} 
\Big |\begin{array}{c} \\ [-17pt]
d/2-2+\Delta \\[-5pt]
d/2-2+\Delta
\end{array} \Big) =\left(p^2\right)^{\frac{d}{2} -2-2 \Delta }\left(\frac{a(1)}{2^2 \pi ^{\frac{d}{2} }}\right)^6\left(\frac{a(\Delta +d/2 -2)}{\pi ^{\mu } 2^{2 (\Delta +d/2 -2)}}\right)^2\frac{\pi ^{\frac{d}{2} } 2^{2 (2 \Delta -\frac{d}{2} +2)}}{a(2 \Delta -\frac{d}{2} +2)} \notag\\
&~~~~~\times \frac{\pi ^{2d }a(2)^3  a(\frac{d}{2} -1)^3 a(d -3)}{\Gamma (\frac{d}{2} )}\Big(\frac{1}{\Delta }+4 B(2)-B(d -3)-3 B(\frac{d}{2} -1)\Big)\,,
\end{align}
where
\begin{align}
a(\alpha)&=\Gamma(\frac{d}{2}-\alpha)/\Gamma(\alpha),\quad \quad B(x)=\psi(x)+\psi(\frac{d}{2}-x)\,.
\end{align}
We consider this integral as the master integral.  All other diagrams  of this type can be related to this master integral using a non-trivial recursion relation\footnote{Notice that for some $\alpha$ and $\beta$ in order to correctly apply this recursion relation one has to take into account $\mathcal{O}(\Delta)$ terms in the integrals. }:
\begin{align}
&L\Big( \begin{array}{c} \\ [-17pt]
1\, 1\, 1 \\[-5pt]
1\, 1\, 1 
\end{array} 
\Big |\begin{array}{c} \\ [-17pt]
\alpha \\[-5pt]
\beta
\end{array} \Big)=\frac{(d-2-\alpha-\beta)(3d/2-4-\alpha-\beta)}{(d-3-\alpha)(d/2-1-\alpha)p^2}L\Big( \begin{array}{c} \\ [-17pt]
1\, 1\, 1 \\[-5pt]
1\, 1\, 1 
\end{array} 
\Big |\begin{array}{c} \\ [-17pt]
\alpha-1 \\[-5pt]
\beta
\end{array} \Big)\notag \\
&~~~~~~~~~~~~-\frac{(d-2-\alpha-\beta)(d-3)}{(d-3-\alpha)(d/2-1-\alpha)p^2}L\Big( \begin{array}{c} \\ [-17pt]
1\, 1\, 0 \\[-5pt]
1\, 1\, 1 
\end{array} 
\Big |\begin{array}{c} \\ [-17pt]
\alpha \\[-5pt]
\beta
\end{array} \Big)\notag \\
&~~~~~~~~~~~+\frac{(2d-5-2\alpha-\beta)(d-3)}{(d-3-\alpha)(d/2-1-\alpha)p^2}L\Big( \begin{array}{c} \\ [-17pt]
0\, 1\, 1 \\[-5pt]
1\, 1\, 1 
\end{array} 
\Big |\begin{array}{c} \\ [-17pt]
\alpha \\[-5pt]
\beta
\end{array} \Big)-\frac{(d-3)}{(d/2-1-\alpha)p^2}L\Big( \begin{array}{c} \\ [-17pt]
1\, 0\, 1 \\[-5pt]
1\, 1\, 1 
\end{array} 
\Big |\begin{array}{c} \\ [-17pt]
\alpha \\[-5pt]
\beta
\end{array} \Big)\,,  \label{rel2}
\end{align}
where $\alpha$, $\beta$ can be arbitrary non-integer and the integrals of the type $L\Big( \begin{array}{c} \\ [-17pt]
0\, 1\, 1 \\[-7pt]
1\, 1\, 1 
\end{array} 
\Big |\begin{array}{c} \\ [-17pt]
\alpha \\[-5pt]
\beta
\end{array} \Big)$ and  etc can be reduced to the $K(a_{1},...,a_{5})$ integrals.

\section{$Z_{T}$ factor calculation for the critical fermion. } \label{apc}
\begin{figure}[h!]
   \centering
\includegraphics[width=3.7cm]{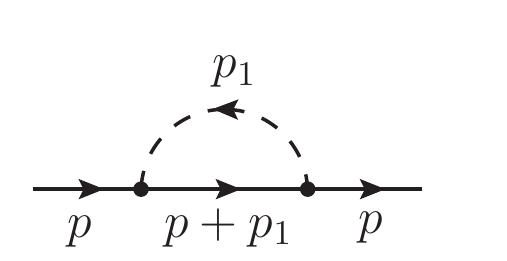}
\caption{One loop correction to the $\langle \psi^{i}(p) \bar{\psi}_{j}(-p)\rangle$ propagator.}
\label{phiphidiag}
\end{figure}

\begin{figure}[h!]
   \centering
\includegraphics[width=8.5cm]{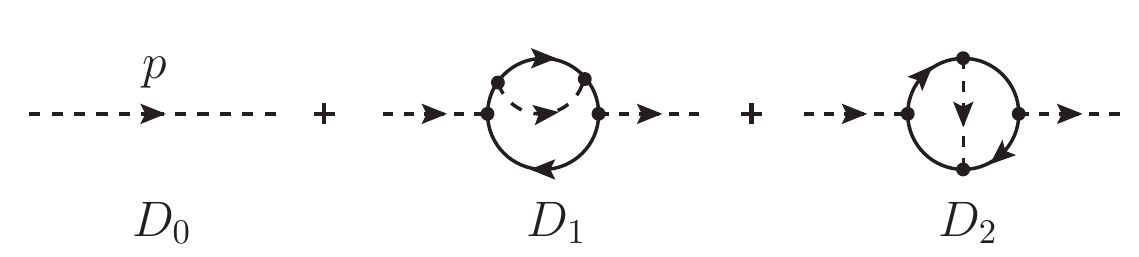}
\caption{Diagrams contributing to   $\langle \sigma_{0}(p)\sigma_{0}(-p)\rangle$ up to order $1/N$.}
\label{sigsigFerm}
\end{figure}

In this appendix we present different methods for the computation of the $Z_{T}$ factor for the stress-energy tensor in the Gross-Neveu model. For what follows, it is important for us to know $\tilde{C}_{\psi 1}$, $\tilde{C}_{\sigma 1}$ and $\eta^{\textrm{GN}}_{1}$, $\kappa^{\textrm{GN}}_{1}$ and $Z_{\psi 1},Z_{\sigma 1}$ defined in (\ref{zpsizsig2}), (\ref{dphidsigma}) and (\ref{cphicsigma}). 
To compute  $\tilde{C}_{\psi 1}$, $\eta^{\textrm{GN}}_{1}$ and $Z_{\psi 1}$ we have to consider the
one loop diagram for the renormalization of the $\langle \psi \bar{\psi} \rangle$ propagator. 
The diagram is depicted in figure \ref{phiphidiag}  
and reads
\begin{align}
D_{1} = \delta^{i}_{j}\frac{\mu^{2\Delta}}{N}\int \frac{d^{d}p_{1}}{(2\pi)^{d}}\frac{\tilde{C}_{\sigma 0}\,i(\slashed{p}+\slashed{p}_{1}) \;}{(p_{1}^{2})^{\frac{d}{2}-1+\Delta}(p+p_{1})^{2}}\,.
\end{align}
Using  the integral  (\ref{intl}) we find 
\begin{align}
\eta_{1}^{\textrm{GN}}= Z_{\psi 1}=\frac{ \Gamma (d -1)(\frac{d}{2} -1)^2}{\Gamma (2-\frac{d}{2} ) \Gamma (\frac{d}{2} +1) \Gamma (\frac{d}{2} )^2} \, \label{eta1fermap}
\end{align}
and 
\begin{align}
\tilde{C}_{\psi 1}= \frac{2^{d -1} \sin \big(\frac{\pi d}{2} \big) \Gamma \left( \frac{d+1}{2}\right)}{\pi ^{3/2} (d/2)^2 \Gamma (\frac{d}{2} )}\,.\label{cpsi1}
\end{align}
To find $\tilde{C}_{\sigma}$ and $\Delta_{\sigma}$ and $Z_{\sigma}$ to the $1/N$ order we have to compute the  diagrams for the $\langle \sigma_{0}(p) \sigma_{0}(-p) \rangle$ propagator represented in figure \ref{sigsigFerm}.

\noindent The expressions for the diagrams are\footnote{Note that it is very important  that we do not shift the power in the $\langle \sigma\sigma\rangle$-external lines by $\Delta$! 
One can explain this by noticing that the $\Delta$ shift in a $\langle \sigma\sigma\rangle$-propagator under a Feynman integral is analogous to changing the dimension of the integral $d\to d' =d-2\Delta$, while  keeping intact the power of  $\langle \sigma\sigma\rangle$-propagator \cite{Muta:1976js, Kazakov:2007su}.   } 
\begin{align}
&D_{0}= \frac{\tilde{C}_{\sigma 0}}{(p^{2})^{\frac{d}{2}-1}}\,, \notag\\
&D_{1}= 2\Big(\frac{\tilde{C}_{\sigma 0}}{(p^{2})^{\frac{d}{2}-1}}\Big)^{2}\mu^{2\Delta}\int \frac{d^dp_1d^dp_2}{(2\pi)^{2d}} 
\frac{(-1)\text{Tr}\big((\slashed{p}+\slashed{p}_1)\slashed{p}_1\slashed{p}_2\slashed{p}_1\big)\tilde{C}_{\sigma0}}
{(p+p_1)^2(p_1^2)^2(p_1-p_2)^{2(\frac{d}{2}-1+\Delta)}p_2^2}\,, \notag \\
&D_2 =\Big(\frac{\tilde{C}_{\sigma 0}}{(p^{2})^{\frac{d}{2}-1}}\Big)^{2} \mu^{2\Delta}\int \frac{d^dp_1d^dp_2}{(2\pi)^{2d}}
\frac{ (-1)\text{Tr}\big((\slashed{p}+\slashed{p}_1)(\slashed{p}+\slashed{p}_2)\slashed{p}_2\slashed{p}_1\big)\tilde{C}_{\sigma0}}
{(p+p_1)^2(p+p_2)^2p_1^2(p_1-p_2)^{2(\frac{d}{2}-1+\Delta)}p_2^2}\,
\end{align}
and 
\begin{align}
\langle \sigma(p) \sigma(-p)\rangle=Z_{\sigma}\langle \sigma_{0}(p) \sigma_{0}(-p)\rangle = Z_{\sigma}\big(D_{0}+D_{1}+D_{2}+\mathcal{O}(1/N^{2})\big)\,.
\end{align}
Computing these diagrams one finds
\begin{align}
Z_{\sigma1}=\frac{4^{\frac{d}{2} } \sin (\pi d/2 ) \Gamma \left(\frac{d+1}{2} \right)}{\pi ^{3/2} \Gamma (\frac{d}{2} +1)}, \quad \Delta_{\sigma}=1+ \frac{4^{\frac{d}{2} } \sin (\pi d/2 ) \Gamma \left(\frac{d+1}{2} \right)}{\pi ^{3/2} \Gamma (\frac{d}{2} +1)}\frac{1}{N}+\mathcal{O}(1/N^{2})\,
\end{align}
and 
\begin{align}
\tilde{C}_{\sigma}=\tilde{C}_{\sigma0} \bigg(1-\frac{1}{N}\eta_{\sigma1}\Big(\mathcal{C}_{\textrm{GN}}(d)+\frac{4(d-1)}{d(d-2)}\Big)+\mathcal{O}(1/N^{2})\bigg)\,,
\end{align}
 where $\Delta_{\sigma}=1+\eta_{\sigma}$ and $\eta_{\sigma}=\eta_{\sigma1}/N+\mathcal{O}(1/N^{2})$ and $\eta_{\sigma}=-\eta^{\textrm{GN}}-\kappa^{\textrm{GN}}$.

\begin{figure}[h!]
   \centering
\includegraphics[width=15cm]{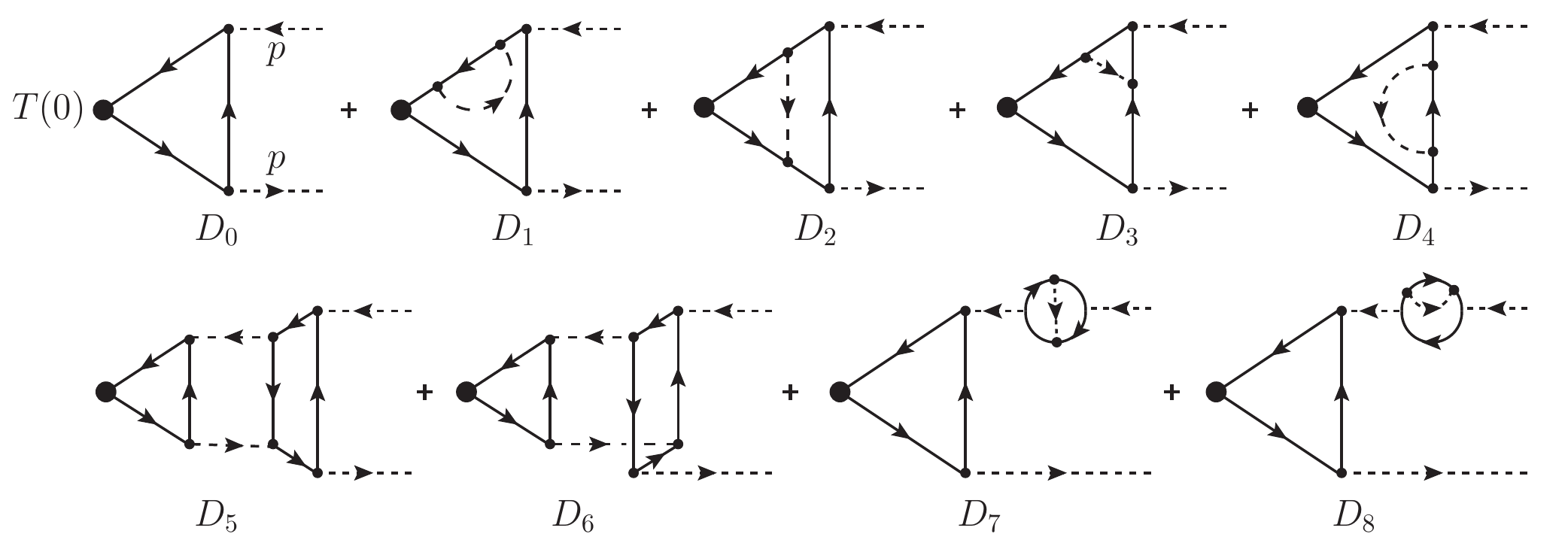}
\caption{Diagrams contributing to   $\langle T(0)\sigma_{0}(p)\sigma_{0}(-p)\rangle$ up to order $1/N$.}
\label{TsigsigFerm}
\end{figure}

We recall that 
the ``bare'' stress-energy tensor $T_{\mu\nu}$ is related to ``renormalized" one $T^{\textrm{ren}}_{\mu\nu}$ as
\begin{align}
T^{\textrm{ren}}_{\mu\nu}(x) =Z_{T}T_{\mu\nu}(x)\,,
\end{align}
where  $Z_{T}= 1+(Z_{T1}/\Delta+Z'_{T1})/N+\mathcal{O}(1/N^{2})$.
Let us first use the three-point function $\langle T^{\textrm{ren}}_{\mu\nu}(x_{1})\sigma(x_{2})\sigma(x_{3})\rangle$ to determine $Z_{T}$ at $1/N$ order. Using conformal invariance  and stress-energy tensor conservation one has the general expression for the three-point  function 
\begin{align}
\langle T^{\textrm{ren}}_{\mu\nu}(x_{1}) \sigma(x_{2})\sigma(x_{3}) \rangle =\frac{-C_{T\sigma \sigma}}{(x_{12}^{2}x_{13}^{2})^{\frac{d}{2}-1} (x_{23}^{2})^{\Delta_{\sigma}-\frac{d}{2}+1}} \Big((X_{23})_{\mu}(X_{23})_{\nu}-\frac{1}{d}\delta_{\mu\nu}(X_{23})^{2}\Big)\,, \label{Tsigsig}
\end{align}
where 
\begin{align}
(X_{23})_{\nu} = \frac{(x_{12})_{\nu}}{x_{12}^{2}} - \frac{(x_{13})_{\nu}}{x_{13}^{2}} \,
\end{align}
and the Ward identity can be used to  relate $C_{T\sigma\sigma}$ with $C_{\sigma}$
\begin{align}
C_{T \sigma \sigma } = \frac{1}{S_{d}} \frac{d\Delta_{\sigma}}{d-1}C_{\sigma}\,, \label{TsigsigiCsig}
\end{align}
where $\Delta_{\sigma}$ is the anomalous dimension of the filed $\sigma$ and $C_{\sigma}$ is the two-point constant of $\langle \sigma \sigma \rangle$-propagator in the coordinate space.
Taking the Fourier transform of (\ref{Tsigsig})  and setting the momentum of the stress-energy tensor to zero one finds, in terms of $T=z^{\mu}z^{\nu}T_{\mu\nu}$
\begin{align}
\langle T^{\textrm{ren}}(0) \sigma(p)\sigma(-p) \rangle= (d-2\Delta_{\sigma})\tilde{C}_{\sigma} \frac{p_{z}^{2}}{(p^{2})^{\frac{d}{2}-\Delta_{\sigma}+1}}\,, \label{TsigsigWard}
\end{align}
where $\tilde{C}_{\sigma}$ is the normalization of the two-point function $\langle \sigma \sigma \rangle$ in momentum space. Now we 
can compute the three-point function 
 $\langle T^{\textrm{ren}}(0) \sigma(p)\sigma(-p) \rangle$ directly using Feynman diagrams. We write
  \begin{align}
\langle T^{\textrm{ren}}(0) \sigma(p)\sigma(-p) \rangle = Z_{T}Z_{\sigma}\langle T(0) \sigma_{0}(p)\sigma_{0}(-p) \rangle \label{Tsigsigdiag}
\end{align}
and the diagrams contributing to $\langle T(0) \sigma_{0}(p)\sigma_{0}(-p) \rangle$ up to order $1/N$  are shown in figure \ref{TsigsigFerm}. Note that for some topologies we did not draw explicitly diagrams with the opposite fermion loop direction, but they have to be included.

\bigskip

\noindent Computing these diagrams and equating the expression (\ref{TsigsigWard}) with the diagrammatic result for the expression (\ref{Tsigsigdiag}) we find 
\begin{align}
Z_{T1} =\frac{2\eta^{\textrm{GN}}_{1}}{d +2},\qquad Z'_{T1}=\frac{8\eta^{\textrm{GN}}_{1}}{(d +2) (d -2)} \,,  \label{Ztfermap}
\end{align}
where $\eta^{\textrm{GN}}_{1}$ is given in (\ref{eta1fermap}).

\begin{figure}[h!]
   \centering
\includegraphics[width=16cm]{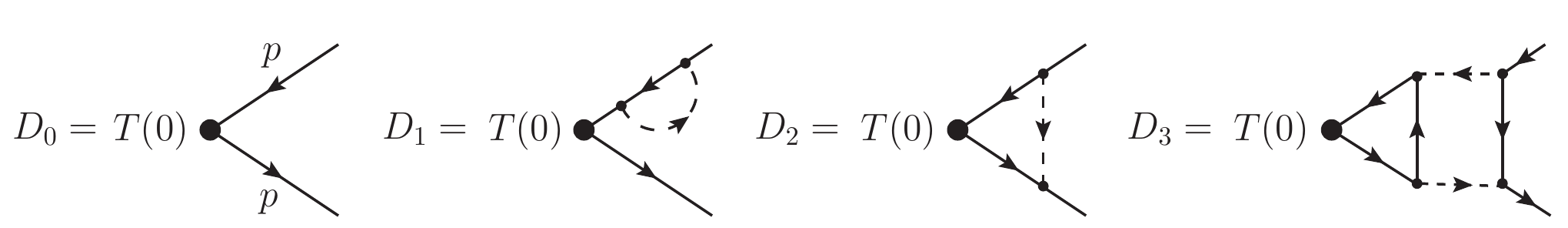}
\caption{Diagrams contributing to $ \langle T(0) \psi_{0}(p)\bar{\psi}_{0}(-p) \rangle$ up to order $1/N$.}
\label{Tpsipsidiag}
\end{figure}

Alternatively, we can consider the three-point function  $\langle T^{\textrm{ren}}_{\mu\nu} \psi^{i} \bar{\psi}_{j} \rangle $. 
Unfortunately, as far as we know, the general form of it in the coordinate space in general $d$ is not known. But from general analysis and from our diagrammatic results we argue that in momentum space and setting $T$ at zero momentum, it has the form\footnote{Here we fix some field, say $\psi =\psi^{1}$ and  don't  write the flavor index explicitly.} :
\begin{align}
\langle T^{\textrm{ren}}(0) \psi(p)\bar{\psi}(-p) \rangle= i\tilde{C}_{\psi}\Big(\frac{\gamma_{z}p_{z}}{(p^{2})^{\frac{d}{2}-\Delta_{\psi}+\frac{1}{2}}}-(d-2\Delta_{\psi}+1) \frac{\slashed{p}\,p_{z}^{2}}{(p^{2})^{\frac{d}{2}-\Delta_{\psi}+\frac{3}{2}}}\Big)\,. \label{TpsipsiWard}
\end{align}
On the other hand we can compute $\langle T(0) \psi_{0}(p)\bar{\psi}_{0}(-p) \rangle$ directly by
Feynman diagrams 
\begin{align}
\langle T^{\textrm{ren}}(0) \psi(p)\bar{\psi}(-p) \rangle = Z_{T} Z_{\psi} \langle T(0) \psi_{0}(p)\bar{\psi}_{0}(-p) \rangle\,, \label{TpsipsiTpsipsi}
\end{align}
where the diagrams contributing to $\langle T(0) \psi_{0}(p)\psi_{0}(-p) \rangle  = D_{0}+D_{1}+D_{2}+D_{3} +\mathcal{O}(1/N^{2})$ are given in figure \ref{Tpsipsidiag} and read
\begin{align}
&D_{0}= \frac{i\slashed{p}}{p^{2}} \frac{i}{2}(2p_{z})\gamma_{z} \frac{i\slashed{p}}{p^{2}} = i\big(\frac{\gamma_{z}p_{z}}{p^{2}}-\frac{2\slashed{p} p_{z}^{2}}{(p^{2})^{2}}\big)\,, \notag\\
&D_{1} = \frac{2(i)^{5}\mu^{2\Delta}}{N}\int \frac{d^{d}p_{1}}{(2\pi)^{d}} \frac{\slashed{p} \,\slashed{p}_{1}\,\slashed{p}\, p_{z}\gamma_{z}\,\slashed{p}\,\tilde{C}_{\sigma0}}{(p^{2})^{3}p_{1}^{2}(p-p_{1})^{2(\frac{d}{2}-1+\Delta)}}\,,  \notag\\
&D_{2}= \frac{(i)^{5}\mu^{2\Delta}}{N} \int \frac{d^{d}p_{1}}{(2\pi)^{d}} \frac{\slashed{p} \,\slashed{p}_{1}\, p_{1z}\gamma_{z}\,\slashed{p}_{1}\,\slashed{p}\,\tilde{C}_{\sigma 0} }{(p^{2})^{2}(p_{1}^{2})^{2} (p-p_{1})^{2(\frac{d}{2}-1+\Delta)} } \,,\notag \\
&D_{3}= \frac{(-1)(i)^{7}\mu^{4\Delta}}{N} \int \frac{d^{d}p_{1}d^{d}p_{2}}{(2\pi)^{2d}} \frac{\slashed{p} \,(\slashed{p} -\slashed{p}_{2})\slashed{p} \,\tilde{C}_{\sigma 0}^{2}\Tr(\slashed{p}_{1}\, \gamma_{z}p_{z} \slashed{p}_{1} (\slashed{p} _{1}-\slashed{p}_{2}))}{(p^{2})^{2} (p-p_{2})^{2}(p_{2}^{2})^{2(\frac{d}{2}-1+\Delta)}(p_{1}^{2})^{2}(p_{1}-p_{2})^{2}}\,.
\end{align}
Computing these diagrams and using (\ref{TpsipsiWard}) and (\ref{TpsipsiTpsipsi}), we find the same result (\ref{Ztfermap}) obtained above.

\section{$Z_{J}$ factor calculation for the critical fermion. } \label{apd}

\begin{figure}[h!]
   \centering
\includegraphics[width=13cm]{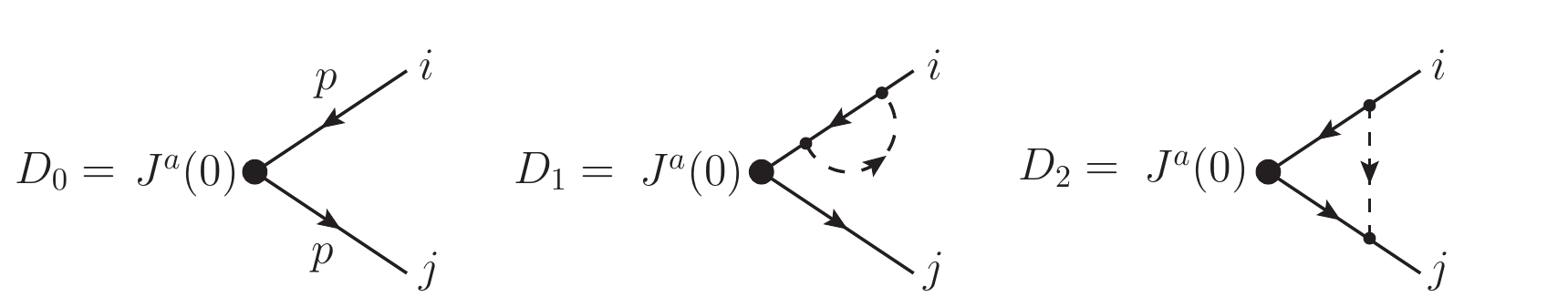}
\caption{Diagrams contributing to $ \langle J^{a}(0) \psi^{i}_{0}(p)\bar{\psi}_{0j}(-p) \rangle$ up to order $1/N$.}
\label{Jphiphibos}
\end{figure}

We can  consider the three-point function $\langle J^{a}_{\mu} \psi^{i}\bar{\psi}_{j}\rangle$, which is fixed by conformal invariance and current conservation \cite{Nobili:1973yu, Petkova:1985uj}
\begin{align}
\langle J^{a}_{\mu}&(x_{1}) \psi^{i}(x_{2}) \bar{\psi}_{j}(x_{3})  \rangle =\notag\\
&= -\bigg( C^{(1)}_{J\psi\bar{\psi}}\frac{ ({\not\,} x_{12} \gamma_{\mu} {\not\,} x_{13})}{(x_{12}^{2})^{\frac{d}{2}}(x_{13}^{2})^{\frac{d}{2}}}+C^{(2)}_{J\psi\bar{\psi}}\frac{(X_{23})_{\mu}({\not\,} x_{23})}{(x^{2}_{12}x^{2}_{13})^{\frac{d}{2}-1}x_{23}^{2}}\bigg)\frac{(t^{a})^{i}_{j}}{(x^{2}_{23})^{\Delta_{\psi}+\frac{1}{2}-\frac{d}{2}}}\,. \label{Jpsipsi}
\end{align}
The Ward identity gives a relation between the structure constants $C^{(1)}_{J\psi\bar{\psi}}$ and $C^{(2)}_{J\psi\bar{\psi}}$
\begin{align}
\langle \delta_{\epsilon} \psi^{i}(x_{2})\bar{\psi}_{j}(x_{3})\rangle = -\epsilon \int d^{d}\Omega r^{d-2}r_{\mu} \langle  J_{\mu}^{a}(x_{1})
\psi^{i}(x_{2})\bar{\psi}_{j}(x_{3}) \rangle, 
\end{align}
where $r_{\mu}= (x_{1}-x_{2})_{\mu}$,  $\int d^{d}\Omega= S_{d} = 2\pi^{d/2}/\Gamma(d/2)$    and $\delta_{\epsilon}\psi^{i}=\epsilon(t^{a})^{i}_{k}\psi^{k}$. 
Performing the integral in the limit $r\to 0$ we find
\begin{align}
C^{(1)}_{J\psi\bar{\psi}} +C^{(2)}_{J\psi\bar{\psi}}= \frac{C_{\psi}}{S_{d}}. \label{JpsipsiC}
\end{align}
Taking the Fourier transform of (\ref{Jpsipsi}) and using (\ref{JpsipsiC}) we get for $J^{\textrm{ren},a}$ at zero momentum
\begin{align}
\langle J^{\textrm{ren},a}(0) \psi^{i}(p)\bar{\psi}_{j}(-p) \rangle=\tilde{C}_{\psi}\Big((d-2\Delta_{\psi}+1)\frac{\slashed{p}p_{z}}{(p^{2})^{\frac{d}{2}-\Delta_{\psi}+\frac{3}{2}}}-\frac{\gamma_{z}}{(p^{2})^{\frac{d}{2}-\Delta_{\psi}+\frac{1}{2}}}\Big)(t^{a})^{i}_{j} \,. \label{JpsipsiWard}
\end{align}
Now to fix $Z_{J}$ we compute $\langle J^{a} \psi \bar{\psi}\rangle$ using Feynman diagrams
\begin{align}
\langle J^{\textrm{ren}, a}(0) \psi^{i}(p)\bar{\psi}_{j}(-p) \rangle = Z_{J} Z_{\psi} \langle J^{a}(0) \psi^{i}_{0}(p)\bar{\psi}_{0j}(-p) \rangle\,, \label{JpsipsiJpsipsi}
\end{align}
and  to $1/N$ order we have three diagrams contibuting to (\ref{JpsipsiJpsipsi}), which are shown in figure \ref{Jphiphibos}. The diagrams read
\begin{align}
&D_{0}= \frac{i\slashed{p}}{p^{2}}(-\gamma_{z})\frac{i\slashed{p}}{p^{2}}(t^{a})^{i}_{j} = \big(\frac{2\slashed{p}p_{z}}{(p^{2})^{2}}-\frac{\gamma_{z}}{p^{2}}\big)(t^{a})^{i}_{j}\,, \notag\\
&D_{1} = \frac{2(i)^{4}\mu^{2\Delta}}{N}\int \frac{d^{d}p_{1}}{(2\pi)^{d}} \frac{\slashed{p} \,\slashed{p}_{1}\,\slashed{p}\, (-\gamma_{z})\,\slashed{p}\,\tilde{C}_{\sigma0}}{(p^{2})^{3}p_{1}^{2}(p-p_{1})^{2(\frac{d}{2}-1+\Delta)}}(t^{a})^{i}_{j}\,,  \notag\\
&D_{2}= \frac{(i)^{4}\mu^{2\Delta}}{N} \int \frac{d^{d}p_{1}}{(2\pi)^{d}} \frac{\slashed{p} \,\slashed{p}_{1}\, (-\gamma_{z})\,\slashed{p}_{1}\,\slashed{p}\,\tilde{C}_{\sigma 0} }{(p^{2})^{2}(p_{1}^{2})^{2} (p-p_{1})^{2(\frac{d}{2}-1+\Delta)} }(t^{a})^{i}_{j}
\end{align}
and
\begin{align}
\langle J^{a}(0) \psi^{i}_{0}(p)\bar{\psi}_{0j}(-p) \rangle  = D_{0}+D_{1}+D_{2} +\mathcal{O}(1/N^{2})\,. 
\end{align}
Computing the diagrams and using (\ref{JpsipsiWard}) and (\ref{JpsipsiJpsipsi}) we find 
\begin{align}
Z_{J} =1+\mathcal{O}(1/N^{2}) \,. \label{zJ}
\end{align}

\newpage 
\section{Integrals and results}  \label{ape}

\subsubsection*{Integrals  for  $C_{J}$ for the $O(N)$ scalar theory in $6-\eps$ (figure \ref{CJepsdiags})}
Explicitly, the diagrams are: 
\begin{align}
D_0 &= \int \frac{d^dp_1}{(2\pi)^d} \frac{(p_{1z} + 2p_z)^2}{p_1^2(p+p_1)^2} \notag\\
&= \frac{\pi^{-\frac{d}{2}}\Gamma(2-\frac{d}{2})\Gamma(\frac{d}{2}-1)^2}{2^d(d-1)\Gamma(d-2)}\frac{p_z^2}{(p^2)^{2-\frac{d}{2}}}\,,\notag\\
D_1 &= 2g_1^2\int \frac{d^dp_1d^dp_2}{(2\pi)^{2d}} \frac{(2p_{1z} + p_z)^2}{(p_1^2)^2(p+p_1)^2(p_1-p_2)^2p_2^2}\notag\\
&=2 g_1^2\frac{16-6d+d^2}{(d-6)(d-4)}\frac{p_z^2}{p^4}I_1 \,, \notag\\
D_2 &= g_1^2\int \frac{d^dp_1d^dp_2}{(2\pi)^{2d}} \frac{(p_{1z} + 2 p_z)(p_{2z} + 2 p_z)}{p_1^2p_2^2(p+p_1)^2(p+p_2)^2(p_1-p_2)^2}\notag\\
&=g_1^2\frac{(32-60d+11d^2-2d^3) I_1+(8-2d)p^2 I_2}{(d-4)^2(d-1)} \frac{p_z^2}{p^4} \,.
\end{align}
We perform tensor reduction to get rid of the $z$ indices, converting each integral into a sum of many scalar integrals with integer indices. Using FIRE \cite{Smirnov:2008iw}, which implements integration by parts relations, we can reduce these into a small number of ``master integrals". In the two loop case, the master integrals $I_1$, and $I_2$ can be easily evaluated:
\begin{equation}
I_1 = l(1,1)l(1,2-\frac{d}{2})\frac{1}{(p^2)^{3-d}}\,, \qquad I_2=l(1,1)l(1,1)\frac{1}{(p^2)^{4-d}}\,,
\end{equation}
where $l(\alpha,\beta)$ is the integral  defined in (\ref{intl}).

\subsubsection*{Integrals  for  $C_{T}$ for the $O(N)$ scalar theory in $d=6-\eps$ (figure \ref{CTscalareps})}
Explicitly, the diagrams are: 
\begin{align}
D_{0}&= \int \frac{d^{d}p_{1}}{(2\pi)^{d}} \frac{(2 p_{1z}( p_{1z}+ p_{z})+c  p_{z}^{2})^{2}}{(p_{1})^{2}(p_{1}+p)^{2}} \notag\\
&=\frac{(d-2)\pi^{-\frac{d}{2}}\Gamma(2-\frac{d}{2})\Gamma(\frac{d}{2}-1)^2}{2^{d+1}(d-1)^2(d+1)\Gamma(d-2)}\frac{p_z^4}{(p^2)^{2-\frac{d}{2}}}\,,\notag\\
D_{1}&= 2 \int \frac{d^{d}p_{1}d^{d}p_{2}}{(2\pi)^{2d}}\frac{(2  p_{1z}( p_{1z}+ p_{z})+c  p_{z}^{2})^{2}}{ (p_1+p)^2(p_1^{2})^2(p_1-p_2)^2p_{2}^{2}} \notag\\
&=\frac{-768+864d-232d^2+36d^3-6d^4+d^5}{6(d-6)(d-4)(d-1)^2(3d-4)}\frac{p_z^4}{p^4}I_1 \,, \notag\\
D_{2}&=\int \frac{d^{d}p_{1}d^{d}p_{2}}{(2\pi)^{2d}}\frac{(2  p_{1z}( p_{1z}+ p_{z})+c  p_{z}^{2})
(2  p_{2z}( p_{2z}+ p_{z})+c  p_{z}^{2})}{ (p_1+p)^2p_{1}^{2}(p_2+p)^2p_{2}^{2}(p_1-p_2)^2} \notag\\
&=\frac{(-768+608d+536d^2-700d^3+238d^4-29d^5+d^6)I_1+(-192+192d-36d^2)p^2I_2}{6(d-4)^2(d-1)^2(d+1)(3d-4)}\frac{p_z^4}{p^4}\,.
\end{align}

\subsubsection*{Integral for the anomalous dimension $\eta$ of $\phi$-field  (figure \ref{phiphidiagON})}
The diagram reads
\begin{align}
D_{1} = \delta^{ij}\frac{\mu^{2\Delta}}{N}\int \frac{d^{d}p_{1}}{(2\pi)^{d}}\frac{\tilde{C}_{\sigma 0}}{(p_{1}^{2})^{\frac{d}{2}-2+\Delta}(p+p_{1})^{2}}\,
\end{align}
and can be easily computed using  the integral (\ref{intl}). 

\subsubsection*{Integrals for $Z_{T}$-factor for the critical scalar  (figure \ref{Tphiphidiagbos})}
\begin{align}
&D_{0}=  \frac{2p_{z}^{2}}{(p^{2})^{2}}\,, \notag\\
&D_{1} = \frac{2\mu^{2\Delta}}{N}\int \frac{d^{d}p_{1}}{(2\pi)^{d}} \frac{\tilde{C}_{\sigma0}2p_{z}^{2}}{(p^{2})^{2}(p+p_{1})^{2}(p_{1}^{2})^{\frac{d}{2}-2+\Delta}}\,,  \notag\\
&D_{2}= \frac{\mu^{2\Delta}}{N} \int \frac{d^{d}p_{1}}{(2\pi)^{d}} \frac{\tilde{C}_{\sigma 0} 2p_{1z}^{2}}{(p^{2})^{2}(p_{1}^{2})^{2} (p-p_{1})^{2(\frac{d}{2}-2+\Delta)} } \,,\notag \\
&D_{3}= \frac{\mu^{4\Delta}}{N} \int \frac{d^{d}p_{1}d^{d}p_{2}}{(2\pi)^{2d}} \frac{\tilde{C}_{\sigma 0}^{2} 2p_{1z}^{2}}{(p^{2})^{2} (p-p_{2})^{2}(p_{2}^{2})^{2(\frac{d}{2}-2+\Delta)}(p_{2}-p_{1})^{2}(p_{1}^{2})^{2}}\,.
\end{align}
These diagrams can be easily calculated with the use of elementary integral (\ref{intl}).

\subsubsection*{Integrals for $Z_{J}$-factor for the critical scalar for  (figure \ref{Jphiphidiag})}
\begin{align}
&D_{0}=  \frac{i2p_{z}}{(p^{2})^{2}}(t^{a})^{ij}\,, \notag\\
&D_{1} = \frac{2\mu^{2\Delta}}{N}\int \frac{d^{d}p_{1}}{(2\pi)^{d}} \frac{\tilde{C}_{\sigma0}i2p_{z}}{(p^{2})^{2}(p+p_{1})^{2}(p_{1}^{2})^{\frac{d}{2}-2+\Delta}}(t^{a})^{ij}\,,  \notag\\
&D_{2}= \frac{\mu^{2\Delta}}{N} \int \frac{d^{d}p_{1}}{(2\pi)^{d}} \frac{\tilde{C}_{\sigma 0} i2p_{1z}}{(p^{2})^{2}(p_{1}^{2})^{2} (p-p_{1})^{2(\frac{d}{2}-2+\Delta)} }(t^{a})^{ij} \,.
\end{align}
These diagrams can be easily calculated with the use of elementary integral (\ref{intl}).

\subsubsection*{Integrals for $C_{J}$ for the critical scalar  (figure \ref{CJdiagsbosons})}
Explicitly, the diagrams are 
\begin{align}
&D_{0} =
\frac{1}{2}\tr(t^{a}t^{b})(i)^{2}\int \frac{d^{d}p_{1}}{(2\pi)^{d}} \frac{(2p_{1z}+p_{z})^{2}}{p_{1}^{2}(p_{1}+p)^{2}}= \frac{1}{2}\tr(t^{a}t^{b})\frac{4^{1-d } \pi ^{\frac{3-d}{2} }}{\sin (\pi d/2 ) \Gamma \left(\frac{d+1}{2}\right)}  \frac{p_z^2}{(p^{2})^{2-\frac{d}{2}}}\,, \notag\\
&D_{1}=2\cdot\frac{1}{2}\tr(t^{a}t^{b})(i)^{2} \frac{\mu^{2\Delta}}{N}\int \frac{d^{d}p_{1}d^{d}p_{2}}{(2\pi)^{2d}}\frac{\tilde{C}_{\sigma 0}(2p_{1z}+p_{z})^{2}}{ (p_1+p)^2(p_{1}^{2})^{2}(p_1-p_2)^2(p_2^2)^{\frac{d}{2}-2+\Delta}} \notag \\
&~~~=\frac{1}{N}\eta_{1}^{\textrm{O(N)}}D_{0}\bigg(-2\Big(\frac{1}{\Delta}-\log(p^{2}/\mu^{2})\Big)-\Big(2\mathcal{C}_{\textrm{
O(N)}}(d)+\frac{2 \left(10 d^3-47 d^2+56 d-16\right)}{(d-4) (d-2) (d-1) d}\Big)\bigg)\,,   \\
&D_{2}=\frac{1}{2}\tr(t^{a}t^{b})(i)^{2} \frac{\mu^{2\Delta}}{N}\int \frac{d^{d}p_{1}d^{d}p_{2}}{(2\pi)^{2d}}\frac{\tilde{C}_{\sigma 0}(2p_{1z}+p_{z})(2p_{2z}+p_{z})}{ (p_1+p)^{2}p_{1}^{2}(p_2+p)^{2}p_{2}^{2}(p_1-p_2)^{2(\frac{d}{2}-2+\Delta)}} 
\notag \\
&~~~=\frac{1}{N}\eta_{1}^{\textrm{O(N)}}D_{0}\bigg(2\Big(\frac{1}{\Delta}-\log(p^{2}/\mu^{2})\Big)+\Big(2\mathcal{C}_{\textrm{
O(N)}}(d)+\frac{2 (2 d-5) (3 d-4)}{(d-4) (d-2) (d-1)}\Big)\bigg)\,. \notag
\end{align}

\subsubsection*{Integrals for $C_{T}$ for the critical scalar (figure \ref{CTdiagsbosons})}
Explicitly, the diagrams are: 
\begin{align}
&D_{0} =\frac{N}{2}\int \frac{d^{d}p_{1}}{(2\pi)^{d}} \frac{(2p_{1z}(p_{1z}+ p_{z})+c  p_{z}^{2})^{2}}{p_{1}^{2}(p_{1}+p)^{2}} \notag\\
&~~~=N\frac{ (\frac{d}{2} -1) \Gamma (2-\frac{d}{2} ) \Gamma (\frac{d}{2} -1)^2 }{ 2(4\pi)^{\frac{d}{2} }(d -1)^2 (d +1)  \Gamma (d -2)}\frac{p_z^4}{(p^{2})^{2-\frac{d}{2}}}\,, \notag\\
&D_{1}= \mu^{2\Delta}\int \frac{d^{d}p_{1}d^{d}p_{2}}{(2\pi)^{2d}}\frac{\tilde{C}_{\sigma 0}(2p_{1z}(p_{1z}+ p_{z})+c  p_{z}^{2})^{2}}{ (p_1+p)^2(p_1^{2})^{2}(p_1-p_2)^2(p_2^2)^{\frac{d}{2}-2+\Delta}} \notag \\
&~~~=\eta_{1}^{\textrm{O(N)}}D_{0}\bigg(-2\Big(\frac{1}{\Delta}-\log(p^{2}/\mu^{2})\Big)-2\Big(\mathcal{C}_{\textrm{
O(N)}}(d)+\frac{11 d^4-45 d^3+26 d^2+36 d-16}{(d-4) (d-2) (d-1) d (d+1)}\Big)\bigg)\,,   \notag\\
&D_{2}= \frac{\mu^{2\Delta}}{2}\int \frac{d^{d}p_{1}d^{d}p_{2}}{(2\pi)^{2d}}\frac{\tilde{C}_{\sigma 0}(2p_{1z}(p_{1z}+ p_{z})+c  p_{z}^{2})(2p_{2z}(p_{2z}+ p_{z})+c  p_{z}^{2})}{ (p_1+p)^{2} p_{1}^{2}(p_2+p)^2p_{2}^{2}(p_1-p_2)^{2(\frac{d}{2}-2+\Delta)}} \notag  \\
&~~~=\eta_{1}^{\textrm{O(N)}}D_{0}\bigg(2\Big(\frac{1}{\Delta}-\log(p^{2}/\mu^{2})\Big)\Big(\frac{d-2}{d+2}\Big)+2\Big(\frac{d-2}{d+2}\mathcal{C}_{\textrm{
O(N)}}(d)+\frac{3 \left(3 d^3-11 d^2+4 d+8\right)}{(d-4) (d-1) (d+1) (d+2)}\Big)\bigg)\,,   
\end{align}
and
\begin{align}
&D_{3}= \frac{\mu^{4\Delta}}{2}\int \frac{d^{d}p_{1}d^{d}p_{2}d^{d}p_{3}}{(2\pi)^{3d}}\frac{\tilde{C}^{2}_{\sigma 0}(2p_{1z}(p_{1z}+ p_{z})+c  p_{z}^{2})(2p_{2z}(p_{2z}+ p_{z})+c  p_{z}^{2})}{ p_1^2(p+p_1)^2(p_1-p_3)^2(p_{3}^{2})^{\frac{d}{2}-2+\Delta}(p_3+p)^{2(\frac{d}{2}-2+\Delta)}p_2^2(p+p_2)^2(p_2-p_3)^2}  \notag \\
&~~~=\eta_{1}^{\textrm{O(N)}}D_{0}\bigg(\Big(\frac{1}{\Delta}-2\log(p^{2}/\mu^{2})\Big)\Big(\frac{4}{d+2}\Big)+\Big(\frac{4}{d+2}\mathcal{C}_{\textrm{
O(N)}}(d)+\frac{2 \left(d^4+18 d^3-93 d^2+66 d+56\right)}{(d-4) (d-2) (d-1) (d+1) (d+2)}\Big)\bigg)\,, \notag
\end{align}
where $c\equiv \frac{d-2}{2(d-1)}$.

\subsubsection*{Integrals for $C_{J}$ for the critical fermion  (figure \ref{CJdiagsfermion})}
The integrals are
\begin{align}
&D_0 = (-1)(i)^{2}\tr(t^{a}t^{b})\int
\frac{d^dp_{1}}{(2\pi)^d}\frac{\text{Tr}\big(\slashed{p}\gamma_z(\slashed{p}+\slashed{p}_1)\gamma_z\big)}{p^2(p+p_1)^2}=\frac{\tr(t^{a}t^{b}){\rm Tr}{\bf 1} \pi ^{1-\frac{d}{2} }  \Gamma (\frac{d}{2} )}{4^{\frac{d}{2} }(d -1) \Gamma (d -2)\sin (\pi d/2)}  \frac{p_z^2}{(p^{2})^{2-\frac{d}{2}}}\,, \notag\\
&D_1 =\frac{2(-1)(i)^{4}\tr(t^{a}t^{b})\mu^{2\Delta}}{N} \int \frac{d^dp_1d^dp_2}{(2\pi)^{2d}} 
\frac{\text{Tr}\big(\gamma_z(\slashed{p}+\slashed{p}_1)\gamma_z\slashed{p}_1\slashed{p}_2\slashed{p}_1\big)\tilde{C}_{\sigma0}}
{(p+p_1)^2(p_1^2)^2(p_1-p_2)^{2(\frac{d}{2}-1+\Delta)}p_2^2} \notag \\
&~~~=\frac{1}{N}\eta_{1}^{\textrm{GN}}D_{0}\bigg(-2\Big(\frac{1}{\Delta}-\log(p^{2}/\mu^{2})\Big)-2\Big(\mathcal{C}_{\textrm{GN}}(d)+\frac{5 d^2-10 d+4}{(d-2) (d-1) d}\Big)\bigg)\,, \\
&D_2 =\frac{(-1)(i)^{4}\tr(t^{a}t^{b})\mu^{2\Delta}}{N} \int \frac{d^dp_1d^dp_2}{(2\pi)^{2d}}
\frac{ \text{Tr}\big(\gamma_z(\slashed{p}+\slashed{p}_1)(\slashed{p}+\slashed{p}_2)\gamma_z\slashed{p}_2\slashed{p}_1\big)\tilde{C}_{\sigma0}}
{(p+p_1)^2(p+p_2)^2p_1^2(p_1-p_2)^{2(\frac{d}{2}-1+\Delta)}p_2^2} \notag\\
&~~~=\frac{1}{N}\eta_{1}^{\textrm{GN}}D_{0}\bigg(2\Big(\frac{1}{\Delta}-\log(p^{2}/\mu^{2})\Big)+2\Big(\mathcal{C}_{\textrm{GN}}(d)+\frac{1}{d-1}\Big)\bigg)\,. \notag
\end{align}

\subsubsection*{Integrals for $C_{T}$ for the critical fermion  (figure \ref{CTdiagsfermions})}
The integrals are
\begin{align}
&D_{0} =\frac{(-1)(i)^{4}\tilde{N}}{2^{2}}\int \frac{d^{d}p_{1}}{(2\pi)^{d}} \frac{(2p_{1z}+p_{z})^{2}\Tr(\gamma_{z} (\slashed{p}+\slashed{p}_{1})\gamma_{z} \slashed{p}_{1})}{p_{1}^{2}(p+p_{1})^{2}} \notag\\
&~~~~=\frac{ -N\pi ^{1-\frac{d}{2} }  \Gamma (\frac{d}{2} )}{4^{\frac{d}{2}+1 }(d -1) (d +1) \Gamma (d -2)\sin(\pi d/2)}\frac{p_z^4}{(p^{2})^{2-\frac{d}{2}}}\,, \notag\\
&D_1 =\frac{(-1)(i)^{6}\tilde{N}\mu^{2\Delta}}{2N} \int \frac{d^{d}p_{1}d^{d}p_{2}}{(2\pi)^{2d}} 
\frac{(2p_{1z}+p_{z})^{2}\text{Tr}\big(\gamma_{z}(\slashed{p}+\slashed{p}_1)\gamma_z\slashed{p}_1\slashed{p}_2\slashed{p}_1\big)\tilde{C}_{\sigma0}}
{(p+p_1)^2(p_1^2)^2(p_1-p_2)^{2(\frac{d}{2}-1+\Delta)}p_2^2} \notag\\
&~~~~=\eta_{1}^{\textrm{GN}}D_{0}\bigg(-2\Big(\frac{1}{\Delta}-\log(p^{2}/\mu^{2})\Big)-2\Big(\mathcal{C}_{\textrm{GN}}(d)+\frac{2 \left(3 d^3-4 d^2-2 d+2\right)}{(d-2) (d-1) d (d+1)}\Big)\bigg)\,,  \\
&D_2 =\frac{(-1)(i)^{6}\tilde{N}\mu^{2\Delta}}{2^{2}N}\int \frac{d^{d}p_{1}d^{d}p_{2}}{(2\pi)^{2d}}
\frac{ (2p_{1z}+p_{z})(2p_{2z}+p_{z})\text{Tr}\big(\gamma_{z}(\slashed{p}+\slashed{p}_1)(\slashed{p}+\slashed{p}_2)\gamma_z\slashed{p}_2\slashed{p}_1\big)\tilde{C}_{\sigma0}}
{(p+p_1)^2(p+p_2)^2p_1^2(p_1-p_2)^{2(\frac{d}{2}-1+\Delta)}p_2^2} \notag\\
&~~~~=\eta_{1}^{\textrm{GN}}D_{0}\bigg(2\Big(\frac{1}{\Delta}-\log(p^{2}/\mu^{2})\Big)\Big(\frac{d-2}{d+2}\Big)+2\Big(\frac{d-2}{d+2}\mathcal{C}_{\textrm{GN}}(d)+\frac{2 \left(2 d^2-2 d-1\right)}{(d-1) (d+1) (d+2)}\Big)\bigg)\,. \notag
\end{align}
The three-loop Aslamazov-Larkin contribution is \footnote{We used the fact that in  $D_{3}$ the two diagrams with different orientation of the fermion loop are equal due to the identity $\Tr(\slashed{A}\,\slashed{B}\slashed{C}\slashed{D})=\Tr(\slashed{A}\,\slashed{D}\slashed{C}\slashed{B})={\rm Tr}{\bf 1}(A\cdot B\, C\cdot D+A\cdot D\, B\cdot C -A\cdot C\, B\cdot D)$.} 
\begin{align}
D_3 =&\frac{(-1)^{2}(i)^{8}\tilde{N}^{2}\mu^{4\Delta}}{2N^{2}}\int \frac{d^{d}p_{1}d^{d}p_{2}d^{d}p_{3}}{(2\pi)^{3d}} \notag\\
&\times \frac{ (2p_{1z}+p_{z})(2p_{2z}+p_{z})\text{Tr}\big(\gamma_{z}(\slashed{p}+\slashed{p}_1)(\slashed{p}_{1}-\slashed{p}_3)\slashed{p}_{1})\Tr(\gamma_z\slashed{p}_2(\slashed{p}_{2}-\slashed{p}_3)(\slashed{p}+\slashed{p}_2)\big)\tilde{C}^{2}_{\sigma0}}
{p_{1}^{2}(p+p_1)^2(p_{1}-p_3)^2p_{3}^{2(\frac{d}{2}-1+\Delta)}(p+p_3)^{2(\frac{d}{2}-1+\Delta)}(p_{2}-p_{3})^{2}(p+p_{2})^{2}p_2^2}
\\
=&\eta_{1}^{\textrm{GN}}D_{0}\bigg(2\Big(\frac{1}{\Delta}-2\log(p^{2}/\mu^{2})\Big)\Big(\frac{2}{d+2}\Big)+2\Big(\frac{2}{d+2}\mathcal{C}_{\textrm{GN}}(d)+\frac{2 d (d+5)}{(d-1) (d+1) (d+2)}\Big)\bigg)\,. \notag
\end{align}

\subsubsection*{Integrals for $C_{J}$ for the GNY model in $d=4-\epsilon$  (figure \ref{CJepsdiags2})}
\begin{align}
D_0 &= -\tilde{N}\int
\frac{d^dk}{(2\pi)^d}\frac{\Tr(\slashed{p_1}\gamma_z(\slashed{p}+\slashed{p}_1)\gamma_z)}{p^2(p+p_1)^2} \notag\\
&=N\frac{(d-2)\pi^{-\frac{d}{2}}\Gamma(2-\frac{d}{2})\Gamma(\frac{d}{2}-1)^2}{2^{d+1}(d-1)\Gamma(d-2)}\frac{p_z^2}{(p^2)^{2-\frac{d}{2}}}\,,\notag\\
D_1 &= 2\tilde{N} g_1^2\int \frac{d^dp_1d^dp_2}{(2\pi)^{2d}}
\frac{\Tr(\gamma_z(\slashed{p}+\slashed{p}_1)\gamma_z\slashed{p}_1(\slashed{p}_1-\slashed{p}_2)\slashed{p}_1)}
{(p+p_1)^2(p_1^2)^2(p_1-p_2)^2p_2^2}\notag\\
&=-N g_1^2\frac{4(d-3)I_1}{3(d-4)}\frac{p_z^2}{p^2}\,, \notag\\
D_2 &= \tilde{N} g_1^2\int \frac{d^dp_1d^dp_2}{(2\pi)^{2d}}
\frac{\Tr(\gamma_z(\slashed{p}+\slashed{p}_1)(\slashed{p}+\slashed{p}_2)\gamma_z\slashed{p}_2\slashed{p}_1)}
{(p+p_1)^2(p+p_2)^2p_1^2(p_1-p_2)^2p_2^2}\notag\\
&=N g_1^2\frac{(d-3)(-4 I_1+3p^2 I_2 )}{6(d-1)}\frac{p_z^2}{p^2}\,.
\end{align}

\subsubsection*{Integrals  for $C_{T}$ for the GNY model in $d=4-\epsilon$   (figure \ref{CTGNYeps})}
These integrals are equal to:
\begin{align}
D_1 &= 2\tilde{N} g_1^2 \int
\frac{d^dp_1d^dp_2}{(2\pi)^{2d}}\frac{\frac{1}{4}(2p_{1z}+p_{z})^2\text{Tr}(\gamma_{z}(\slashed{p}+\slashed{p}_1)
\gamma_{z}\,\slashed{p}_1\,\slashed{p}_2\,\slashed{p}_1)}{(p+p_1)^2(p_1^2)^2p_2^2(p_1-p_2)^2}\notag \\
&=-N \frac{(d-3)(8-2d+d^2)I_1}{3(d-4)(3d-4)(3d-2)}\frac{p_z^4}{p^2}\,,\notag\\
D_2 &= \tilde{N} g_1^2 \int
\frac{d^dp_1d^dp_2}{(2\pi)^{2d}}\frac{\frac{1}{4}(2p_{1z}+p_{z})(2p_{2z}+p_{z})\text{Tr}(\gamma_{z}(\slashed{p}+\slashed{p}_1)(\slashed{p}+\slashed{p}_2)
\gamma_{z}\,\slashed{p}_2 \,\slashed{p}_1)}{(p+p_1)^2p_1^2p_2^2(p_1-p_2)^2p_1^2}\notag \\
&=N \frac{(d-3)\left((-32+40d+12d^2-24d^3+4d^4)I_1 + (24-54d+27d^2)p^2I_2 \right)}{24(d-1)^2(d+1)(3d-4)(3d-2)}\frac{p_z^4}{p^2} \,,\notag\\
D_3 &= 2\tilde{N} g_1^2 \int
\frac{d^dp_1d^dp_2}{(2\pi)^{2d}}\frac{\frac{1}{2}(2p_{1z}(p_{1z}+p_{z})+c p_{z}^{2})(2p_{2z}+p_{z})
\text{Tr}(\gamma_{z}\slashed{p}_2(-\slashed{p}_1+\slashed{p}_2)(\slashed{p}+\slashed{p}_2))}
{(p+p_1)^2p_1^2(p_1-p_2)^2(p+p_2)^2p_1^2}\notag\\
&=N\frac{\left( (-48+76d+16d^2-57d^3+18d^4-d^5)I_1+(24-54d+27d^2)p^2I_2 \right)}{12(d-1)^2(d+1)(3d-4)(3d-2)}\frac{p_z^4}{p^2} \,, \notag \\
D_4 &= -\tilde{N} g_1^2 \int
\frac{d^dp_1d^dp_2}{(2\pi)^{2d}}\frac{(2p_{1z}(p_{1z}+p_{z})+c p_{z}^{2})^2
\text{Tr}(\slashed{p}_2(\slashed{p}_1-\slashed{p}_2))}
{(p+p_1)^2(p_1^2)^2p_2^2(p_1-p_2)^2}\notag\\
&=-N \frac{(d-2)^2d(d+2)(d+4)I_1}{24(d-4)(d-1)^2(3d-4)(3d-2)}\frac{p_z^4}{p^2}\,.
\end{align}
As before, we have a factor of $2$  in the diagram $D_1$ to account for the fact
that the loops may renormalize either the top or bottom line.

\subsubsection*{Integrals for $C_{J}$ for the GN model in $d=2+\epsilon$  (figure  \ref{GNCJ12loop} and \ref{GNCJdiag3loop})}
We have:
\begin{align}
D_0 &= \tilde{N} \int
\frac{d^dk}{(2\pi)^d}\frac{\Tr(\slashed{p}\gamma_{z}(\slashed{p}+\slashed{p}_1)\gamma_{z})}{p^2(p+p_1)^2}= - N \frac{\pi^{1-\frac{d}{2}}\csc{(\pi\frac{d}{2})}
\Gamma(\frac{d}{2})}{2^{d}(d-1)\Gamma(d-2)} \frac{p_z^2}{(p^{2})^{2-\frac{d}{2}}} \,,\notag\\
D_1 &= g \tilde{N} \int \frac{d^dp_1d^dp_2}{(2\pi)^{2d}}
\frac{\Tr(\gamma_{z}(\slashed{p}+\slashed{p}_1)(\slashed{p}+\slashed{p}_2)\gamma_{z}\slashed{p}_2\slashed{p}_1)}
{(p+p_1)^2(p+p_2)^2p_1^2p_2^2}=-gN\frac{\pi^{2-d}\csc^2{(\pi\frac{d}{2})}\Gamma(\frac{d}{2})^2}{4^d(1-d)^2\Gamma(d-2)^2} \frac{ p_z^2}{(p^{2})^{3-d}}\,.
\end{align}
\begin{align}
D_{2} =& g^2 \tilde{N} \int\frac{d^dp_1 d^d p_2 d^d p_3}{(2\pi)^{3d}}\frac{\Tr (\gamma_{z} (\slashed{p}+\slashed{p}_1)   (\slashed{p}+\slashed{p}_3)(\slashed{p}+\slashed{p}_2) \gamma_{z}\, \slashed{p}_2\,  \slashed{p}_3\,\slashed{p}_1)}{p_1^2(p_1+p)^2 p_2^2(p_2+p)^2 p_3^2(p_3+p)^2} = g^2 N \frac{(d-2)^3 p^4 p_z^2}{8 (d-1)^3} M_2 \notag\\
D_{3}=&g^2 \tilde{N}^2 \int\frac{d^dp_1 d^d p_2 d^d p_3}{(2\pi)^{3d}} \frac{\Tr(\gamma_{z} (\slashed{p}+\slashed{p}_1) \gamma_{z} \slashed{p}_1 (\slashed{p}_3-\slashed{p}_2) \slashed{p}_1 )\Tr(\slashed{p}_2 (\slashed{p}_3 - \slashed{p}_1)  )}{(p_1^{2})^{2} (p_1+p)^2(p_1-p_3)^2 p_2^2 (p_2-p_3)^2} \notag\\
&-g^2 \tilde{N} \int\frac{d^dp_1 d^d p_2 d^d p_3}{(2\pi)^{3d}} \frac{\Tr(\gamma_{z} (\slashed{p}+\slashed{p}_1) \gamma_{z} \slashed{p}_1  (\slashed{p}_3-\slashed{p}_2) (\slashed{p}_3-\slashed{p}_1)\slashed{p}_2 \slashed{p}_1 )}{(p_1^{2})^{2} (p_1+p)^2(p_1-p_3)^2 p_2^2 (p_2-p_3)^2}\notag\\
=&g^2\frac{N(N-1)(d-2)^2 p_z^2}{4(3d-4)(2d-3)}M_1
\end{align}
and 
\begin{align}
D_{4}=&g^2 \tilde{N}^2 \int\frac{d^dp_1 d^d p_2 d^d p_3}{(2\pi)^{3d}} \frac{\Tr(\gamma_{z} (\slashed{p}+\slashed{p}_1) (\slashed{p}+\slashed{p}_2) \gamma_{z} \slashed{p}_2\, \slashed{p}_1 ) \Tr((\slashed{p}_1-\slashed{p}_3) (\slashed{p}_2-\slashed{p}_3))}{p_1^2(p_1+p)^2(p_1-p_3)^2 p_2^2(p_2+p)^2 (p_2-p_3)^2} \notag\\
&-2g^2 \tilde{N} \int\frac{d^dp_1 d^d p_2 d^d p_3}{(2\pi)^{3d}} \frac{\Tr(\gamma_{z} (\slashed{p}+\slashed{p}_1) (\slashed{p}_1-\slashed{p}_3) (\slashed{p}_2-\slashed{p}_3) (\slashed{p}+\slashed{p}_2)\gamma_{z} \slashed{p}_2 \, \slashed{p}_1)}{p_1^2(p_1+p)^2(p_1-p_3)^2 p_2^2(p_2+p)^2 (p_2-p_3)^2} \notag\\
&-g^2 \tilde{N} \int\frac{d^dp_1 d^d p_2 d^d p_3}{(2\pi)^{3d}}\frac{\Tr(\gamma_{z} (\slashed{p}+\slashed{p}_1) (\slashed{p}_1-\slashed{p}_3)(-\slashed{p}_2) \gamma_{z} (-\slashed{p}-\slashed{p}_2) (-\slashed{p}_2+\slashed{p}_3) \slashed{p}_1 )}{p_1^2(p_1+p)^2(p_1-p_3)^2 p_2^2(p_2+p)^2 (p_2-p_3)^2}\notag\\
=& g^2\frac{N(d-2)^2\left(\left(f_1(d)+N f_2(d)\right)M_1 -12 p^4 (d-3)^2(12-17d+6d^2)(11-9d+N(3d-3))M_3\right)p_z^2}{36(3d-4)^2(d-3)(d-1)^2(2d-3)(3d-8)} \,. \label{GNCJD4}
\end{align}
where:
\begin{align}
f_1(d)&= 86112-260472d+307525d^2-176601d^3+49203d^4-5319d^5\notag\\
f_2(d)&=3(-7776+24912d-30833d^2+18395d^3-5283d^4+585d^5)
\end{align}

\begin{figure}[h!]
   \centering
\includegraphics[width=12cm]{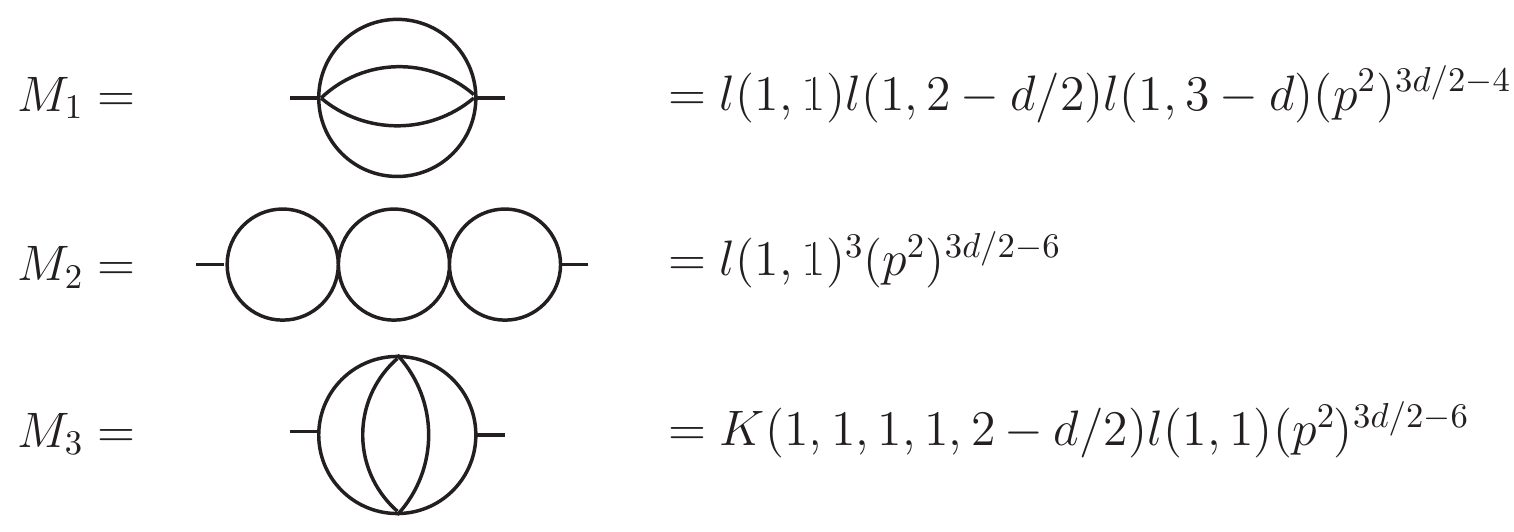}
\caption{Master integrals}
\label{MasterIntegrals}
\end{figure}

After evaluating the traces and performing tensor reduction, each integral becomes a sum of many scalar integrals of the ladder-type with integer indices. Using FIRE \cite{Smirnov:2008iw} to apply integration by parts relations, we can convert all of them into a sum of three master integrals, $M_1$, $M_2$, and $M_3$ as shown in \ref{MasterIntegrals}.

\noindent The first two master integrals are primitive and can be readily evaluated with the use of
the integral (\ref{intl}). The integral $K(1,1,1,1,2-d/2)$ (defined in Appendix B (\ref{Kdef})) in
the third master integral can be evaluated using the Gegenbauer Polynomial technique
\cite{Chetyrkin:1980pr, Kotikov:1995cw}.  Its expansion in $d=2+\eps$ is\footnote{Using the Gegenbauer Polynomial technique  for the integral $K(1,1,1,1,2-d/2)$ we obtain an analytic expression for any $d$. This expression includes a hypergeometric function. To expand the hypergeometric function in $d=2+\eps$ we used the program HypExp \cite{Huber:2005yg}.}:
\begin{align}
&K(1,1,1,1,2-d/2)= \notag\\
&~~~~=\frac{7}{6 \pi ^2 \epsilon ^2}+\frac{14 (\gamma -\log (4 \pi ))-25}{12 \pi ^2 \epsilon }+\frac{84 (\gamma -\log (4 \pi ))^2-300 (\gamma -\log (4 \pi ))-7 \pi ^2-228}{144 \pi ^2}+\mathcal{O}(\eps)\,.
\end{align}

\subsubsection*{Integrals for $C_{T}$ for the GN model in $d=2+\epsilon$ ( figure \ref{GNCTdiag3loop})}
The integrals are
\begin{align}
D_{0}&= \tilde{N} \intdd{p_1} \frac{1}{4}(2p_{1z}+p_{z})^2\frac{\Tr (\gamma_{z} (\slashed{p}+\slashed{p}_1) \gamma_{z} \slashed{p}_1)}{p_1^2(p_1+p)^2}= - N \frac{\pi^{1-\frac{d}{2}}\csc{(\pi\frac{d}{2})}\Gamma(\frac{d}{2})}{4^{\frac{d}{2}+1}(d^2-1)\Gamma(d-2)} \frac{p_z^4}{(p^{2})^{2-\frac{d}{2}}} \,,\notag\\
D_{1}&=-g \tilde{N} \int \frac{d^d p_1 d^d p_2}{(2\pi)^{2d}}\frac{1}{4}(2p_{1z}+p_{z})(2p_{2z}+p_{z}) \frac{\Tr (\gamma_{z} (\slashed{p}+\slashed{p}_1)  (\slashed{p}+\slashed{p}_2) \gamma_{z} \slashed{p}_2 \,\slashed{p}_1)}{p_1^2(p_1+p)^2 p_2^2(p_2+p)^2}=0\,.
\label{D1-GN}
\end{align}
and 
\begin{align}
D_{2} =& g^2 \tilde{N} \int\frac{d^dp_1 d^d p_2 d^d p_3}{(2\pi)^{3d}}\frac{1}{4}(2p_{1z}+p_{z})(2p_{2z}+p_{z})\frac{\Tr (\gamma_{z} (\slashed{p}+\slashed{p}_1)   (\slashed{p}+\slashed{p}_3)(\slashed{p}+\slashed{p}_2) \gamma_{z} \slashed{p}_2 \, \slashed{p}_3\, \slashed{p}_1)}{p_1^2(p_1+p)^2 p_2^2(p_2+p)^2 p_3^2(p_3+p)^2}=0\,, \notag\\
D_{3}=&g^2 \tilde{N}^2 \int\frac{d^dp_1 d^d p_2 d^d p_3}{(2\pi)^{3d}} \frac{1}{4}(2p_{1z}+p_{z})^{2}\frac{\Tr(\gamma_{z} (\slashed{p}+\slashed{p}_1) \gamma_{z} \slashed{p}_1 (\slashed{p}_3-\slashed{p}_2) \slashed{p}_1 )\Tr(\slashed{p}_2 (\slashed{p}_3 - \slashed{p}_1)  )}{(p_1^{2})^{2} (p_1+p)^2(p_1-p_3)^2 p_2^2 (p_2-p_3)^2} \notag\\
&-g^2 \tilde{N} \int\frac{d^dp_1 d^d p_2 d^d p_3}{(2\pi)^{3d}}\frac{1}{4}(2p_{1z}+p_{z})^{2} \frac{\Tr(\gamma_{z} (\slashed{p}+\slashed{p}_1) \gamma_{z} \slashed{p}_1  (\slashed{p}_3-\slashed{p}_2) (\slashed{p}_3-\slashed{p}_1)\slashed{p}_2 \slashed{p}_1 )}{(p_1^{2})^{2} (p_1+p)^2(p_1-p_3)^2 p_2^2 (p_2-p_3)^2}\notag\\
=& g^2\frac{N(N-1)(d-2)^2(2-2d+d^2)}{32(d-1)(2d-3)(2d-1)(3d-4)}M_{1}p_z^4
\end{align}
and 
\begin{align}
D_{4}=&\frac{g^2 \tilde{N}^2}{4} \int\frac{d^dp_1 d^d p_2 d^d p_3}{(2\pi)^{3d}}\frac{(2p_{1z}+p_{z})(2p_{2z}+p_{z})\Tr(\gamma_{z} (\slashed{p}+\slashed{p}_1) (\slashed{p}+\slashed{p}_2) \gamma_{z} \slashed{p}_2\, \slashed{p}_1 ) \Tr((\slashed{p}_1-\slashed{p}_3) (\slashed{p}_2-\slashed{p}_3))}{p_1^2(p_1+p)^2(p_1-p_3)^2 p_2^2(p_2+p)^2 (p_2-p_3)^2} \notag\\
&+\frac{2g^2 \tilde{N}^2}{4} \int\frac{d^dp_1 d^d p_2 d^d p_3}{(2\pi)^{3d}}\frac{(2p_{1z}+p_{z})(2p_{2z}+p_{z})\Tr(\gamma_{z} (\slashed{p}+\slashed{p}_1) (\slashed{p}_{1}-\slashed{p}_3) \slashed{p}_1 ) \Tr(\gamma_{z} \slashed{p}_{2} (\slashed{p}_{2}-\slashed{p}_3) (\slashed{p}+\slashed{p}_2) )}{p_1^2(p_1+p)^2(p_1-p_3)^2 p_2^2(p_2+p)^2 (p_2-p_3)^2} \notag\\
&-\frac{2g^2 \tilde{N}}{4} \int\frac{d^dp_1 d^d p_2 d^d p_3}{(2\pi)^{3d}} \frac{(2p_{1z}+p_{z})(2p_{2z}+p_{z})\Tr(\gamma_{z} (\slashed{p}+\slashed{p}_1) (\slashed{p}_1-\slashed{p}_3) (\slashed{p}_2-\slashed{p}_3) (\slashed{p}+\slashed{p}_2)\gamma_{z} \slashed{p}_2 \, \slashed{p}_1)}{p_1^2(p_1+p)^2(p_1-p_3)^2 p_2^2(p_2+p)^2 (p_2-p_3)^2} \notag\\
&+\frac{g^2 \tilde{N}}{4} \int\frac{d^dp_1 d^d p_2 d^d p_3}{(2\pi)^{3d}}\frac{(2p_{1z}+p_{z})(2p_{2z}+p_{z})\Tr(\gamma_{z} (\slashed{p}+\slashed{p}_1) (\slashed{p}_1-\slashed{p}_3)(-\slashed{p}_2) \gamma_{z} (-\slashed{p}-\slashed{p}_2) (-\slashed{p}_2+\slashed{p}_3) \slashed{p}_1 )}{p_1^2(p_1+p)^2(p_1-p_3)^2 p_2^2(p_2+p)^2 (p_2-p_3)^2}\notag\\
=& g^2 \frac{N(N-1)(d-2)^2 \Big( f_3(d) M_1 + 24p^4(d-3)^2(24-118d+203d^2-144d^3+36d^4)M_3
\Big)}{288(3d-4)^2(d-3)(d-1)(d+1)(2d-3)(2d-1)(3d-8)(3d-2)}p_z^4\,, \label{GNCTD4}
\end{align}
where
\begin{equation}
f_3(d)=-28800+164832d-368444d^2+406366d^3-234072d^4+67473d^5-7884d^6+81d^7
\end{equation}

\bibliographystyle{ssg}
\bibliography{cj-ct}

\end{document}